\documentclass[useAMS,usenatbib,usegraphicx,onecolumn]{mn2e}
\usepackage{times}
\usepackage{amssymb}

\title[On the X-ray emission from massive  star clusters  and their evolving   superbubbles II]{On the X-ray emission from massive star clusters  and their evolving   superbubbles II.   Detailed analytics and observational effects}
\author[G.A. A\~norve-Zeferino, G. Tenorio-Tagle  and S. Silich]{G.A. A\~norve-Zeferino$^{1}$\thanks{E-mail:
ganiorve@inaoep.mx},   G. Tenorio-Tagle$^{1}$  and S. Silich$^{1}$\\
$^1$Instituto Nacional de Astrof\'isica, \'Optica y Electr\'onica,  Apdo. Postal 51 y 216,  72000, Puebla, Pue.,  M\'exico \\ }
\begin{document}

\date{Accepted date. Received date; in original form  date}

\pagerange{\pageref{firstpage}--\pageref{lastpage}} \pubyear{2008}

\maketitle

\label{firstpage}

\begin{abstract}
In this work,  we  present a comprehensive X-ray  picture of the interaction between   a super star   cluster  and the ISM. In order to do that,   we compare and combine the X-ray emission  from the superwind driven by the cluster  with the emission from the wind-blown bubble. Detailed  analytical  models for the hydrodynamics and X-ray luminosity  of  fast polytropic  superwinds are presented.  The superwind X-ray luminosity  models are an extension of the results  obtained  in Paper I.  Here,  the superwind polytropic character   allows to parameterize a wide variety of effects, for instance, radiative cooling. Additionally, X-ray properties  that are valid for all bubble models taking thermal evaporation into account are derived.    The final X-ray picture is obtained   by calculating analytically the  expected  surface brightness and weighted temperature of each component.  All of our  X-ray models have an   explicit dependence on  metallicity  and  admit   general  emissivities as functions of the  hydrodynamical variables.  We consider a realistic X-ray emissivity that separates the contributions from hydrogen and metals.  The paper ends with a comparison  of the models with observational data.  
\end{abstract}


\section{INTRODUCTION}

\begin{figure}
\centering
\includegraphics{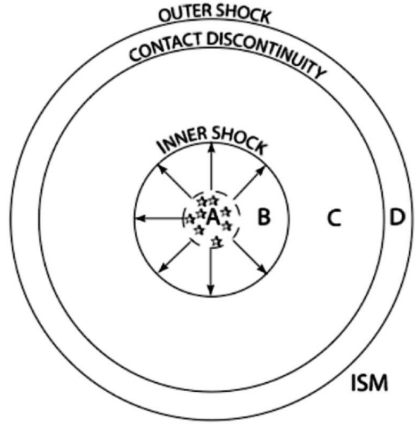}
\caption{Structure of a superbubble.  Each zone  is defined according to its contents: region A contains  the thermalised SNe and stellar wind ejecta,  region B  contains the  free-expanding wind,   region C represents the hot superbubble where the wind thermalised at the inner  shock is confined, and region D is the shell that contains the  swept-up ambient gas. An unperturbed ISM  surrounds the structure.  \label{fig:fig0}}
\end{figure}

Young super star clusters (SSCs) are compact   stellar aggregates resulting from   intense, concentrated, violent star formation episodes. With ages between $\sim 10^6$--$10^7$  yr and masses in the range of  $\sim10^4$--$10^7$ M$_\odot$, they represent the current analogues of globular clusters  and have been identified as  the building blocks of stellar formation in many galaxies (O'connell et al. 1995;  Whitmore \& Schweizer 1995;  Ho 1997; Whitmore et al. 1999; Melo et al. 2005 and references therein). Within its effective radius of up to a few parsecs ($\lesssim 10$-pc, Meurer et al. 1995; Melo et al. 2005), a SSC  contains many  thousands or tens of thousands of early-type  stars (Leitherer \& Heckman 1995) that dominate the injection of mass and energy  ($\sim10^{38}$--$10^{41}$ erg s$^{-1}$)   to the ISM  through the overall contribution of their individual stellar winds and  supernovae explosions.   The interactions that take place among these inner flows  lead to the  thermalization of their kinetic energy through strong shock waves that  generate a high central over-pressure.   As a result, a  powerful superwind is driven out  of the cluster at  large speed ($\gtrsim 1000$ km s$^{-1}$). Outside of the cluster, the superwind expands almost freely until  the presence of the ISM becomes important.  The  subsequent interaction of this supersonic and metal-rich  outflow with the ambient interstellar gas  generates a  structure with a more complex dynamics, namely,  a superbubble (Castor, McCray \& Weaver 1975; Weaver et al. 1977, W77 hereafter; Koo \& McKee 1992a,b; Bisnovatyi-Kogan \& Silich 1995 and references therein). The configuration of a superbubble is traced by two  shock fronts that are inherent to its hydrodynamical evolution: a leading (outer) shock  that sweeps up, compresses   and confines the ambient ISM to an  external shell;  and  a secondary  (inner) shock  that propagates backwards (in the leading shock reference frame) and thermalises the kinetic energy of the free superwind.    The resulting shocked gases  are  separated by a contact discontinuity and the  thermal pressure of the hot, shocked  superwind   drives the outer shell.  Soon, the  mass  of the shocked ambient gas becomes progressively larger than that of the shocked superwind. As this tendency continues, the outer shock becomes radiative, and the  shocked ISM  collapses  into a thin, dense shell.  Being  more tenuous, the bubble hot interior has a larger cooling time scale and thus it remains quasi-adiabatic.    This stage is called the  snowplow phase. Here,  in the same fashion as in Paper I,   the  resulting structure   is  split into  the four zones shown in  Fig. \ref{fig:fig0}:  region A corresponds to  the star cluster volume itself, region B  is the zone where  the  wind  freely expands, region C  is the superbubble that contains the shocked superwind, and   region  D is the outer shell where the leading shock accumulates  ambient gas.

Due to the high temperatures produced by the multiple shock waves and the large amount of mass deposited by the sources, both superwinds and superbubbles are potential diffuse X-ray emitters (Chevalier 1992;  Chu et al. 1995, C95 henceforth; Cant\'o, Raga \& Rodr\'iguez 2000; Stevens \& Hartwell 2003). This represents an opportunity to study  the X-ray emission associated with massive stellar clusters to better understand and constrain the ongoing hydrodynamics and its energetics. With the aid of \emph{Chandra}, it has been found that several clusters in the Local Group emit an important amount of diffuse  X-rays: near the centre of our Galaxy, the X-ray emissions of the Arches and Quintuplet clusters have been detected and studied by Yusef-Zadeh et al. (2002) and Law \& Yusef-Zadeh (2004); that of NGC 3603 --one of the most luminous  clusters in the Milky Way-- has been discussed  by Moffat et al. (2002); and Stevens \& Hartwell (2003) have analysed R136 and NGC 346 in the Magellanic Clouds (LMC and SMC, respectively). On the other hand, Chu \& Mac Low (1990) examined the \emph{Einstein}  archives for LMC and found seven superbubbles  that were related to OB associations. Repeating later  the same analysis using \emph{ROSAT} data, they found four additional but dimmer superbubbles    (C95). More recently, Townsley et al. (2003) have reported diffuse emission from a superbubble associated to M 17, also known as the Omega Nebula; similarly,  Smith \& Wang (2004) have used \emph{XMM-Newton} data to study  the emission from a 100-pc superbubble related to the dense OB association LH90 in 30 Dorados. In distant galaxies the X-ray emission presents two components: a  system of point-like  non-resolved sources, and an extended diffuse component possibly associated to a collection of superbubbles generated by individual SSCs (Summers et al. 2004; Smith, Struck \& Nowak 2005).

This theoretical work  has a  twofold objective. The first one is  to present detailed  hydrodynamic and  X-ray  luminosity  models for fast polytropic superwinds, models that can  be easily modified  to account for  general emissivity functions  (Section \ref{Polywinds}).  This property allowed us  to explicitly  separate  the X-ray  contributions from  hydrogen and metals and to obtain the  plasma weighted temperature.   These results could be of particular use when comparing with synthetic or observed  spectra, since  we obtained them theoretically   taking into consideration bands suitable to  the  instruments on board  \emph{Chandra} and \emph{XMM-Newton}.  Additionally,  we also present analytical formulas for the corresponding    X-ray surface brightness. The second objective is to     undertake the task of  obtaining thoroughly all of the above for  the  standard  analytical bubble models (Section \ref{Bubbles}).  Three properties that are universal for bubbles accounting for shell evaporation  are derived:  an X-ray luminosity scaling factor that is independent  of the particular dynamical evolution,  the shape of the related surface brightness profile, and  the projected X-ray temperature. We examine also the case of when the evaporation of material from the outer shell is prohibited  (Section \ref{Bnev}).   Later, we assemble our models and present  the  complete X-ray panorama (Sections \ref{Pic}). Finally,   we make a comparison  with  observational data  (Section \ref{Comparison}). 

All of our X-ray models are analytical. Moreover, they are independent of the actual shape of the emissivity function  because they involve hypergeometric functions. In the more general case, they can be evaluated using the code of Colavecchia \& Gasaneo (2004); however, that would not be necessary here because  we have constructed   closed-form expansions in terms of  elementary algebraic expressions.   From the theoretical point of view, the freedom of selecting the emissivity function  and the separation of the contributions from hydrogen and metals are advantages, since  it is usual in the literature to assume a  constant X-ray emissivity (C95; Garc\'ia-Segura \& Mac Low 1995) and that the X-ray luminosity scales linearly with metallicity ($\propto Z$). The first  advantage  enables   a more realistic modeling,  and  the second allows  to discriminate the importance of each contribution at a given temperature and metal abundance. Our formalism  also permits  a straightforward calculation of  other   integral properties  such as the  luminosity of the plasma in other bands and the mass of the emitting gas (see Appendix \ref{TLaws}).  On the hydrodynamical side, the polytropic character of the winds makes  it possible to implicitly parameterize the effect of radiative cooling in a simple manner.

Throughout the Paper,  astrophysical units are used when convenient: $t_6$ is the time in Myr, $\dot E_{38}$  is the mechanical luminosity in units of $10^{38}$ erg s$^{-1}$, $R_{\rm sc,pc}$  is the star cluster radius in parsecs, $V_{8}$ is  the terminal speed in units of 1000 km s$^{-1}$,  $T_{\rm keV}$ is the  plasma temperature at the star cluster centre in units of keV and $\Lambda_{\rm X,-23}$ is the  X-ray  emissivity    in units of $ 10^{-23} $ erg s$^{-1}$cm$^3$.  A comprehensive list of  symbols is given in Table \ref{Table4}. The reader not interested in the detailed theoretical derivations may wish to proceed directly to Sections \ref{Pic} and \ref{Comparison}.

\section{POLYTROPIC SUPERWINDS}\label{Polywinds}

{{Stationary superwind models have proven valuable in reproducing the average behaviour of the star-cluster outflows  obtained from   full 3D hydrodynamical simulations. Although the continuous wind-source distributions assumed in the  analytical models do not  produce the detailed features  that might result from the local interaction of   stellar winds and supernovae explosions, they provide a good approximation for the radially averaged hydrodynamical profiles even for clusters with  only $\sim$100 stars and  become an  increasingly   accurate approximation as the cluster mass and stellar density also increase. Furthermore, not to far  away from the cluster the agreement is better than in the inner zone  because the inhomogeneities introduced by the discrete stellar population are quickly smoothed out  (Cant\'o et al.  2000).

After a short  initial transitory period,  both the averaged hydrodynamical profiles and the X-ray luminosity reach almost steady values. The transitory period  lasts until  the cluster core is filled with shocked gas,  which  depends on the cluster radius and   mass and energy injection rates. For instance, in their simulations, Rockefeller et al. (2005) have found this time to be $\sim 2\times 10^4$ yr for the Arches cluster ($R_{\rm sc,pc}\approx 0.2$ pc), whereas for the $\sim$ 5 times larger Quintuplet cluster they found it  to be larger than  $\sim 10^5$  yr. They also found small temporal variations of the X-ray luminosity: $\sim 1$ percent for the Arches cluster and  around $4$--$7$ per cent for the Quintuplet cluster over a period of $\sim 50$ yr. 

All of the  above implies that  stationary models can be used to sequentially approximate the hydrodynamical evolution of a cluster as well as the corresponding diffuse X-ray emission, given that the self-adjustment times between transitions are short and not too frequent in comparison with the cluster age and   the intrinsic temporal  variations of the quasi-steady states are large in comparison with the observing time.  Here we present a new set of stationary wind models and obtain their associated  X-ray luminosity, surface brightness and spectroscopical temperature. Certainly, the presented X-ray models do not  include short-scale variations associated with interacting binaries. These are associated with the point-like X-ray sources, whose luminosity must be removed from the diffuse X-ray component (Moffat et al. 2002). }}

Consider the Chevalier \& Clegg  (1985, CC85 henceforth)    \emph{fast}  superwind model, which  assumes that the  flow is steady, spherically-symmetric,   adiabatic and ideal. Each star cluster can be  defined by a set  $\Pi_\star=\{ R_{\rm sc}, q_{\rm e}, \{q_{\rm m},  V_{\rm \infty A}\}\}$  containing 3  parameters: the cluster radius, $R_{\rm sc}$;   the energy deposition rate per unit volume, $ q_{ \rm e}={\dot E}/{V_{\rm sc}}$; and  the mass deposition rate per unit volume, $ q_{\rm m}={\dot M}/{V_{\rm sc}}$. Alternatively,  the adiabatic terminal speed $V_{\rm \infty A}=({{2 q_{\rm e}}/{q_{\rm m}}})^{1/2}$ may be used. For the interior of the cluster, $r < R_{\rm sc}$ (region A), the  equations of conservation of mass and momentum are

  \begin{equation}
\rho = \frac{q_{\rm m}r}{3u} \label{rhoin} 
\end{equation}
\noindent and
 \begin{equation} \rho u \frac{{\rm{d}} u}{{\rm{d}}r}=-\frac{{\rm{d}}P}{{\rm{d}}r} - q_{\rm m} u,
 \end{equation}

\noindent  respectively. Above, only models with   finite  central densities ($\rho_{\rm c}>0$) are considered, \rm{i.e.}  it is  a requirement that $u_{\rm c}=u(0)=0$ km s$^{-1}$. For region B, the corresponding equations are

 \begin{equation}
\rho = \frac{\dot M}{4\pi u r^2}
\end{equation}

\noindent and
 \begin{equation}
\rho  u \frac{{\rm{d}}u}{{\rm{d}}r }= -\frac{{\rm{d}}P}{{\rm{d}}r}. \label{p_out}
\end{equation}

For both regions, the conservation of energy can be stated as a  steady state Bernoulli-like equation
\begin{equation}
\epsilon=\frac{1}{2} u^2 + \frac{\gamma}{\gamma-1} \frac{P}{\rho} =\frac{1}{2}V_{\rm \infty A}^2,\label{Ber}
\end{equation}

\noindent where $\epsilon$ is the total energy per unit mass. Since the gas is adiabatic and perfect, $\gamma={5}/{3}$.  Let's consider now   a polytropic perfect gas with equation of state

 \begin{equation}P=K\rho^{\frac{\eta+1}{\eta}} =nkT, \label{Poly}
 \end{equation}

\noindent where $\eta$ is the polytropic index and $K>0$  is a proportionality constant.  This equation replaces (\ref{Ber}) in the  polytropic case, where the entropy variations are parametrized through a constant $\eta$. 

The combination of  (\ref{p_out}) and (\ref{Poly}) yields the explicit form of the energy conservation per unit mass for region B

\begin{equation}
\epsilon_{\rm BP}=\frac{1}{2}u^2+(\eta+1) \frac{P}{\rho}=\rm constant. \label{eP}
\end{equation}

\noindent As expected, the polytropic and the CC85 solutions  have the  same algebraic structure on $r\ge R_{\rm sc}$. They are isomorphic through   the transformations $\gamma \rightarrow ( {\eta+1})/{\eta}$ and
$\epsilon_{\rm BP}= ({1}/{2})V_{\rm \infty P}^2 \; \; (V_{\rm \infty A}\rightarrow V_{\rm \infty P})$. The constant $V_{\rm \infty P}$  is arbitrary and is related to the temperature at  $r=R_{\rm sc}$, or, equivalently,  to the speed at which the superwind leaves the star cluster surface. However, this arbitrariness vanishes when additional constraints or boundary conditions    related to the parameterized effects are imposed on the model,  for example,    $V_{\rm \infty P}< V_{\rm \infty A}$ when just dissipative processes are involved.  When more specific information about the  nature of the processes is available,   explicit expressions for the constraints can be given as threshold lines.  For region A,  the isomorphism  is not exactly the case; nevertheless,  as discussed below, equation (\ref{eP}) remains a good approximation there.

A continuous gas acceleration requires of the following condition at $r=R_{\rm sc}$

\begin{equation}
  u_{\rm sc}=\left[\frac{(\eta+1)}{\eta}\frac{P_{\rm sc}}{\rho_{\rm sc}}\right]^{1/2}<V_{\rm \infty P}.\label{uscc}
\end{equation}

The \emph{isomorphic} solutions  are self-similar if written in terms of the dimensionless variables $R={r}/{R_{\rm sc}}$ and $U={u^2}/{V_{\rm \infty P}^2}$

  \begin{equation}
 R = D_1 U^{1/2} \left[ 1+ (6\eta+5) U \right]^{-\frac{4\eta+3}{6\eta+5}},\; \;  r<R_{\rm sc}, \label{polyadA}
 \end{equation}
 \begin{equation}
 R= D_2U^{-1/4}\left( 1-U  \right)^{-\frac{\eta}{2}},\; \;  \; r\ge R_{\rm sc}.\label{polyout}
  \end{equation}

   \begin{figure*}
\centering

\includegraphics[height=50mm]{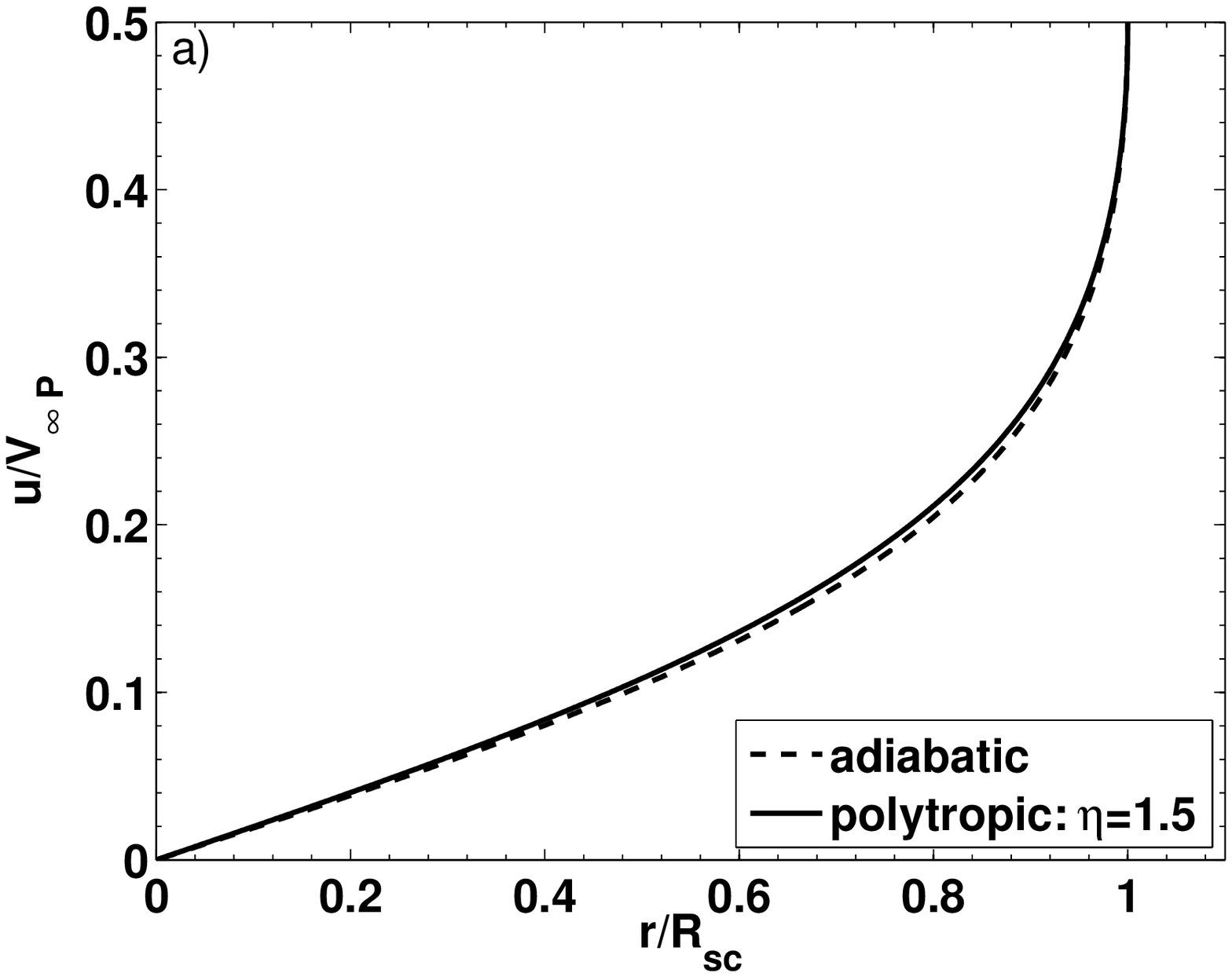}
\hfill
\includegraphics[height=50mm]{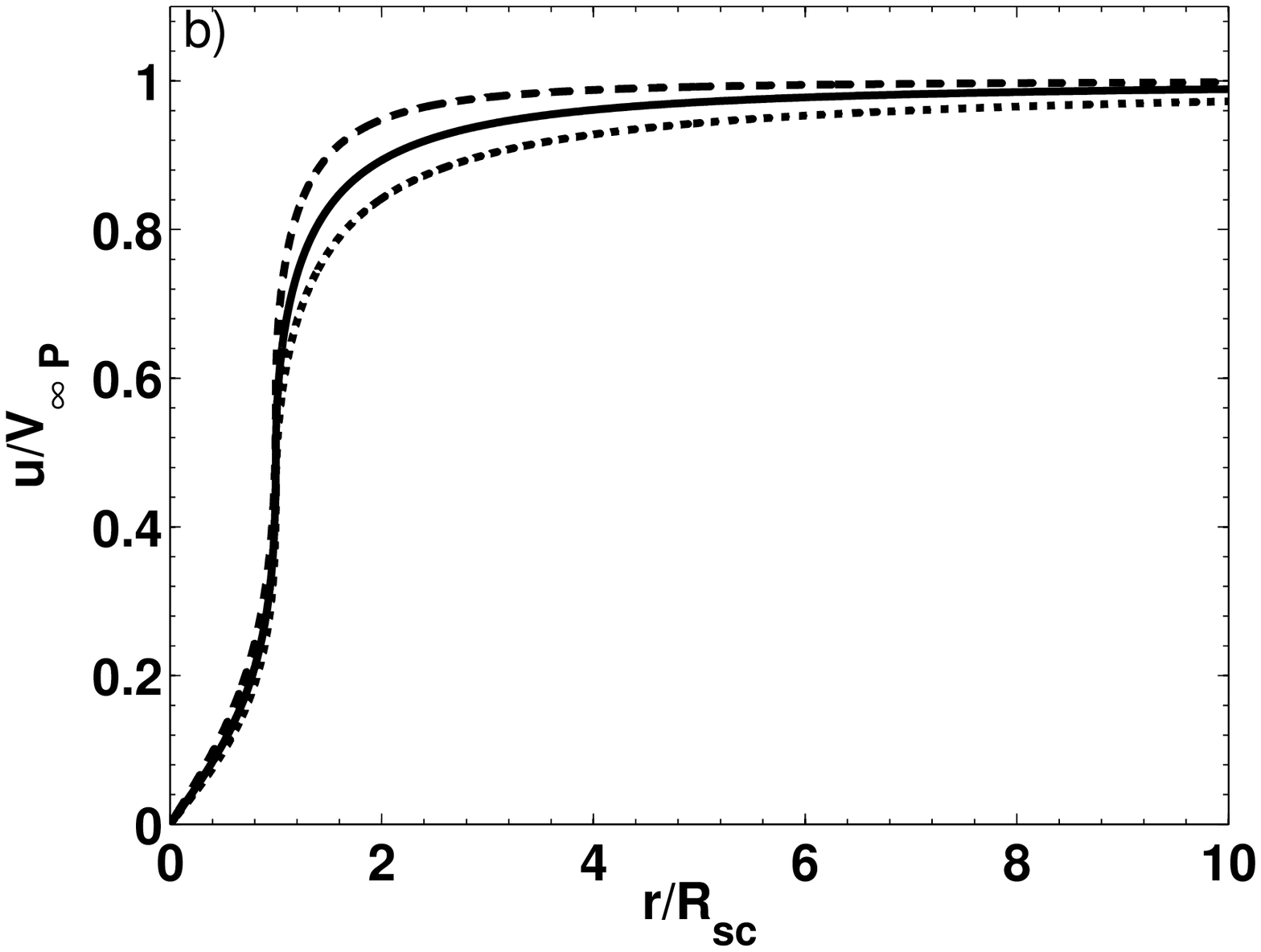}\\
\vfill
\includegraphics[height=50mm]{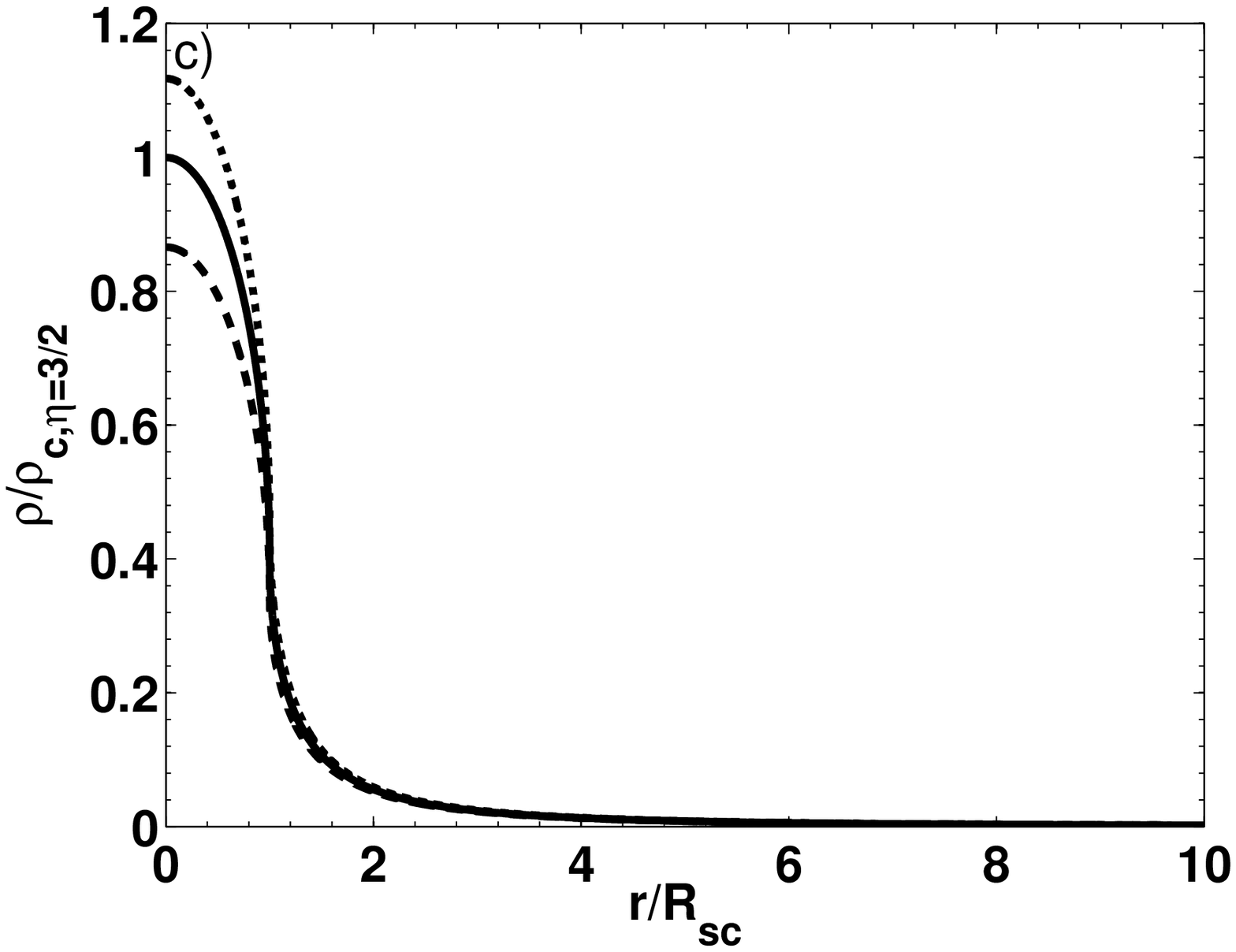}
\hfill
\includegraphics[height=50mm]{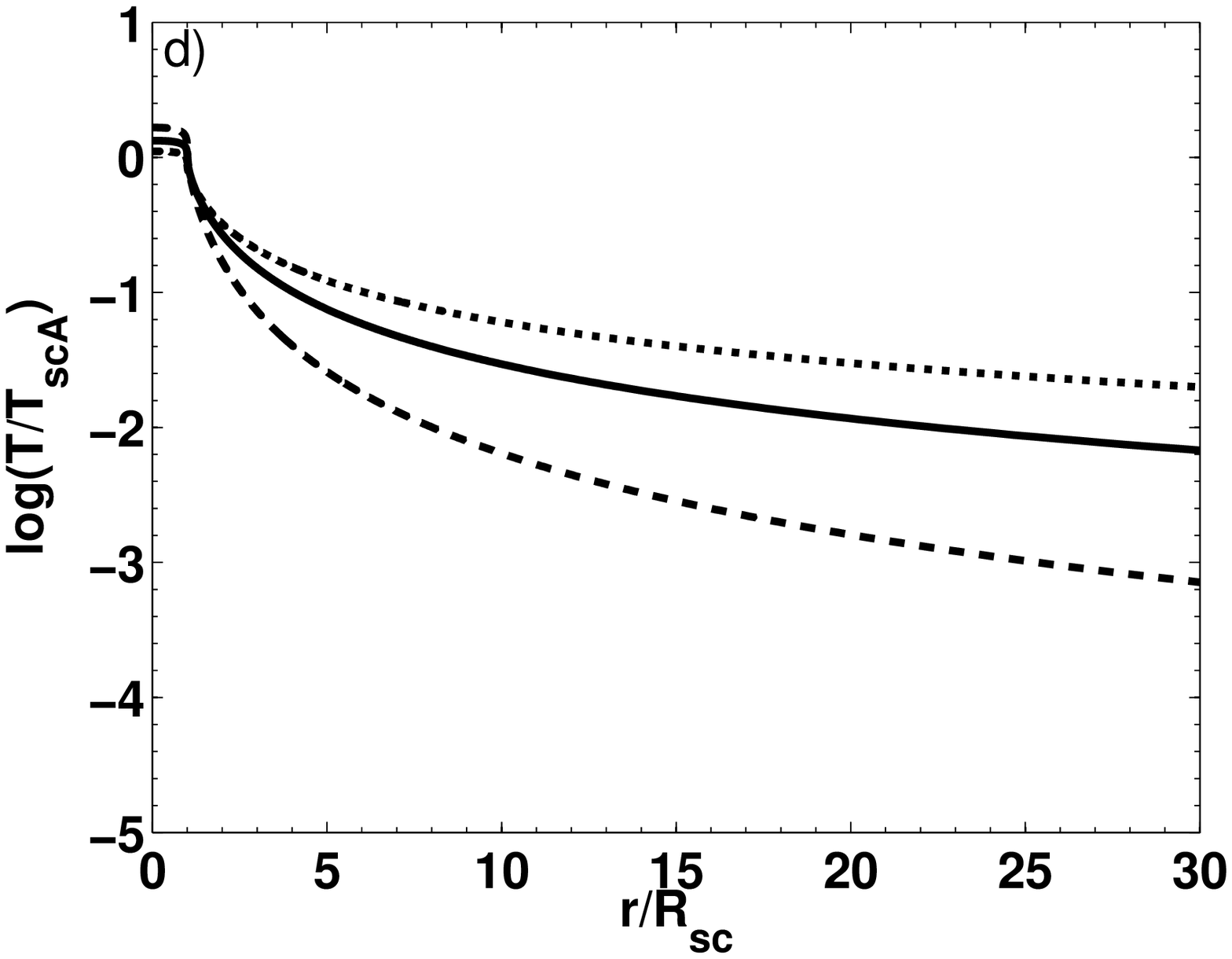}
\caption{ Panel (a):  comparison of the  adiabatic isomorphism  velocity profile for region A, equation  (\ref{polyadA}), with that  of the pseudo-adiabatic  solution, equation (\ref{polyin}).  In general, the solutions are alike for the same set of parameters.  Here, $\eta={3}/{2}$ and $V_{\rm \infty P}=V_{\rm \infty A}$. Panels (b), (c) and (d): normalised velocity, density and temperature profiles for pseudo-adiabatic polytropic solutions with $\eta={3}/{2}$ (solid lines), $\eta=1$ (dashed lines) and $\eta=2$ (dotted lines), respectively. The temperature profile is by far the most sensitive to  changes of $\eta$.   For superwinds with terminal speeds of the order of $V_{8} \ga1$, $T_{\rm scA}  \ga 10^7$ K. \label{fig:fig1}}
\end{figure*}
\noindent Accordingly, the boundary condition (\ref{uscc})   becomes 

\begin{equation}
U_{\rm sc}=\frac{1}{2\eta+1}. \label{eq:BCsc}
\end{equation}

  From equations (\ref{polyadA}) and (\ref{polyout}) and condition (\ref{eq:BCsc}),  one can   find that  for an accelerating solution 

   \begin{equation}
 D_1=  (2\eta+1)^{1/2} \left( 1+ \frac{6\eta+5}{2\eta+1}\right)^{\frac{4\eta+3}{6\eta+5}}
 \end{equation}
 \noindent and
 \begin{equation}
 D_2=(2\eta+1)^{-1/4} \left(1-\frac{1}{2\eta+1}\right)^{\eta/2}.
  \end{equation}

The actual polytropic solution for the innermost  region  ($r<R_{\rm sc}$)  can be   obtained from  one of the pfaffian forms  related to the  conservation laws.  The simplest pfaffian form  is
 
 \begin{equation}
\left( A  \rho^{2+\frac{1}{\eta}} -r ^2\right){ \rm{ d}\rho} + 4\rho r \,{\rm{d}}r =0,
 \end{equation}
 
 \noindent with $A=9(\eta+1)K/(q_{\rm m}^2\eta)$. The associated ordinary differential equation is inexact,  but it has a unique integrating factor: $I_F=\rho^{-\frac{1}{2}}$. After multiplying   by it,   the next solution can be found with ease
 
 \begin{equation}
 r^2+\frac{ A \eta}{3\eta+2} \rho^{\frac{2\eta+1}{\eta}} +C \rho^{1/2} =0, \label{trueP}
 \end{equation}

\noindent where $C$  is an arbitrary constant.  Such a solution has several branches according to the   values that $C$, $K$ and $\eta$ take. By using equation (\ref{eP})  and the transformation $\epsilon_{\rm BP}= \frac{1}{2}V_{\rm \infty P}^2$  as  closure relations for the  the true polytropic solution given by equation (\ref{trueP}),    a simple parameterization that we call the  \emph{pseudo-adiabatic} solution  can be obtained
  
  \begin{equation}
R= D_{\rm 1P} U^{1/2}\left(1+3(2\eta+1) U\right)^{-2/3}\;\; (r<R_{\rm sc}).\label{polyin}
\end{equation}

In what follows, we shall consider    the polytropic solutions   (\ref{polyin}) and (\ref{polyout})  satisfying  the    boundary condition  (\ref{eq:BCsc}).  Thus,

  \begin{equation}
D_{\rm 1P} =2^{4/3}(2\eta+1)^{1/2}
\end{equation}
\noindent and the values of the hydrodynamical variables at $r=R_{\rm sc}$ are
\begin{equation}
T_{\rm sc}=  \left(\frac{\eta}{\eta+1} \right)\left(\frac{1}{2\eta+1}\right) \frac{\mu_{\rm i}}{k}V_{\rm \infty P}^2, \label{TscRA}
\end{equation}
\begin{equation}
\rho_{\rm sc}=(2\eta+1)^{1/2}\frac{q_{\rm m}R_{\rm sc}}{3V_{\rm \infty P}},
\end{equation}
\noindent and

\begin{equation}
P_{\rm sc}= \left(\frac{\eta}{\eta+1} \right)\left(\frac{1}{2\eta+1}\right)^{1/2} \frac{q_{\rm m}R_{\rm sc}}{3}V_{\rm \infty P},
 \end{equation}

\noindent where $\mu_{\rm i}$ is the mean mass per particle for a completely ionized interstellar plasma and $k$ is the Boltzmann constant.  For solar abundance $\mu_{\rm i}=14/23 \mu_{\rm H}$. The central  values  are

  \begin{equation}
T_{\rm c}= \frac{1}{2(\eta+1)} \frac{\mu_{\rm i}}{k} V_{\rm \infty P}^2, \label{polyTc}
\end{equation}
\begin{equation}
\rho_{\rm c}=D_{\rm 1P} \frac{q_{\rm m}R_{\rm sc}}{3V_{\rm \infty P}},
\end{equation}
\noindent and
\begin{equation}
 P_{\rm c}= D_{\rm 1P}\left(\frac{1}{2\eta+1}\right)  \frac{q_{\rm m}R_{\rm sc}}{3} V_{\rm \infty P}.
 \end{equation}
 
Notice that   the models are  valid just for $\eta>0$. For simplicity and  in order to cover the polytropic states of astrophysical interest, we shall further  consider that $\eta\ge{3}/{10}$ in region A.  The isomorphism approaches the pseudo-adiabatic solution  for large $\eta$ (the  isothermal case). For intermediate  values of  $\eta$ they are similar  [see panel (a) of Fig. \ref{fig:fig1}] since they predict   the same values for the hydrodynamical variables at $R_{\rm sc}$ and the same central temperature and velocity.  Their  densities  and pressures just differ by  factors of $D_{\rm 1}/D_{\rm 1P}$ at the centre and their ratios reduce towards $R_{\rm sc}$.

\begin{figure}
\centering
\includegraphics[height=60mm]{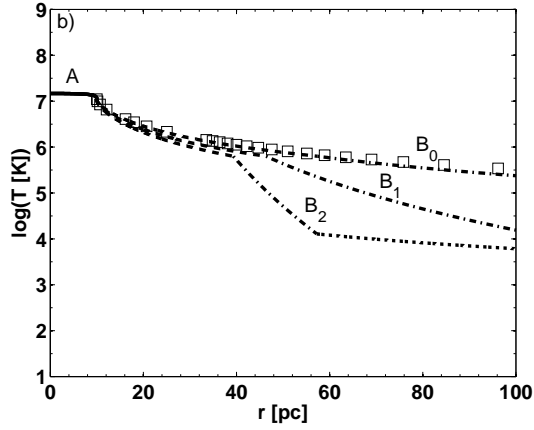}
\caption{ Silich et al. (2004, fig. 4) radiative temperature profiles for clusters with $\Pi_*=\{ R_{\rm sc,pc}=10,\dot E_{38}=\{500,2000,3000\}, V_8=1\}$.  They were here constructed using piecewise continuous polytropic laws. For the cluster cores (region A),  $\eta\approx 3/2$ for the three mechanical luminosities. The respective  free winds  have different thermodynamical configurations:   $B_0$ represents a quasi-adiabatic wind with  $\dot E_{38}=500$,  $\eta_1\approx1.45$  (dashed-line) and  $\eta_2\approx 1.15$ (dash-dotted line); likewise, $B_1$  is  a  radiative wind with  $\dot E_{38}= 2000$, $\eta_1\approx 1.3$ and  $\eta_2\approx 0.42$; and  $B_2$ is a strongly radiative wind with $\dot E_{38}=3000$, $\eta_1\approx 1.19$ and  $\eta_2\approx 0.22$. In the later case, the velocity-radius relationship adopts  $\eta=3/2$ for $T>10^4$ K (dotted line). The assumed metallicity is $Z=$ Z$_\odot$. The squares show the  external temperature profile expected from the adiabatic solution, which better applies  for clusters with smaller masses (mechanical luminosities).\label{fig:fig2}}
\end{figure}

  Since the effective polytropic index depends on the balance of heating and cooling among other factors, our models can parameterise these effects when complemented with acceptable boundary conditions or constraints, as we discuss next. In principle,  zones with distinct entropy variations  and properties can be modeled  using  a piecewise polytropic index, or, in the limit, a slow varying one.  For the case of a continuous accelerating solution,   equation (\ref{eP}) implies that any change in  the available energy per unit mass  is concomitant with  a change of $\eta$ in the same direction.  Nevertheless, notice  that velocity discontinuities produced by shock fronts are not prohibitive.  For a  smooth accelerating  solution with parameters $\Pi_*$,  models with $\eta<{3}/{2}$ have  higher velocity  profiles and  lower density profiles than the corresponding adiabatic  solution  ($\eta=3/2$, $V_{\infty A}=V_{\infty P}$). On the other hand, their temperature profiles have a bimodal behaviour: in region A they are higher, whereas in region B they  are lower and very sensitive to the adopted value of $\eta$. Models with $\eta>{3}/{2}$   have the opposite behaviour\footnote{Notice however  that a very large $\eta$ produces very slow winds which are susceptible to instabilities and might have  different driving mechanisms.} (see Fig.  \ref{fig:fig1}).  In the external zone, all the density profiles fall off  asymptotically  as $\dot M/4\pi V_{\rm \infty P}r^{2}$.  In connection with the radiative case,  Silich, Tenorio-Tagle \& Rodr\'iguez-Gonz\'alez (2004)  have shown that  whilst the velocity and density profiles are barely affected by the small variations of the effective terminal speed promoted by radiative cooling, the temperature profile can be  changed  drastically at distances larger than  a few times the star cluster radius.    Given that for dissipative solutions $V_{\rm \infty P}<V_{\rm \infty A}$ and that the relative  behaviour  of the profiles has to scale accordingly, it  follows that models with $\eta>3/2$ are appropriate  for region A, whereas those with $\eta<3/2$  may be used for region B.   In Fig. \ref{fig:fig2}, we reproduce   the numeric results of  Silich et al. (2004) for radiative winds using the aforesaid method.  At least 3 different $\eta$'s were required for each case presented.  The profiles were constructed     using the X-ray luminosity  (Section \ref{WX}) to estimate  the energy radiated by the cluster core,  since $\Lambda_{\rm X}\approx \Lambda_{\rm total}$ for  temperatures in the order of  $\sim 10^7$ K and $Z\approx$Z$_\odot$. The emissions of the external sub-zones were weighed  (see Appendix \ref{SWX}) using the total cooling function tabulated by Plewa (1995). From this, the respective changes in $V_{\rm \infty P}$ were found, and with the aid of (\ref{eP}), also the  corresponding values of $\eta$.  As a conclusion, it can be  pointed out that just  a   change of $\eta$  (adequate for each zone) is  sufficient to  reproduce  and explain the radiative profiles.  For region A,  we found  that $\eta$ is always  close to 3/2. However, a closed  analytical estimation  to    find  explicitly  the  adequate values of $\eta$ for the free wind  is still lacking. In principle,   just    a knowledge of the star cluster parameters and the overall cooling function is sufficient to obtain it.   A general discussion of the radiative case  will be presented on a forthcoming paper.

\subsection{X-ray luminosity}\label{WX}

The cumulative  X-ray luminosity  of a spherically symmetric gas distribution is given by the integral

\begin{equation}
L_{\rm X}(r)=4\pi  \int_{0}^{r} n(r')^2 \Lambda_{\rm X}[T(r'),Z]r'^2 {\rm{d}}r', \label{LX}
\end{equation}

\noindent where $r$ is the radial coordinate from the object centre,  $r'$ the corresponding dummy integration variable, $n$  is  the particle number  density, and $\Lambda_{\rm X} $ is the X-ray emissivity function,  which  in general  depends on the plasma temperature and metallicity.  Here, the later is assumed homogeneous; nevertheless,  tools to deal with metallicity gradients are given.

The actual dependence of the superwind X-ray emission on the star cluster parameters is determined by  the functional relation   of  $\Lambda_{\rm X} $ with the hydrodynamical variables. This functional dependence  synthesizes the physics behind the emission mechanisms.   In Appendix \ref{InitialX}, we  give an initial approximation that corresponds to  a constant emissivity, whereas in Appendix \ref{TLaws},  we give transformation laws that are used  to show how to overcome such a  limitation. Those laws can also be used to incorporate the effect of metallicity gradients {{or  non-collisional  ionization equilibrium}}. Below, we concentrate on region A because, as  is demonstrated in Appendix \ref{InitialX}, region B contributes  less than 20 per cent  to the superwind  total X-ray luminosity.

In Appendix \ref{TLaws}, it is  demonstrated that for a realistic emissivity given as piecewise continuous  power laws, $\Lambda_{\rm X}= \displaystyle{\sum\Lambda_{\rm \alpha} T^{\alpha}}$,  the cumulative X-ray luminosity is given by
\begin{equation}
L_{\rm XA}(\eta,\{\alpha\},U) =X_{A} U^{3/2}\sum_{\rm \alpha} \mathcal{F_{A}}(\eta,\alpha,U)\Lambda_{\rm \alpha}T_{\rm c}^{\alpha},\label{HyperLXA}
\end{equation}

\noindent where  

 \begin{equation}
X_{\rm A}=\frac{D_{\rm 1P}^5\dot E^2}{\pi \mu_{\rm n}^2V_{\rm \infty P}^6 R_{\rm sc}}, \label{XAT}
 \end{equation}

 \begin{equation}
\mathcal{F_{A}}(\eta,\alpha,U)=\frac{350\left[1+\frac{\eta^2(8\eta+1)}{5(2\eta+1)}  U^2\right]^{4/10}}{9\left(\frac{350}{3}+\phi_1 U+\phi_2 U^2\right)}\left(1-U\right)^{\frac{\alpha}{30}},\label{Hyperexp}
\end{equation}
  \begin{equation}
\phi_1(\eta,\alpha)=70(\alpha+28\eta+14),
\end{equation}
\noindent and
\begin{equation}
\phi_2(\eta,\alpha)=[17\alpha^2+\alpha(501+952\eta)+9528\eta(1+\eta)
+2382]. \label{LXAend}
 \end{equation}

\noindent  In these expressions $\alpha\in[-4,4]$ and $U\in[0,{1}/{(2\eta+1)}]$.

To account for a realistic emission, {{we assume collisional ionization equilibrium (CIE). This assumption is valid for the central region, where  the ionization and recombination time-scales  used to assess  the collisional equilibrium (Mewe 1984, 1997) are smaller than the radiative and  adiabatic expansion cooling time-scales  (see Ji, Wang \& Kwan 2006). Outside of the cluster, CIE does not hold for $r>2$--$3 R_{\rm sc}$ ($T\lesssim 3\times 10^6$ K); however, the predicted X-ray emission of such region is negligible. Hence,}}  we   use the Strickland \& Stevens  (2000, SS00 hereafter) X-ray emissivity tables, which are based on a recent version of the Raymond \& Smith  (1977) code.  These tables give the X-ray emissivity as a function of temperature for two energy bands appropriate for \emph{Chandra} and \emph{XMM-Newton}  observations:  a soft band that goes from 0.3--2.0 keV  and a  hard  band from 2.0--8.0 keV.  They  account for  continuum emission by incorporating the free-free (dominant for $T \ga $ 3 keV),   the free-bound and the two photon (significant  for $T\le$ 2 keV in the energy range 1--10 keV) processes. They  also  account for the line emission, which dominates the spectrum at temperatures around  and below 1 keV.   Additionally, the contributions from  hydrogen and   metals were separated to allow the inclusion of arbitrary metallicities.  Therefore, the total  X-ray emissivity can be written as

\begin{equation}
\Lambda_{\rm X}(T,Z)=\Lambda_{\rm XH}(T)+Z\Lambda_{\rm XM}(T).\label{Lambdapar}
\end{equation}

We fitted power laws and rational functions   to the SS00 tables using a trust region method (see Table \ref{Table2}  and Table \ref{Table3}). As an outcome,   it was  found that  for  fast winds (  $V_{\rm \infty P}  \ga 1000$ km s$^{-1}$  and $T_{\rm sc} \ga 9\times10^6$ K)  the   hydrogen component   in both bands  well resembles a bremsstrahlung law, whereas  the emissivity of metals has an almost constant value at high temperatures (2.6--8.6 keV) and a sharp power law behaviour at the other end (0.8--2.6 keV). Using these results  (marked with $\star\star$  in Table \ref{Table2}) and evaluating formulas (\ref{HyperLXA}) and  (\ref{Hyperexp})  at $U_{\rm sc}=1/(2\eta+1)$  we get that

  \begin{equation}
L_{\rm XA,total}= X_{\rm A}U_{\rm sc}^{3/2}\left\{\mathcal{F_A} (\eta,\frac{1}{2},U_{\rm sc})\Lambda_{\rm H} T_{\rm c}^{1/2}+\right. 
Z\left.\left[\mathcal{F_A} (\eta,-\frac{5}{2},U_{\rm sc})\Lambda_{\rm M1} T_{\rm c}^{-5/2}+\mathcal{F_A} (\eta,0,U_{\rm sc})\Lambda_{\rm M2} T_{\rm c}^{0}\right]\right\}.
\end{equation}

\begin{figure*}
\centering

\includegraphics[height=60mm]{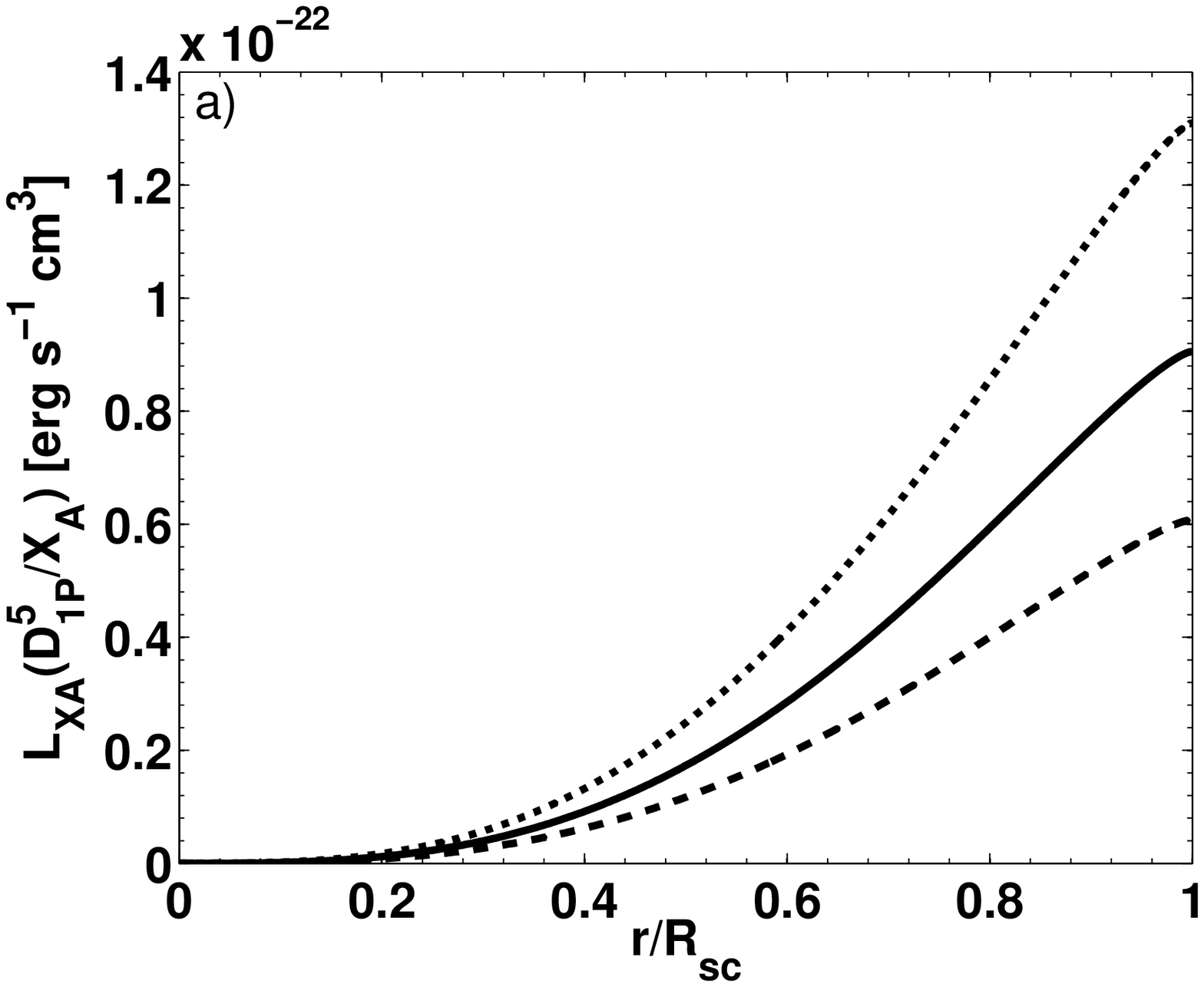}
\hfill
\includegraphics[height=60mm]{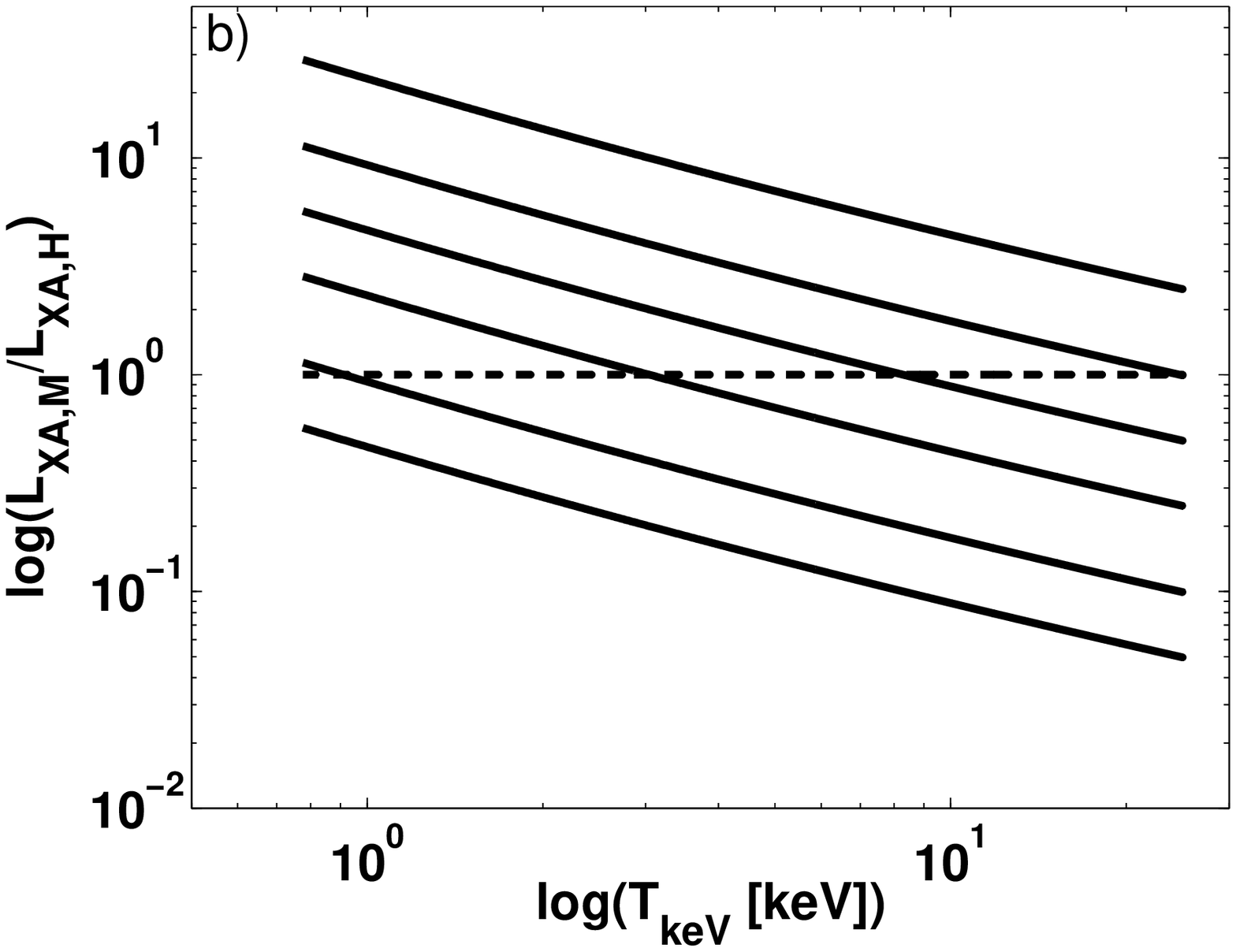}
\caption{Panel (a): region A  cumulative X-ray luminosity normalised  by a factor of  $X_{\rm A}/ D_{\rm 1P}^5$, which contains the dependence on the  cluster parameters.  The values of the polytropic index are  $\eta={3}/{2}$ (solid line), $\eta=1$ (dashed line) and $\eta=2$ (dotted line).  Panel (b): ratio of the X-ray luminosity of  metals (solid lines, region A) to that of hydrogen  for different core temperatures ($T_{\rm keV}$).  Here $\eta=3/2$.  The dashed line is used as a reference. From bottom  to top of the panel, the metallicities are 0.1,0.2,0.5,1.0,2.0 and 5.0 Z$_{\odot}$. The axes marks are logarithmically spaced for a better visualization. \label{fig:fig3}}
\end{figure*}

Bearing in mind equations (\ref{polyTc}) and (\ref{XAT}), and taking $\eta={3}/{2}$ we have  that

 \begin{equation}
L_{\rm XA,total}= (5.12 \times 10^{33} \,\mbox{erg s$^{-1}$})\frac{\dot E_{38}^2}{R_{\rm sc,pc}V_{8}^5}
\left[ 1.00+ 1.32Z\left(\frac{1}{V_{8}^6}+\frac{1.30}{V_{8}}\right)\right], \label{LXA1}
\end{equation}

\noindent or, in terms of $T_{\rm keV}$,

 \begin{equation}
L_{\rm XA,total}= (9.31\times 10^{33}\, \mbox{erg s$^{-1}$}) \frac{\dot E_{38}^2}{R_{\rm sc,pc}T_{\rm keV}^{5/2}} 
\left[1.00+ 1.32Z\left(\frac{2.05}{T_{\rm keV}^3}+\frac{1.47}{T_{\rm keV}^{1/2}}\right)\right],\label{LXA2}
\end{equation}

\noindent where the first  and second terms between  square brackets  give the relative contribution from hydrogen and metals, respectively. Since a term $\propto T^{1/2}$ was factorized, the bracketed quantities show the variations  of the SS00 emissivity function with respect to bremsstrahlung. The X-ray luminosity of warmer ($T_{\rm sc}<9\times 10^6$ K) and   generally slower winds ($V_8<1$) can be obtained in a similar way using  equations (\ref{HyperLXA})--(\ref{LXAend}) together with Table \ref{Table2}.

 In Paper I,  the total X-ray luminosity of regions A and B  was estimated by approximating the   temperature  and density by the constant values that they take at $R_{\rm sc}$ according to the CC85 adiabatic model, i.e. by setting $\rho=\rho_{\rm sc}$ and $T=T_{\rm sc}$ with $\eta=3/2$.  Consequently, it was found that  $L_{\rm XAB,total}=4\pi\alpha_{\rho}^2 \Lambda_X(T_{\rm sc},Z)\rho_{\rm sc}^2 R_{\rm sc}^3/3\mu_{\rm n}^2$.  Explicitly, this is

 \begin{equation}
 L_{\rm X,AB}=\left(1.27\times 10^{34}\mbox{erg s$^{-1}$}\right) \frac{\Lambda_{\rm X,-23}( T_{\rm sc},Z) \dot E_{38}^2}{R_{\rm sc,pc}V_{8}^6} 
=\left( 1.08\times 10^{34}\mbox{erg s$^{-1}$}\right)  \frac{\Lambda_{\rm X,-23}(T_{\rm sc},Z)\dot E_{38}^2}{R_{\rm sc,pc}T_{\rm keV}^3}.\label{LXP1} 
\end{equation}

In the above formula, the effects of the actual thermodynamical structure of zones A and B and the realistic X-ray emissivity  (SS00) were incorporated through the fiducial coefficient  $\alpha_{\rho}\approx2$, which  smoothed the differences between the analytical and the numerical calculations that were carried out  for winds with $V_8\approx 1$.  Because such calculations also showed that the contribution of region B was always less than one-quarter  of that of region A, the former was also  incorporated into (\ref{LXP1}) through $\alpha_\rho$.  These  results agree fairly well with equations (\ref{LXA0t})  and (\ref{LXB/LXA}) which just assume  a constant X-ray emissivity. In contrast, equations (\ref{LXA1}) and (\ref{LXA2}) explicitly show their physical dependence on the thermodynamical structure and the SS00 X-ray emissivity function  through the numerical coefficients  obtained, the separation of the contributions of hydrogen and  metals, and  their functional dependence on the terminal speed and temperature, respectively. Another advantage of these formulae  over that of Paper I is that they  can be used for very fast winds with $V_{\rm \infty P }\sim  $ 2500--4500 km s$^{-1}$.

Additionally, in the current formalism, equations (\ref{HyperLXA}) and (\ref{polyin}) allow  to calculate the   cumulative X-ray luminosity profile of region A for different polytropic indexes. If it were needed, the luminosity of region B could be calculated using our transformation laws.  The  cumulative X-ray luminosity  profile of the cluster core  for different values of $\eta$ is presented in panel (a) of  Fig. \ref{fig:fig3}. In the same figure, panel (b) shows that  for $T_{\rm keV}\ge $ 3 keV the  contribution from  hydrogen  is comparable (within a factor of $\sim 2$) to that from the metals  when   $Z\sim$ Z$_{\odot}$,  whereas for lower metallicities hydrogen can dominate in the same temperature range.  On the other hand,  metals dominate the cluster core emission at lower temperatures    when $Z\sim  0.5$--1  Z$_\odot$ or higher. In general, hydrogen always dominates for $Z< 0.2\,$Z$_\odot$  and  metals for $Z>2\,$ Z$_\odot$. We remark that metallicity gradients or a different X-ray emissivity can be handled using the aforementioned transformation laws and the adequate polytropic index.  {{For instance, when the CIE assumption does not hold  one can use non-equilibrium emissivity functions like the one presented by Gnat \& Sternberg (2007), whom  for  solar metallicity found that  $\Lambda_{\rm, noneq}\propto 5.6\times 10^{-20} T^{-0.46}$ erg s$^{-1}$ cm$^3$. When CIE does not hold inside of the cluster, formula (\ref{HyperLXA}) can handle this new emissivity function with ease. By separating   components and because of their  ability to handle arbitrary emissivities, our formulas round out  the X-ray luminosity model  for the gas contained within the star cluster.}}

 \subsection{\bf{Surface brightness}}

\begin{figure}
\centering
\includegraphics[scale=1]{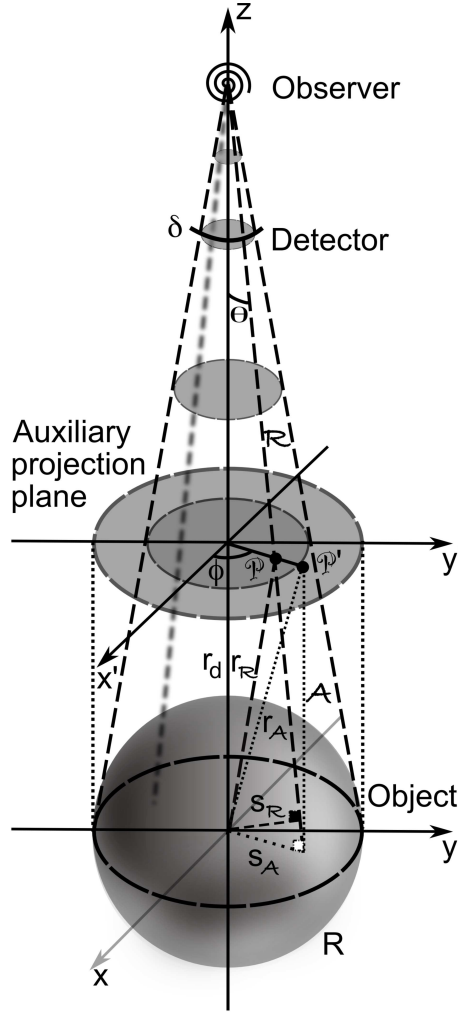} 
\caption{Abel transform of a spherically  symmetric function with compact support on a ball of radius $R$ and projected angular diameter $\delta$. For $r_{\rm d}\gg R$    the Radon Transform lines (dashed), approach the  Abel Transform lines (dotted)  since $s_{\mathcal{R}}=r_{\rm d}\sin(\theta)\approx r_{\rm d} \theta$ and $s_{\mathcal{A}}=r_{\rm d} \tan(\theta) \approx r_{\rm d} \theta$.  The right angles that the projected radii and the respective projection lines subtend are indicated  with black and white squares. The spherical symmetry of the function warrants that  the integral  yields the same value along both lines of sight.  The difference is that in  the astrophysical case the  distribution is projected along the $\mathcal{R}$ lines, i.e.  the values at points $\wp'$ are mapped to points $\wp$. Compare the difference in the auxiliary projection plane.  \label{fig:fig4}}. 
\end{figure}

 To calculate the surface brightness and other projected quantities we  use the Abel transform (see, e.g. Yoshikawa \& Suto 1999)
\begin{equation}
Q( s_{\rm })=2\int_{ s_{\rm }}^{\infty}  q(r) \frac{r}{\sqrt{r^2- s_{\rm }^2}}{\rm{d}}r, \label{AbelT}
\end{equation}

\noindent where $r$ has the same meaning as before, $ s_{\rm }=r_{\rm d}\theta$ is the projected radius,  $r_{\rm d}$ is the angular diameter distance to the object and $\theta$ the projected subtended angle. The factor of 2 takes advantage of the spherical symmetry (see Fig. \ref{fig:fig4}). Since we are dealing with finitely localized gas distributions,  the traditional upper  limit ($\infty$) is replaced by the size of the region of interest. This size is determined by the gas temperature. We consider an X-ray cut-off temperature  of $T_{\rm cut}=5\times 10^5 $ K  for all regions. The corresponding radius is $R_{\rm cut}$.

The surface brightness  can be  obtained by setting $q=n^2 \Lambda_{\rm X}$ in (\ref{AbelT}).  For the superwind, it  is given in terms of elliptical integrals in non-canonical form, the  reduction of which requires the awkward calculation of the roots of several quartics for each $ s_{\rm }$. None the less,  the whole superwind (region A+B) surface brightness can be fitted by a function of the form
\begin{equation}
\sigma_{\rm AB}( s_{\rm })=A_{\rm AB}\left[1+\left(\frac{ s_{\rm }}{R_{\rm sc}}\right)^{\alpha_1}\right]^{\alpha_2}.\label{SXA}
\end{equation}

For $\eta={3}/{2}$  we have that $\alpha_1\approx 3$ and $\alpha_2\approx -5/2$. The associated central brightness  can be found in terms of hypergeometric functions. For  fast winds, $V_8\gtrsim 1$,   it is given by

   \begin{equation}
A_{\rm AB}\approx 1.28 \frac{\Lambda_{\rm X}(T_{\rm sc},Z)\dot E^2}{\mu_{\rm n}^2R_{\rm sc}^3V_{\infty}^6}
=(1.9\times 10^{34}\,\mbox{erg s$^{-1}$ pc$^{-2}$}) \frac{\Lambda_{\rm X,-23}(T_{\rm sc},Z)\dot E_{38}^2}{R_{\rm sc,pc}^3T_{\rm keV}^3}. \label{AXAB}
\end{equation}

\begin{figure}
\centering
\includegraphics[height=60mm]{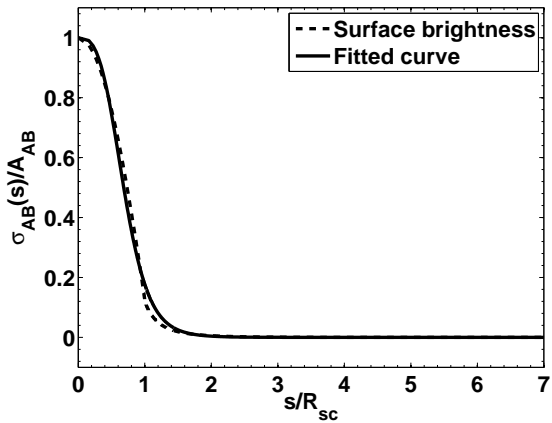}
\caption{Normalised X-ray surface brightness of a  superwind ($\eta={3}/{2}$). Soft line: actual surface brightness, dashed line: fitted profile. Note that although   the outer layers of the core contribute relatively more to the   central X-ray luminosity (Fig. \ref{fig:fig4}), no limb-brightening is produced there. The overall superwind surface brightness sharply drops in the external zone.  \label{fig:fig5}}
\end{figure}

From this formula  and Fig. \ref{fig:fig5},  it becomes clear  that even in the adiabatic case ($\eta=3/2$) the superwind  X-ray surface brightness drops extremely fast ($\propto  s_{\rm }^{-{15}/{2}}$) for $ s_{\rm }\gg R_{\rm sc}$.  Thus, although in principle clusters in the quasi-adiabatic regime per se  can generate   very extended X-ray emitting haloes, see equation (\ref{RTout}), their  surface brightnesses  far away from the cores  could be too weak to be detectable, specially  when the integration times are short (see Lang et al. 2005).

\subsection{\bf{Weighted temperatures}}\label{WindT}

Several definitions of the projected X-ray temperature  exist, that is,  of  the temperature $T_{\rm X}$ by which the spectrum of an  often multi-temperature plasma is fitted as if it would have just one component. The most frequently used ones are  of the form (see \rm{e.g.} Navarro, Frenk \& White 1995)

\begin{equation}
T_{\rm X}= \frac{\int n^2 \Lambda T \rm{d}V}{ \int n^2 \Lambda \rm{d}V },\label{weight}
\end{equation}

\noindent where $\Lambda$ is a weighting function that may  be the X-ray emissivity function  (see Mazzotta et al. 2004 and references therein). The emission-measure weighting scheme uses $\Lambda=1$,  whereas  the so called (bolometric) emission  weighting scheme assumes that $\Lambda=\Lambda(T)\propto T^{{1}/{2}}$,  i.e. it assumes  just free-free emission, notwithstanding the fact  that this later  mechanism dominates just  for $T>$ 3 keV. Mazzotta et al. (2004) have shown  that theoretically,  none of these definitions is  an accurate approximation of the spectroscopic temperature  of  multi-temperature plasmas. However,  they have argued that since  noise, instruments response, and instrumental and cosmic backgrounds among other factors affect and distort the observed spectrum, sometimes it  is possible  to fit  a single-temperature plasma model to the data. In their work, they simulated  the expected spectra of galaxy clusters  ($Z=$Z$_\odot$) taking into account the sensitivity  of  the \emph{Chandra} and \emph{XMM-Newton} instruments in the soft band. They found  that  for $T\ge $ 3 keV   the synthetic spectra    (with temperature $T_{\rm spec}$)   could be reproduced  --at a level better than 5 per cent--  by weighting with a function  of the form   $\Lambda=\Lambda(T)\propto T^{-{3}/{4}}$.  Such a function yields a spectroscopic-like temperature $T_{\rm SL}$ that can be used as an   estimator of the actual   temperatures  of  the isothermal models that might be  fitted to real spectral  data.

None of the above  schemes take into consideration arbitrary  metal abundances. Vikhlinin (2006) sought to extend the previous result for  $T \ga  $ 0.5 keV and arbitrary metallicities. Although his algorithm yields a significant accuracy  gain when lower temperatures are involved, it does not provide analytical estimations because his weighting  function  has to be constructed  ad hoc (it depends on the measurement  device characteristics and individual observational conditions).  We present here a theoretical  weighting scheme based on the SS00  tables.   Since we do not  consider observational artefacts,  our  spectroscopical temperature $T_{\rm X}$ corresponds to the ideal projected  X-ray temperature of the models. Nevertheless, our scheme can incorporate artefacts easily, provided that the  overall weighting function could be parameterized in the same way  as the one found by  Mazzotta el al. (2004), \rm{i.e.} as a power  law (or series) in $T$.

For  bubbles and superbubbles we give preference  to the projected temperature map as a diagnostic tool

\begin{equation}
T_{\rm X}( s_{\rm })=\frac{\int n^2 \Lambda T   (r^2- s_{\rm }^2)^{-1/2} r {\rm{d}}r}{\int n^2 \Lambda    (r^2- s_{\rm }^2)^{-1/2} r {\rm{d}}r}. \label{PTemp}
\end{equation}

For  superwinds,  the emission-measure  and  the \emph{Chandra} and \emph{XMM-Newton}  spectroscopic-like  temperatures  (whether the later applies or not depends on $\eta$; for  $\eta\approx \frac{3}{2}$, it is  a valid  measure when $V_{8} \ga 1.65$)   can be  obtained from  equations  (\ref{HyperLXA})   and (\ref{Hyperexp}) 

\begin{equation}
T_{\rm X1}(\eta,{\alpha},U_{\rm sc})= \frac{ L_{\rm XA}(\eta,\alpha+1,U_{\rm sc})}{L_{\rm XA}(\eta, \alpha,U_{\rm sc})}=\frac{ \mathcal{F_A}(\eta,\alpha+1,U_{\rm sc})}{\mathcal{F_A}(\eta,\alpha,U_{\rm sc})} T_{\rm c}, \label{TX1}
\end{equation}

\noindent  where $\alpha=0$   for the first  estimator  and  $\alpha=-3/4$ for the second one. Finally, our X-ray emission weighted temperature is

\begin{equation}
T_{\rm X2}(\eta,{\alpha},U_{\rm sc},Z)= \frac{ L_{\rm XA}( \eta, \{\alpha+1\},U_{\rm sc})}{L_{\rm XA}(\eta, \{\alpha\},U_{\rm sc})}, \label{TX3}
\end{equation}

\noindent  which depends implicitly on metallicity. For   superwinds with $V_{8}\gtrsim 1$, $\eta \ga{3}/{2}$ and $Z=$Z$_{\odot}$, all the  above estimators yield that $ T_{\rm X}   \approx 0.95\,T_{\rm c}$ and that $T_{\rm X}$ gets  closer to $T_{\rm c}$ as $\eta $ increases.  There is just a slight difference at the other end: for  $\eta\approx 0.3$ we obtained that $T_{\rm X}\approx 0.85\,T_{\rm c}$. This implies that the cluster central temperature is  adequate for representing the spectroscopical one. For superwinds, ${T_{\rm sc}}/{T_{\rm c}} \in[{3}/{8}, 1)$, so in many cases, the temperature difference  between the centre and the edge might not be  significative. However, for superbubbles --as we shall see-- such a difference does matter.

\section{Bubbles with shell  evaporation}  \label{Bubbles}

On the theoretical side,  X-ray models for superbubbles with evaporation  from the cold, dense and  thin outer shell (W77, region C in Fig. 1)    have an interesting  analytical behaviour. During the sweep-up  phase,  thermal conduction fluxes across the contact discontinuity   promote  the injection of  radiatively  cooled  material into the cavity that contains the  hotter (shock-thermalised)   but  much more tenuous superwind.  As a result,  the gas  distribution has very steep density and temperature gradients and to account for the actual shape of the emissivity function becomes  crucial. On the other hand, in some cases  the observations pose another challenge, for  after  comparing  with the  C95 standard X-ray luminosity  model, discrepancies  of up to two orders of magnitude  have been found (Chu, Gruendl \& Guerrero 2003; Dunne et al. 2003).

Most analytical models  for spherical bubbles on the sweep-up phase (Castor et al. 1975; W77; Hanami \& Sakashita 1987; Koo \& McKee 1992a,b; Garc\'ia-Segura \& Mac Low 1995) share the following similarities:

\begin{enumerate}
\item  In all models,  the compact outer shell is driven by the thermal pressure of  region C. There, the  shocked wind is isobaric  (except in a small zone near the secondary shock) and  its pressure equals that of the shocked ISM. Therefore, the pressure of the hot bubble  is only a function of time. 
\item  Similarly,  it is demonstrated that the bubbles have a pseudo-adiabatic interior because the characteristic cooling  time of the shocked superwind is larger than the bubble dynamical time scale (see, e.g.,  Mac Low \& McCray 1988 and Koo \& McKee 1992a,b).   The prospective evaporation of cold gas from the external shell     does not modify this behaviour significantly, even after  the evaporated mass dominates the region.  

\item Under this common premises, but considering different additional physics, the authors determined  the bubble \emph{geometrical evolution}  and its energy budget,  finding explicit expressions for the principal and secondary shock positions, the  pressure,  and the total energy: $R_{\rm p}(t)$, $R_{\rm s}(t)$, $P_{\rm C}(t)$ and $E_{\rm C}(t)$, respectively. 
\item With exception of the Koo \& McKee model, which just considers the intrinsic self-similar evolution of region C,  the rest of the models assume that the \emph{thermodynamical structure}   is determined by   the thermal evaporation of cold material from the external shell. The later process is governed by the  classical thermal conduction theory (Cowie \& McKee, 1977). The predicted  temperature  profile is of the form
\begin{equation}
T_{\rm C}(r)=T_{\rm Cc}\left(1-\frac{r}{R_{\rm p}}\right)^{2/5}. \label{TCc}
\end{equation}
\item The density profile is derived from the isobaric condition stated in (i)
\end{enumerate}

\begin{equation}
n_{\rm C}(r)=n_{\rm Cc}\left(1-\frac{r}{R_{\rm p}}\right)^{-2/5}, \label{nCc}
\end{equation}

\noindent  where $T_{\rm Cc}$ and $n_{\rm Cc}$ are the  extrapolated central temperature and density.  Note that the above profiles are spatially self-similar. 

The points above imply that the dynamical evolution (linked to the motion energetics)  and the  thermodynamical structure are totally independent, regardless of the choice of the bubble model.  Each model assumes different physical conditions that lead  to different  evolutions: W77 considered constant energy injection rates and  ISM densities. They showed that if radiative cooling (when the bubble interior departs from a quasi-adiabatic state) and  distinct  ambient pressures  are  incorporated into their model, the bubble size can change but the self-similarity of the thermodynamical profiles is  not  affected. Hanami \& Sakashita (1987)  and Garc\'ia Segura \& Mac Low (1995)  assumed  different ambient stratifications: they  considered ISM densities that fall off as  $\rho_{\rm 0}\propto r^{-\omega}$ and $\rho_0 \propto r^{-2}$,  respectively. The later authors also incorporated variable mechanical luminosities. Again, the geometrical expansion is different, but  the shapes of the  density and temperature profiles  are not disturbed.

The coupling of the physics resulting from (iii) and (iv)   is done through  $T_{\rm Cc}$ and $n_{\rm Cc}$ which inherit the time dependence of  the   geometrical model, i.e. they depend on the derived $R_{\rm p} (t)$, $R_{\rm s} (t)$, $P_{\rm C}(t)$ and $E_{\rm C}(t)$. They also  enclose the efficiency of the evaporative process given by  the numerical value of the thermal conduction coefficient (Spitzer 1956).  It can be shown  that   as long as conditions (iv)  and (v) are satisfied, the X-ray emission and other integral properties like the evaporated  mass (which dominates  in this zone)  are  independent from the geometrical (dynamical) model. These considerations are taken into account in the X-ray models  here presented,  where they are used to derive properties that are valid for all evaporation dominated models. Such properties can be helpful to constraint the related physics and serve as  auxiliary tools at the time of analysing  the observations.

  For simplicity and in order to explore the parameter space of  a specific model,  we shall consider that of  W77, unless otherwise stated.  It that model,  the leading  and reverse shock positions  and the  central  temperature and density are given by
  
  \begin{equation} R_{\rm p}=(67 \,\mbox{pc}) \dot E_{38}^{1/5} n_0^{-1/5}t_6^{3/5}, \label{Rp}\end{equation}
 \begin{equation} R_{\rm s}=(26 \,\mbox{pc}) \left(\frac{\dot E_{38}^3}{n_0^3V_8^5}\right)^{1/10} t_{6}^{2/5},\label{Rs}
  \end{equation}
   \begin{equation}
 T_{\rm Cc}= (5.25\times 10^6 \, \mbox{K}) \dot E_{38}^{8/35} n_0^{2/35}t_6^{-6/35},\label{TCc}\end{equation}
 \noindent and
  \begin{equation}
 n_{\rm Cc}= (1.63\times 10^{-2} \,\mbox{cm}^{-3}) \dot E_{38}^{6/35}n_0^{19/35}t_6^{-22/35} \label{nCc},\end{equation}

\noindent where $n_0$ is the particle number density of the ISM. {{ For normal OB associations,  McCray \& Kafatos  (1987) have pointed out that initially (at ages up to $ 5\times 10^6$ yr), the injection of energy (a few times $\sim 6\times 10^{35}$  erg s$^{-1}$) is dominated by the most  massive but much less numerous O  type stars  which blow bubbles with sizes $\sim 100$ pc. At later times and up to $5\times 10^7$ yr,  the injection rates are sustained by  the  supernovae explosions of   B type  stars; specially,  after the most massive star have died.  The supernovae inject energy at a steady rate  of $N_{*} \alpha \times 10^{35}$ erg s$^{-1}$ (where $\alpha$ is a constant of order unity and $N_{*}$ is the total number of stars with masses above 7 M$_\odot$) and drive a new outflow over the previously injected gas.  For rich OB  associations, the wind  dominated stage can be as important as the supernovae dominated one, i.e. for associations or clusters with  $E_{38} \gtrsim 1$  (Abbott, Bieging \& Curchwell 1981). When these effect are enhanced,  models like those of Hanami \& Sakashita (1987) and Garc\'ia-Segura \& Mac Low (1995)  might be more adequate to describe the evolution of the corresponding superbubbles. Here however, we follow McCray \& Kafatos (1987) and assume that the energy input rate remains constant through time in order to analyse later the case of M 17.  }}

\subsection{X-ray luminosity}

In its simplest form,  the total X-ray luminosity of a   well-developed bubble  with shell evaporation is  

\begin{equation}
L_{\rm XC, total}=4\pi n_{\rm Cc}^2 R_{\rm p}^3  \overline \Lambda_{\rm X} (\vartheta_{\rm cut}, Z_{\rm C}; T_{\rm Cc}), \label{LXGMI}
\end {equation}
\noindent where
\begin{equation}
\overline \Lambda_{\rm X} =\frac{5}{2}\int_{\vartheta_{\rm cut}}^{1 } \vartheta^{1/2} (1-\vartheta^{5/2})^2 \Lambda_{\rm X}(\vartheta,Z_{\rm C}; T_{\rm Cc}) {\rm d}\vartheta,  \label{LambdaGMI}
\end{equation}

\begin{figure}
\centering
\includegraphics[height=60mm]{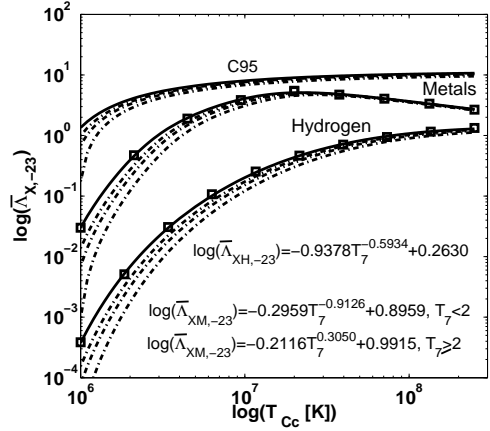}
\caption{Solid lines: evaporatively averaged soft band emissivities  resulting from   the C95 constant value  and the SS00 tables (hydrogen: $\Lambda_{\rm XH}$, metals: $\Lambda_{\rm XM}$). Here,  $T_7$ is  $T_{\rm Cc}$ in units of $10^7$ K and  $Z_{\rm C}=$Z$_{\odot}$.    The lines  for the C95 model and $\Lambda_{\rm XM}$ scale linearly with metallicity.  The former was  extrapolated to the  temperature range $10^6$--$2.5\times 10^{8}$ K. The  best-fitting approximations to $\Lambda_{\rm XH}$ and $\Lambda_{\rm XM}$  are   shown as empty squares.  Significant deviations from the C95 predicted values  are observed as  $T_{\rm Cc}\rightarrow 10^6$ K.   The deviations induced by the reverse shock when  $R_{\rm s}/R_{\rm p} =0.6, 0.7$ and 0.8 are shown as dashed-dotted lines cascading down from the   respective averaged emissivities.   The axes marks are logarithmically spaced.      \label{fig:fig6}}
\end{figure}

\noindent   $\vartheta=T_{\rm C}/T_{\rm Cc}$ is the normalised temperature and $\vartheta_{\rm cut}=T_{\rm cut}/T_{\rm Cc}$  is the normalised cut-off temperature.  Generally, $\Lambda_{\rm X}$ (on the integrand) cannot be put in dimensionless form with respect to $\vartheta$ and, as a consequence, it  will depend on  $T_{\rm Cc}$ \emph{as a parameter}.  The variable $\overline \Lambda_{\rm X} (\vartheta_{\rm cut}, Z_{\rm C}; T_{\rm Cc})$  can be thought as an "evaporatively averaged" X-ray emissivity. It is practically independent of the geometrical evolution when  $R_{\rm s}/R_{\rm p}<0.4$  (see  Fig. \ref{fig:fig6}). This is what we call a well-developed bubble and thus  the upper limit of the  integral in   (\ref{LambdaGMI}) is justified. For $ 0.4<R_{\rm s}/R_{\rm p}\le0.6$ there is just a slight geometrical dependence for bubbles with $T_{\rm Cc}<3\times 10^7$ K. If this ratio increases so does the later effect, specially in the $\sim 10^6$ K temperature range.  Because in all current models  bubbles get well-developed at early times,   we shall just  discuss that case.

In Paper I,   the bubble X-ray luminosity  was estimated   using  the C95  model which assumes that  for the \emph{Einstein} satellite $\Lambda_{\rm X}(T,Z)=Z_{\rm C}\Lambda_0=3Z\times 10^{-23}$ erg cm$^3$ s$^{-1}$ on the temperature range $10^{6}$--$10^7$ K. Therefore, one obtains that  $ \overline \Lambda_{\rm X,C95}= Z_{\rm C}\Lambda_0 I(\vartheta_{\rm cut})$ with $I(\vartheta)=(125/33)-5\vartheta^{1/2}+(5/3)\vartheta^3-(5/11) \vartheta^{11/2}$.  However, in some cases,  the differences  that arise when one considers the more realistic SS00 emissivity  function can be substantial, as  shown in Fig. \ref{fig:fig6}. There, the average of the SS00 emissivity given by equation (\ref{Lambdapar}) displays separated  the expected soft-band X-ray contributions  from hydrogen and  metals. The corresponding X-ray luminosity is  

\begin{equation}
L_{\rm XC, total}=4\pi n_{\rm Cc}^2 R_{\rm p}^3  \left[\overline \Lambda_{\rm XH}(T_{\rm Cc})+Z_{\rm C}\overline \Lambda_{\rm XM}(T_{\rm Cc})\right]. \label{LambdaGMII}
\end{equation}

The above formula is just corrected for  the effect of the shape of $\Lambda_{\rm X}$; the  variables  that depend on the geometrical evolution ($R_{\rm p}$, $T_{\rm Cc}$ and $n_{\rm Cc}$) are still undetermined  as functions of the problem parameters  ($n_0$, $\dot E$, $V_{\rm \infty P}$) and time. Such functional  relation is a consequence of the assumed bubble model.  However, a great advantage of (\ref{LambdaGMII}) is  that  it just depends on  variables that are either directly observable or spectroscopically derivable.  It is clear that the key issue is to find relationships between the central values of temperature and density  and their spectroscopic  estimators (Section \ref{PTC}).

On the purely theoretical side, if one also assumes the  geometrical growth given by W77,  an analogue to the C95  X-ray formula follows

\begin{equation}
L_{\rm XC, W77}=(3.25 \times 10^{35}  \mbox{erg s$^{-1}$}) \overline \Lambda_{\rm X} \dot E_{38}^{33/35}n_{0}^{17/35} t_{6}^{19/35},\label{LXCW77}
\end{equation}

\noindent where  $\overline \Lambda_{\rm X}$  is the average of the  SS00 emissivity. Since it is smaller (for all $T$) than the constant value assumed by C95, the bubbles evolve even more adiabatically and the  use of the   W77  model is a safe assumption. This ensures complete consistency with the emissivity function that we are using and our aim of incorporating additional physics.

\subsection{Surface brightness}\label{Bevsur}

 For a single power-law  X-ray emissivity, $\Lambda_{\rm X}=\Lambda_{\rm \alpha}(Z_{\rm C})T^{\alpha}$, the surface brightness was decomposed  (Appendix \ref{Bev}) into  three components: a temporal component related to an intensity amplitude,  $A_{\rm C}(t)$; a weighting emissivity that implicitly varies with time because of  its dependence  on $T_{\rm Cc}$, $\lambda_{\rm \alpha}=\Lambda_{\alpha,-23}T_{\rm Cc}^{\alpha}$ (though  $\lambda_{\rm \alpha}=\lambda_0=3Z_{\rm C}$ is a constant for the  C95 model); and  a spatial profile that slowly varies with time, $ S_{\rm C}( s_{\rm })$ [or $\Theta(\vartheta)$ as a function of  the normalised temperature].   In terms of these components    $\sigma_{\rm C}( s_{\rm },t)=A_{\rm C}(t)\lambda_{\rm \alpha} S_{\rm C}( s_{\rm })$.   Their respective model-independent expressions are given by equations (\ref{AXCpre})-(\ref{Abubev}). 
 
 For the W77 evolutionary model the amplitude factor is given by

  \begin{equation}
A_{\rm C}(t)=(1.05 \times 10^{31} \mbox{\,erg s$^{-1}$pc$^{-2}$ } )
\dot E_{38}^{19/35} n_0^{31/35}t_{6}^{-23/35}.    \label{AXC}
 \end{equation}

 Because of the thermodynamical structure self-similarity,  the shape of the spatial profile  ($S_{\rm C}$) in terms of the  bubble normalised projected radius ($s/R_{\rm p}$) depends solely  on the the instantaneous value of the central temperature, not on the  dynamical evolution.  This implies that for a fixed $T_{\rm Cc}$  such a shape  is the same for all bubble models accounting for shell evaporation.    Note that $\lambda_{\alpha}$ is also fixed for a given $T_{\rm Cc}$  and  that the prescription  of a particular bubble model merely implies a scaling of the  intensity amplitude. Thus, the results here presented  are  very  general and can be applied to the overall time evolution of a single object as well as to different objects with distinct parameters at a fixed age.  As we shall see,  the form of the assumed X-ray emissivity will play an important role in the  subsequent discussion.  
 
Approximations for $S_{\rm C}$ as a function of temperature were implemented in  equations (\ref{Pro1}) and (\ref{Pro2}) for a single power-law emissivity  and the realistic SS00 emissivity, respectively. The first one  allows a straightforward evaluation, while the last one, more elaborated, is necessary for a more robust discussion. In terms of $ s_{\rm }\in[0,R_{\rm cut}]$ the former becomes

   \begin{equation}
S_{\rm C\triangle}(s,\alpha;\xi_{\rm f},\xi_{\rm i})=\left(1-\frac{s^2}{R_{\rm p}^2}\right)^{-1/2} 
\left[ (1-\xi)^{\frac{1+2\alpha}{5}} \mathcal{F_{C,C}}(1-s,1-\xi;\alpha)\right]_{\xi_{\rm i}=\frac{s_{\rm cut}}{R_{\rm p}}}^{\xi_{\rm f}=\max(\frac{s}{R_{\rm p}},\frac{ R_{\rm s}}{R_{\rm p}})}, \label{Pro3}
\end{equation}

\noindent where the square-bracketed quantity written in terms of the  dummy variable $\xi$ is meant to be evaluated at the indicated limits (as the result of a definite integral would),  $\triangle$ indicates that we used  a pseudo-triangle approximation  (Appendix \ref{HT}),  $\max({ s_{\rm }}/{R_{\rm p}},{ R_{\rm s}}/{R_{\rm p}})$ incorporates  the   reverse shock position  and $\mathcal{F_{C,C}}(\tau,\tau_{\rm f};\alpha)$ given by  equation (\ref{catheti}) is an algebraic expression that involves the arcsin function. The central value of the spatial profile is obtained by evaluating the above equation at $ s_{\rm }=0$. The upper limit  for $\xi$ then becomes $\max(0,{R_{\rm s}}/{R_{\rm p}})={R_{\rm s}}/{R_{\rm p}}$ and we have that
\begin{equation}
 S_{\rm C\triangle}(0,\alpha;...)=\frac{5}{1+2\alpha} \left[\left(\frac{T_{\rm Rs}}{T_{\rm Cc}}\right)^{\frac{1}{2}+\alpha} -\left(\frac{T_{\rm cut}}{T_{\rm Cc}}\right)^{\frac{1}{2}+\alpha} \right],  \label{Proc}
\end{equation}

 \noindent where $T_{\rm Rs}=T_{\rm Cc}(1-{R_{\rm s}}/{R_{\rm p}})^{2/5}$ is the temperature at the reverse shock position according to some specific bubble model.  When $R_{\rm s}$ is ignored (which is equivalent to putting $R_{\rm s}=0$)  as in the C95  X-ray luminosity model,  ${T_{\rm Rs}}/{T_{\rm Cc}}$ in the previous formula becomes unity.

\begin{figure*}
\centering
\includegraphics[height=60mm] {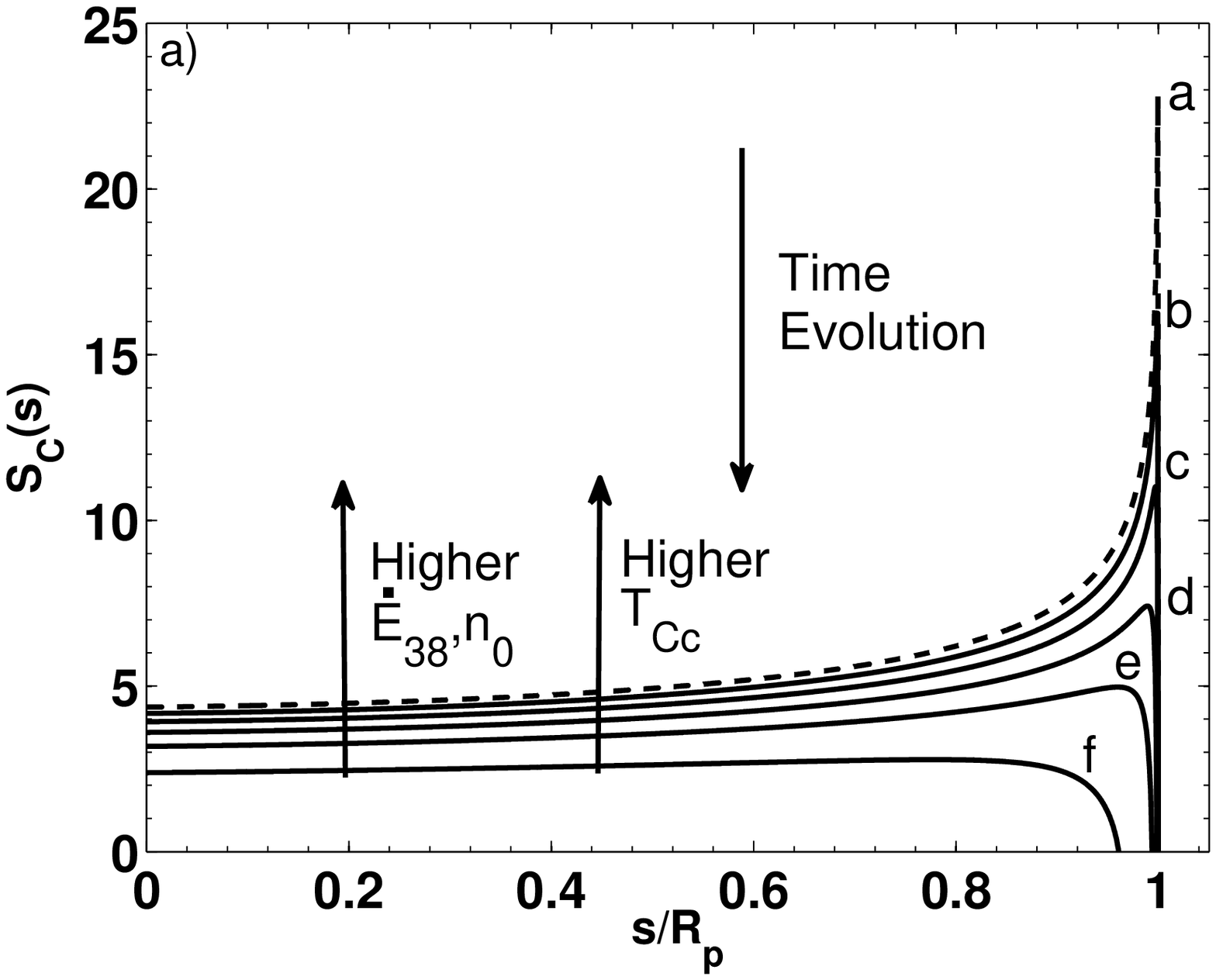}    
\hfill
\includegraphics[height=60mm] {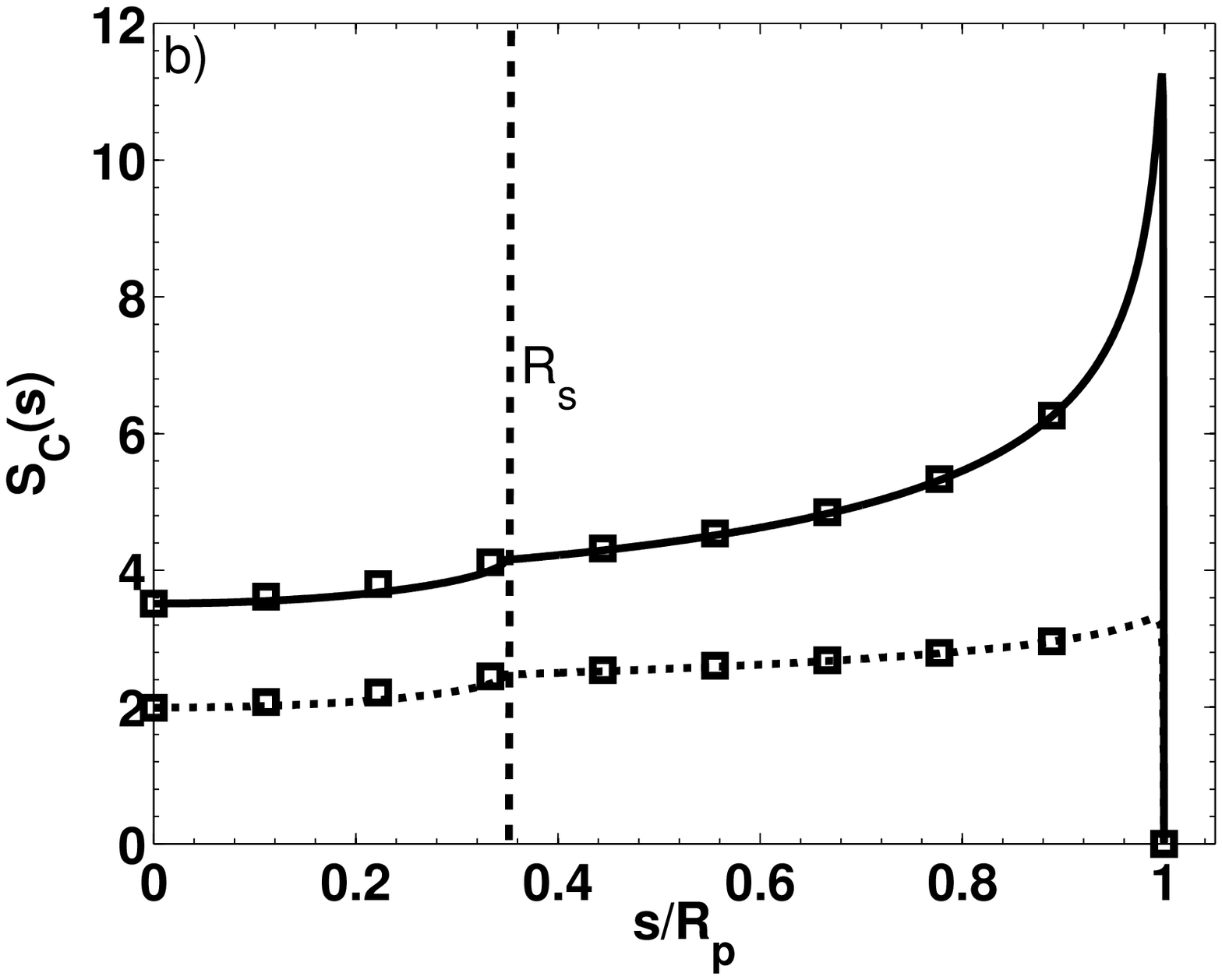}    
\caption{Panel (a): X-ray  spatial profiles for the C95 X-ray model with $Z_{\rm C}=$Z$_\odot$. The lines scale  uniformly with metallicity.  From top to bottom of the panel, the  central temperatures corresponding to the lines are  3.09,1.83,1.08,0.64,0.38, and 0.18, in units of  $10^7$ K.  If one assumes the   W77 model,   these temperatures might correspond to objects with the following  parameters: lines a-d, SSCs  with  masses of $10^7,10^6,10^5$ and $10^4$ M$_{\odot}$ an age   $t=3$-Myr, \rm{i.e.} $\dot E_{38}=3000, 300,30$ and  $3$ (Leitherer \& Heckman 1995), respectively;  line e, OB association with $\dot E_{38}=0.3$; and  line f,  W77  "typical interstellar bubble" blown by a star with $ \dot E_{38}= 0.0126$. For all cases $n_0=10$-cm$^{-3}$ and $V_8=1$. The reverse shock position is not included in the calculation. The top arrow indicates the direction of the profile time evolution for a fixed set of parameters ($E_{38},n_0$).  The bottom arrows    show  how the profile varies  with the energy input rate, the ISM density and the  central temperature. Panel (b): spatial profile for  cluster  c in panel (a)  when a constant  emissivity ( $\alpha=0$, solid line) and a bremsstrahlung dominated one ($\alpha={1}/{2}$, dotted line) are considered. The respective  pseudo-triangle  approximations (Appendix  \ref{HT}) are marked with squares. The reverse shock position was included in these models.    \label{fig:fig7}}
\end{figure*}

\begin{figure*} 
\centering

\includegraphics[height=55mm] {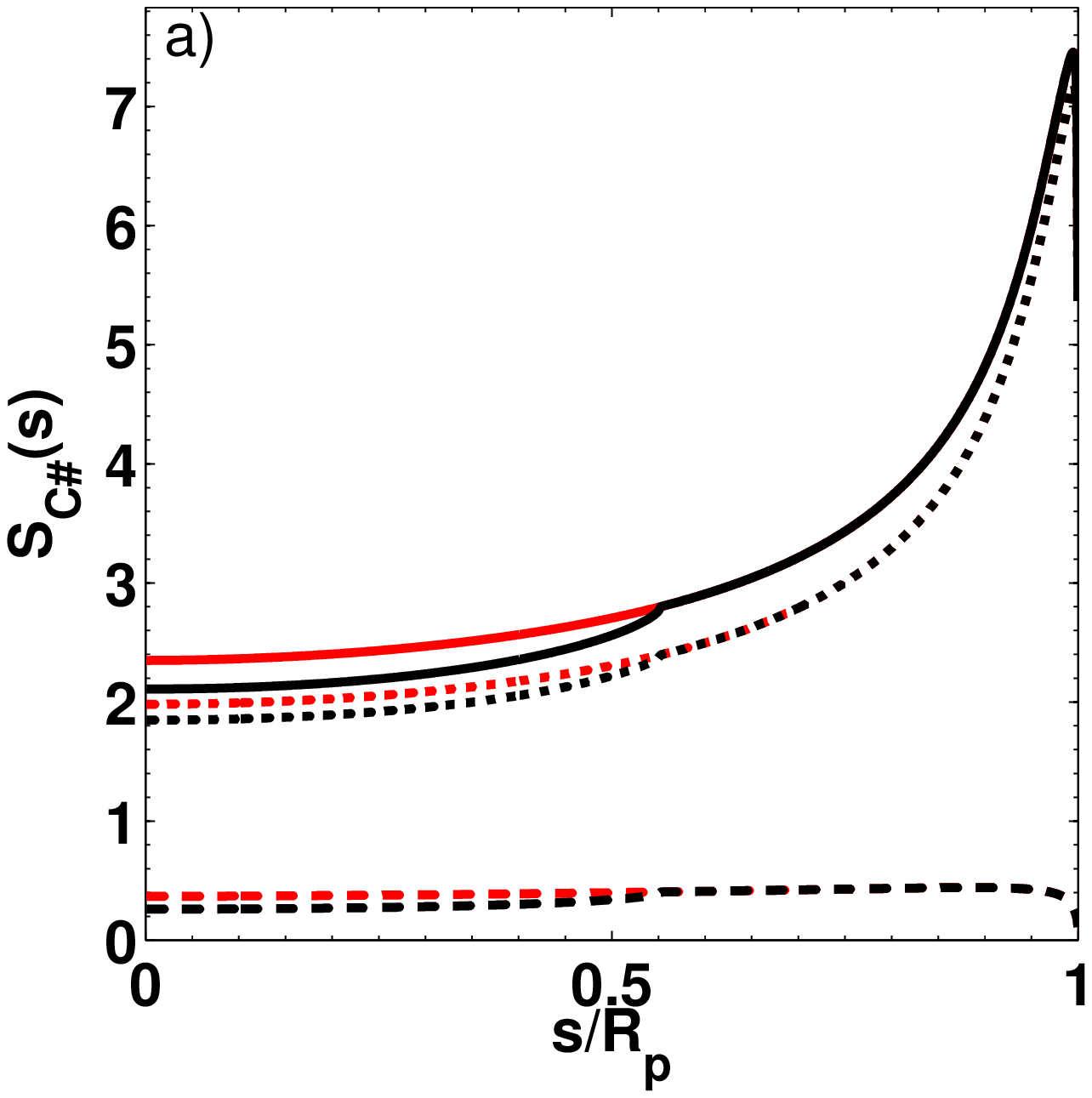}  
\hfill 
\includegraphics[height=55mm] {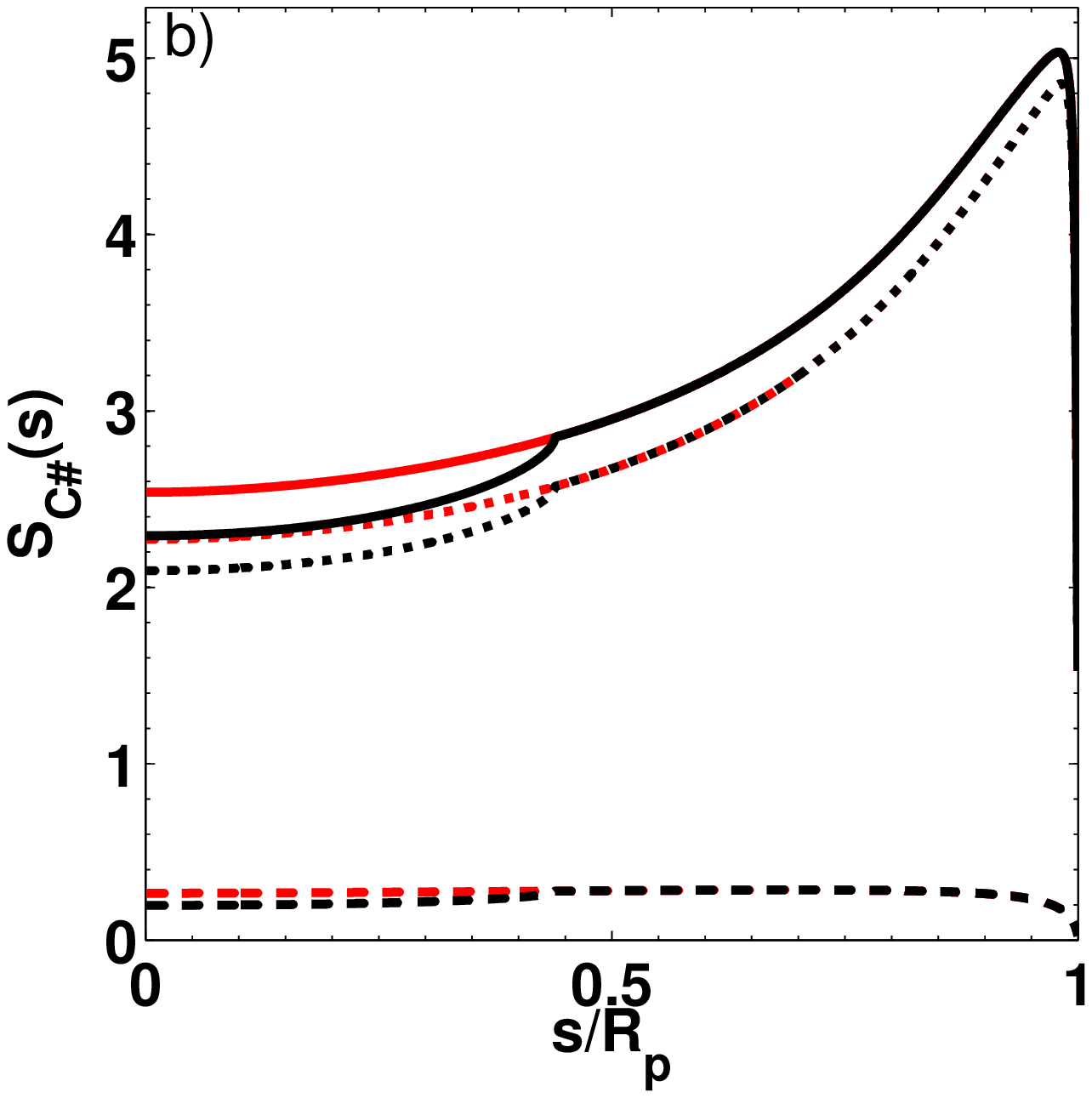}
\hfill 
\includegraphics[height=55mm] {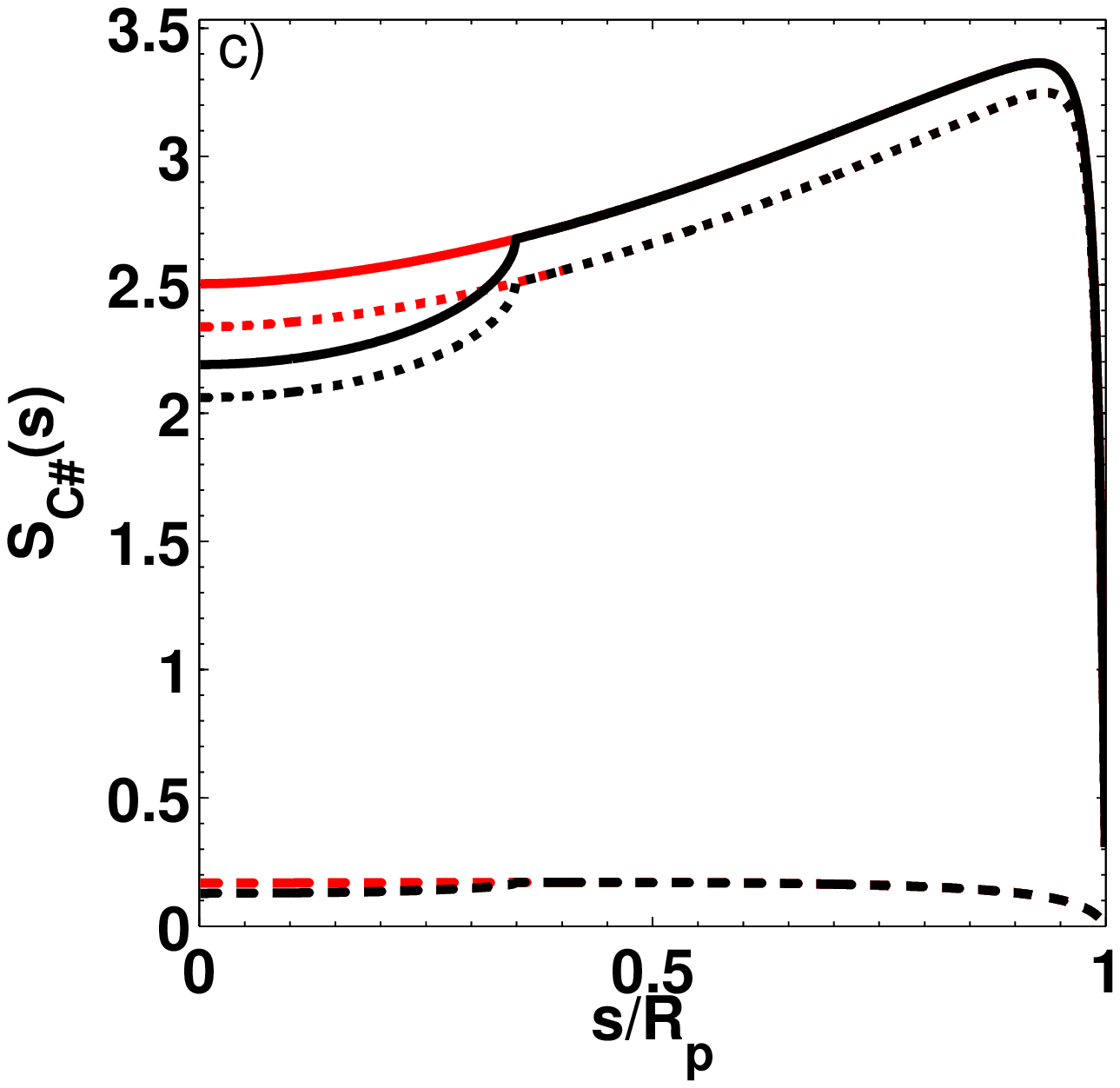}    \\
\vfill
\includegraphics[height=55mm] {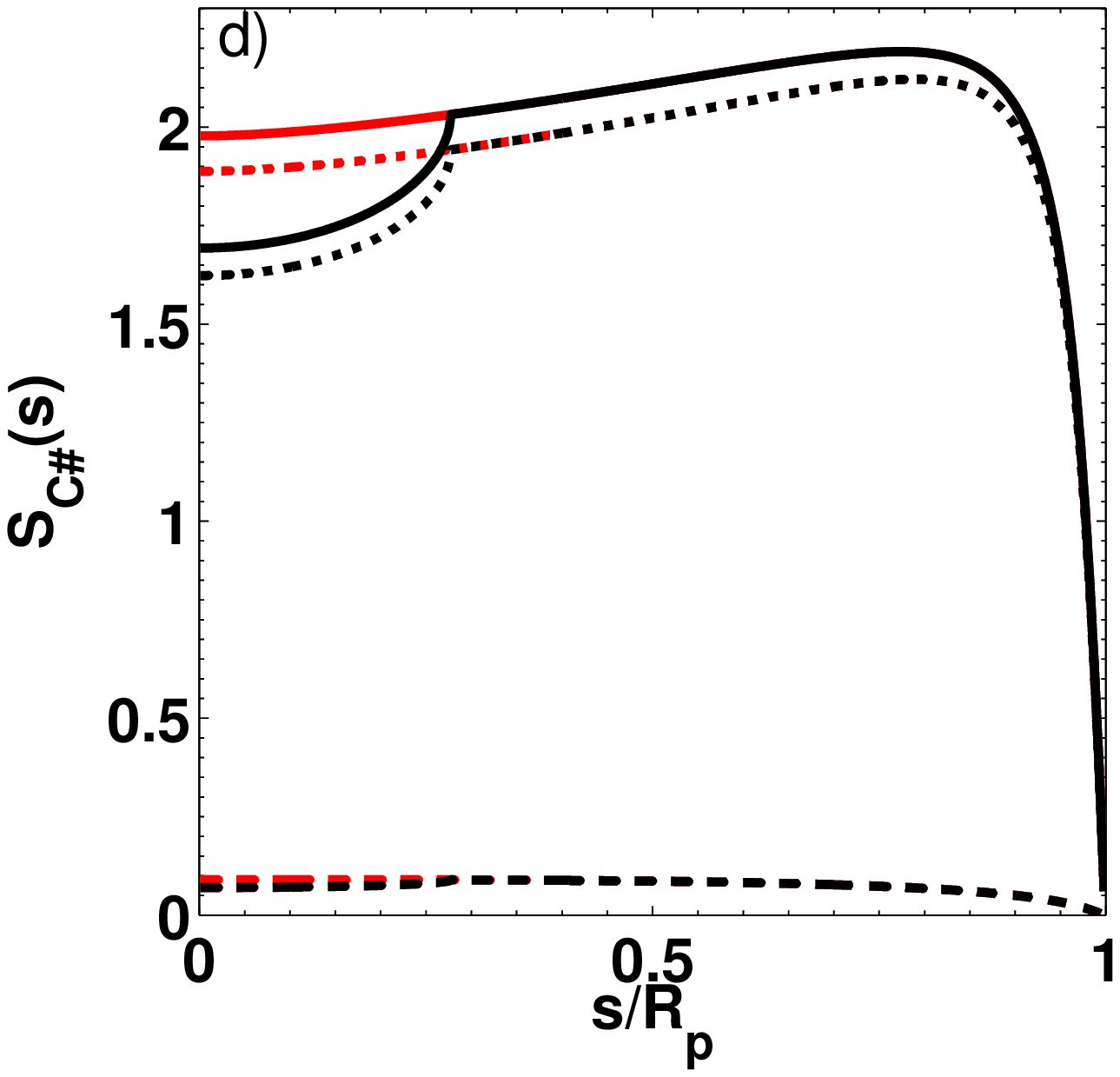}  
\hfill
\includegraphics[height=55mm] {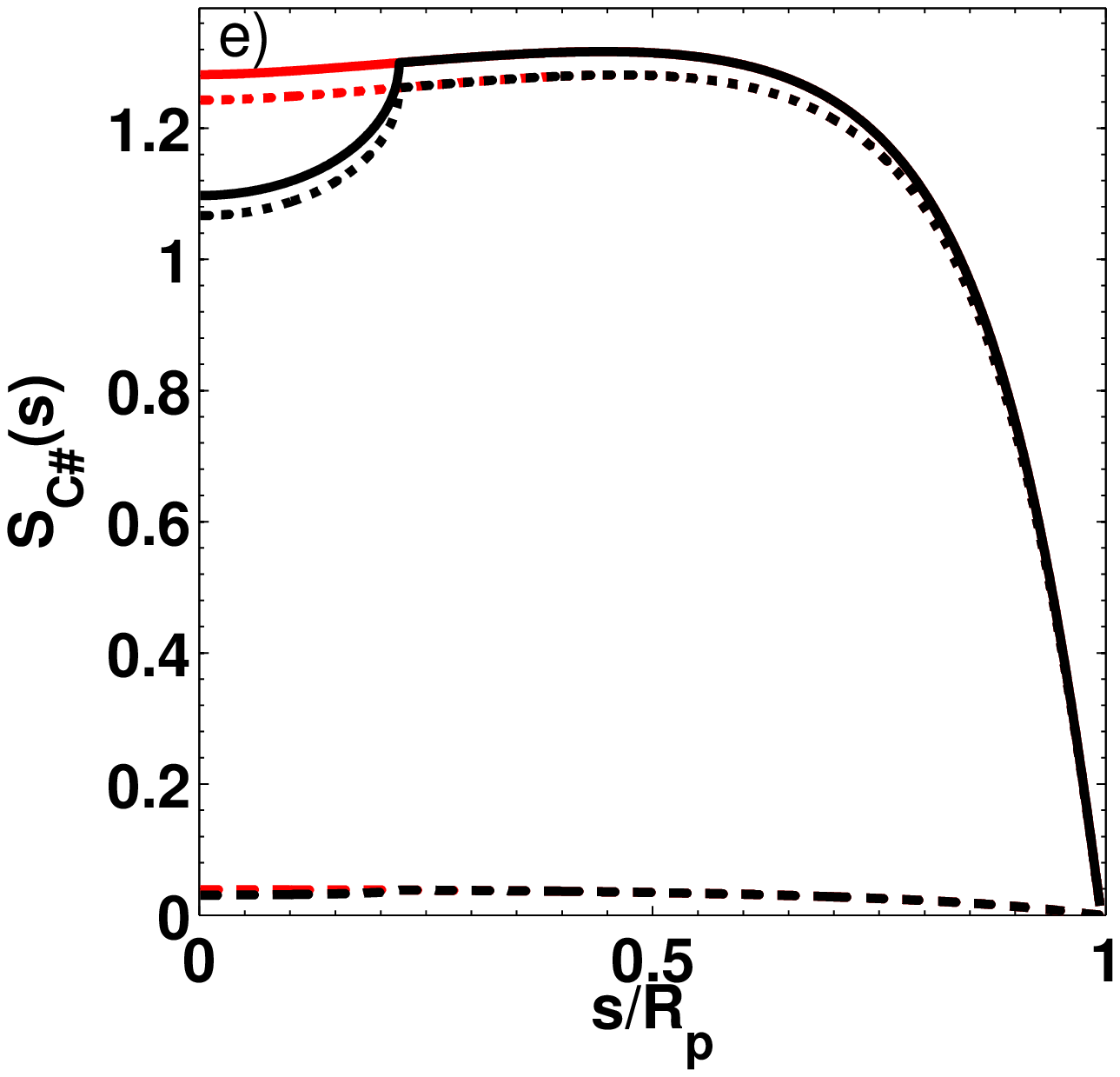}    
\hfill
\includegraphics[height=55mm] {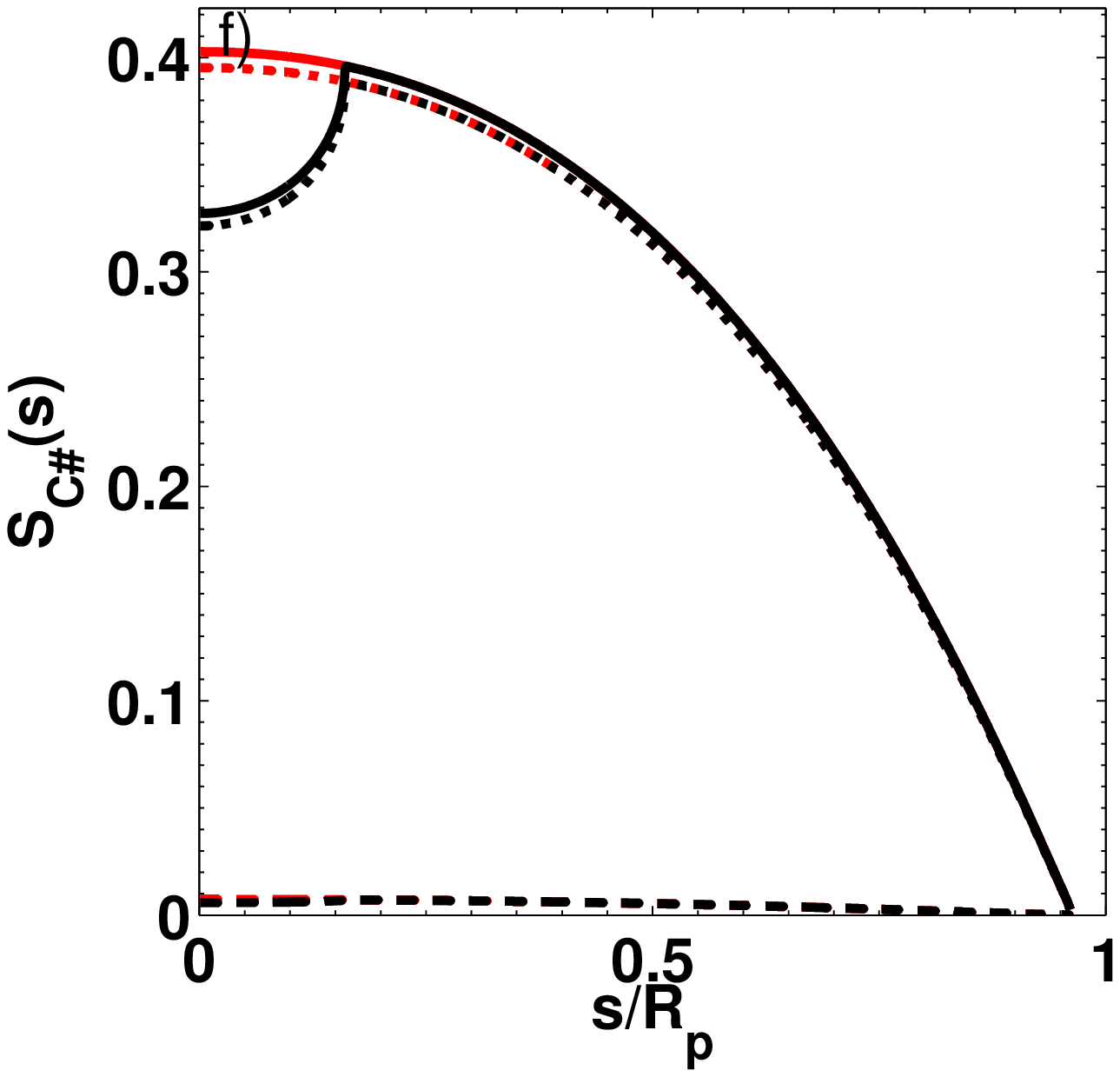}

  \caption{ Spatial profiles in the soft band for the sample shown in Fig. \ref{fig:fig7}.   Here, the SS00 emissivity was considered. The contributions from hydrogen (dashed lines) and metals (dotted lines)  were separated.   The  solid lines represent the sum of both contributions. The  dark lines show the profiles  when  $R_{\rm s}$ is included  in the calculations. The red lines  are obtained when   the reverse shock position is neglected. Notice that by our definition, the bubbles in panels  (a) and (b) are not well-developed (i.e. $R_{\rm s}/R_{\rm p}>0.4$). None the less,  they have high central temperatures;    thus, the position of the reverse shock does not affect their X-ray luminosity significantly (see Fig. \ref{fig:fig6}). \label{fig:fig8}} 
\end{figure*}

As a first step, we obtain the surface brightness corresponding to the C95 X-ray luminosity model. Theoretically,  this model is equivalent  to considering  that the  X-ray emission is proportional to the emission measure, i.e that  $L_{\rm XC}=Z_{\rm C}\Lambda_0 EM_{\rm C}$   ($\alpha=0$, $\Lambda_{\rm X}=Z_{\rm C}\Lambda_0$). According to this,  one has that  while interstellar bubbles present a practically  flat  profile,  superbubbles driven by the most massive SSCs  exhibit a noticeable limb brightening  near the contact discontinuity due to their more powerful mass and energy injection rates, which  translate into  higher central temperatures. This is shown in panel (a) of Fig.  \ref{fig:fig7}. The same panel also shows the direction of the   spatial profile expected time evolution   for a fixed set of model parameters. As time progresses, the profile becomes flatter and weaker due to a  continuous and slow\footnote{It is slow except for exceptionally  short times ($t\ll1$ Myr) and low densities ($n_0<1$ cm$^{-3}$). \label{foot}} decrease of the central temperature. However, the energy injection rate is not fixed as it increases by a factor of ~3 after 3-Myr (Leitherer \& Heckman 1995). This implies that a slight   limb-brightening  increment is expected for  both bubbles and star clusters at this age,   when the first supernovae start to explode.  These results derived from our  analytics and the ones presented below (Fig. \ref{fig:fig8}) can also  be used to reconstruct   the  simulated X-ray brightness profiles of Wrigge et al. (2005) for the Wolf-Rayet bubble NGC 6888. In their work,  the Garc\'ia-Segura \& Mac Low (1995) bubble model was used and the required X-ray properties were obtained by  ad hoc adjusting  the thermal conductivity coefficient in order to vary  $T_{\rm Cc}$. We remark that  the different stratification  of the ISM that they considered does not produce a divergence from the  analytical shape of our spatial profiles, since as it was previously stated,  the thermodynamical configuration of  region C is  essentially the same. There are however  two  conspicuous differences with respect to the W77 model: in their case  a)  the spatial profile evolves faster ($T_{\rm Cc}\propto t^{-2/7}$ and $n_{\rm Cc} \propto t^{-12/7}$) and b) the predicted   size of the bubble is not the same.

In panel (b) of Fig. \ref{fig:fig7},  the C95 model is compared with one that assumes   an emissivity of  the form $\Lambda_{\rm X}=Z_{\rm C}\Lambda_{\rm \alpha}T^{{1}/{2}}$, \rm{i.e.} a bremsstrahlung dominated emission, which is a  commonly used approximation. The main difference between them is that the second one practically does not show any increment in surface brightness near the contact discontinuity.  This is in better   agreement with  observations  of  the  superbubbles  associated to the Rosette and Omega Nebulae  (Chu, Gruendl \& Guerrero 2003; Townsley et al. 2001, 2003) for which no limb-brightening was detected. One then might suspect that the sharp peak predicted  from a constant emissivity could be an artefact. If   an emissivity that decreases  towards the cut-off temperature is assumed, equation (\ref{Pro3})  reveals that the  brightness  is more similar to the central one. Furthermore, this also applies to  the projected temperature (Section \ref{PTC}).

To constrain the effect of $\Lambda_{\rm X}$, we considered the  SS00 X-ray emissivity function and   obtained  the surface brightness profile   for the set of objects shown   in Fig.  \ref{fig:fig7}.   This  test sample well covers  the temperature  range in which bubbles are observed, from a to f  the  central temperatures are   $T_{\rm Cc}=\{3.09,1.83,1.08,0.64,0.38,0.18\} \times 10^7$ K.   For the current analysis, it  is convenient to use formula (\ref{Pro2})  --which gives the spatial profile as a function of temperature--  together with Table \ref{Table2}. Alternatively, one can use Table \ref{Table3} to  numerically  integrate   instead of  summing up.  For this option, the sum symbols in (\ref{Pro2})  are replaced by $\lambda_0 S_{\rm C\#}( s_{\rm })$  with $\lambda_0=3$, $ S_{\rm C\#}( s)=({5}/{2\Lambda_0}) \int_{\min( \vartheta _{\rm  s},\vartheta_{\rm Rs})}^{\vartheta_{\rm cut}} (\vartheta^{2}-\vartheta^{-{1}/{2}})\Lambda_{\rm X}(\vartheta,Z_{\rm C}) [(1-\vartheta^{{5}/{2}})^2-{ s_{\rm }^2}/{R_{\rm p}^2}]^{-1/2} \rm{d}\vartheta$  and $\Lambda_0=3\times 10^{-23}$ erg s$^{-1}$.  The results  are shown in Fig. \ref{fig:fig8}.  We have that in the soft band  the surface brightness profile evolves a follows: it  begins in  a limb-brightening mode  for the hottest bubbles ($T_{\rm Cc}>10^7$ K), later it passes through an 'unstable' almost flat mode that occurs when the central temperature is  close to   $T_{\rm Cc}\approx 6.5 \times 10^6$ K (where the X-ray emissivity has an absolute maximum for $Z \gtrsim$Z$_\odot$), and  it ends in  a limb-darkening mode  (centre-brightening)   for lower central temperatures ($T_{\rm Cc}\approx$ 1--2 $\times 10^6$ K). At early ages, $t_6\ll 1$, the limb-brightening  can be extremely strong due to both an increase of the main peak nominal value and the visual effect produced by the  relative closeness of the reverse and leading shocks.  This  last effect would  give the central region a void appearance, if it were not for the contributions of regions  A and B.

Finally, we found  that the  surface brightness  was  much smaller in the hard band.  For  stellar blown bubbles,  the hard band emission was practically negligible. In shape, the profiles were  similar to the one  shown in  Fig. \ref{fig:fig11}. 

  \subsection{Projected temperature}\label{PTC}
\begin{figure*}
 \centering

\includegraphics[height=55mm] {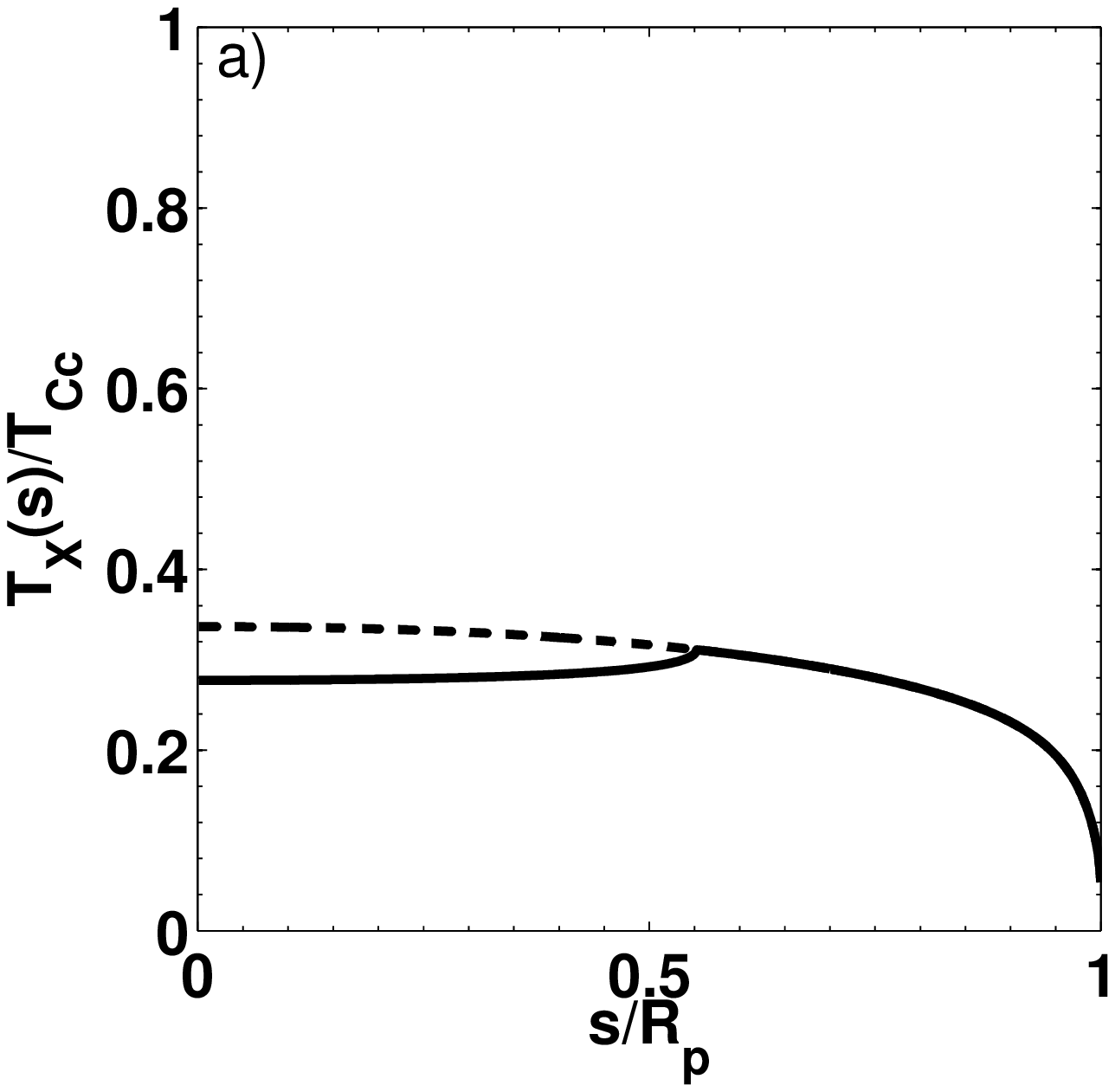}  
\hfill 
\includegraphics[height=55mm] {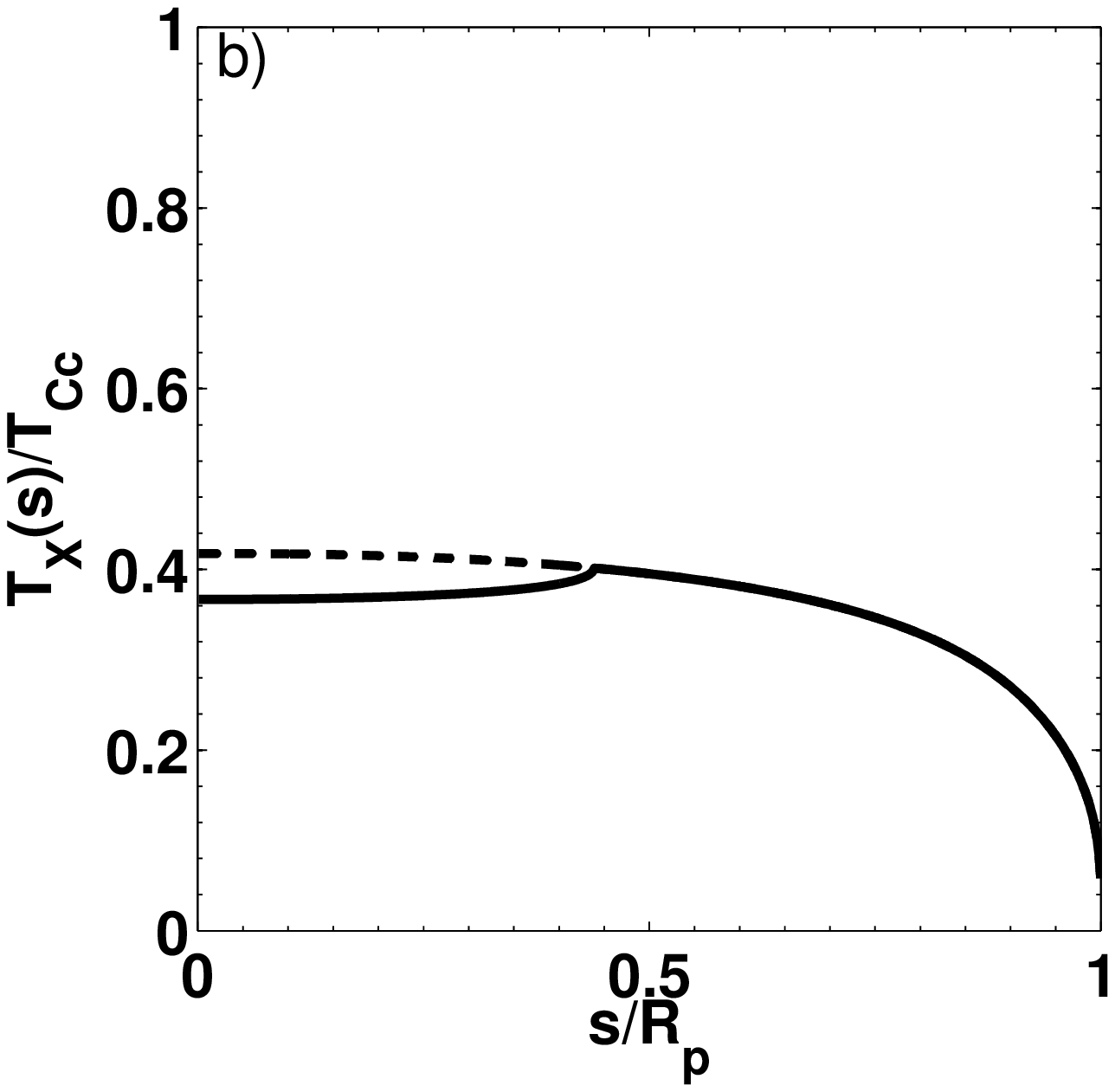} 
\hfill
\includegraphics[height=55mm] {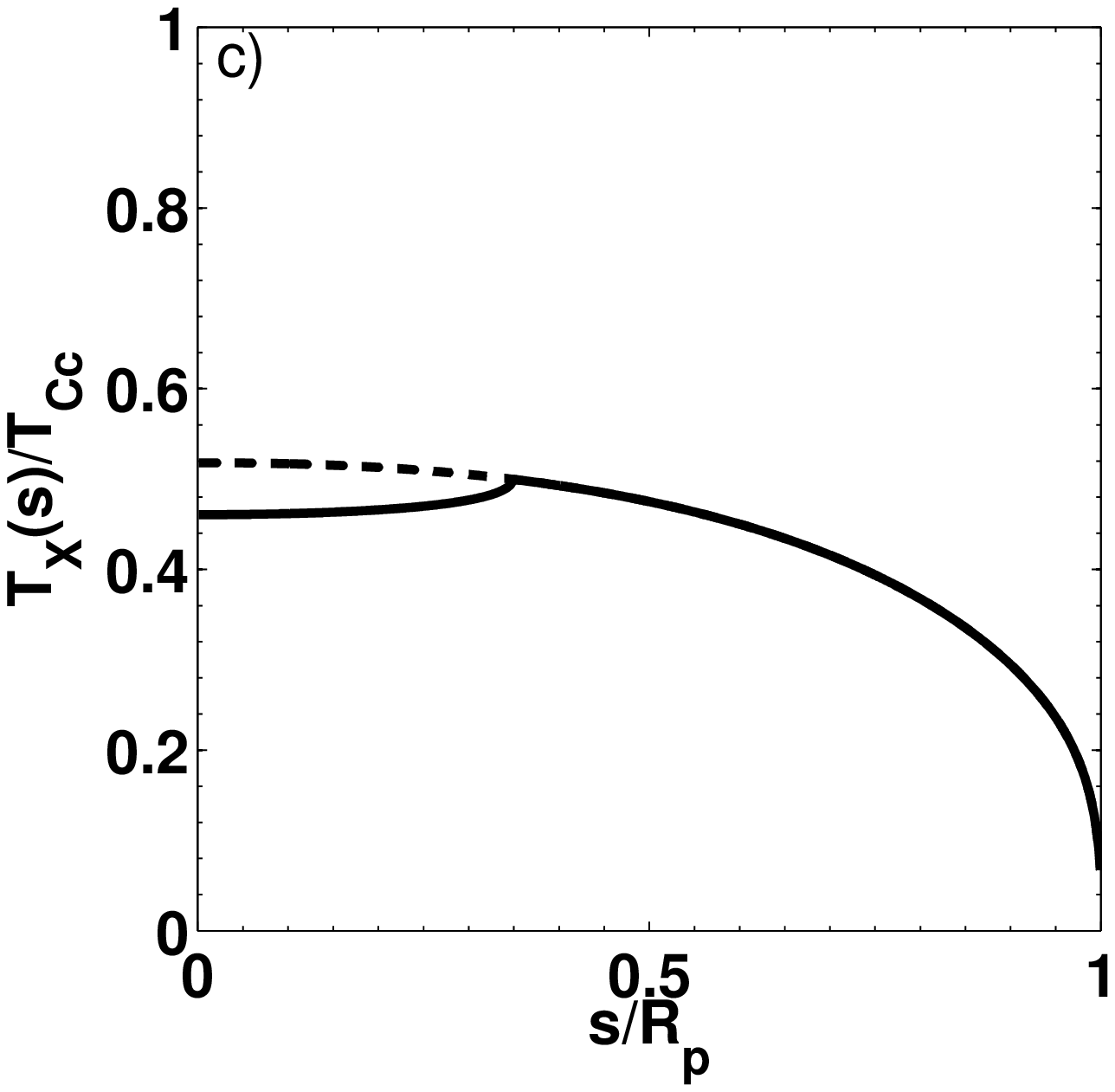} \\
\vfill
\includegraphics[height=55mm] {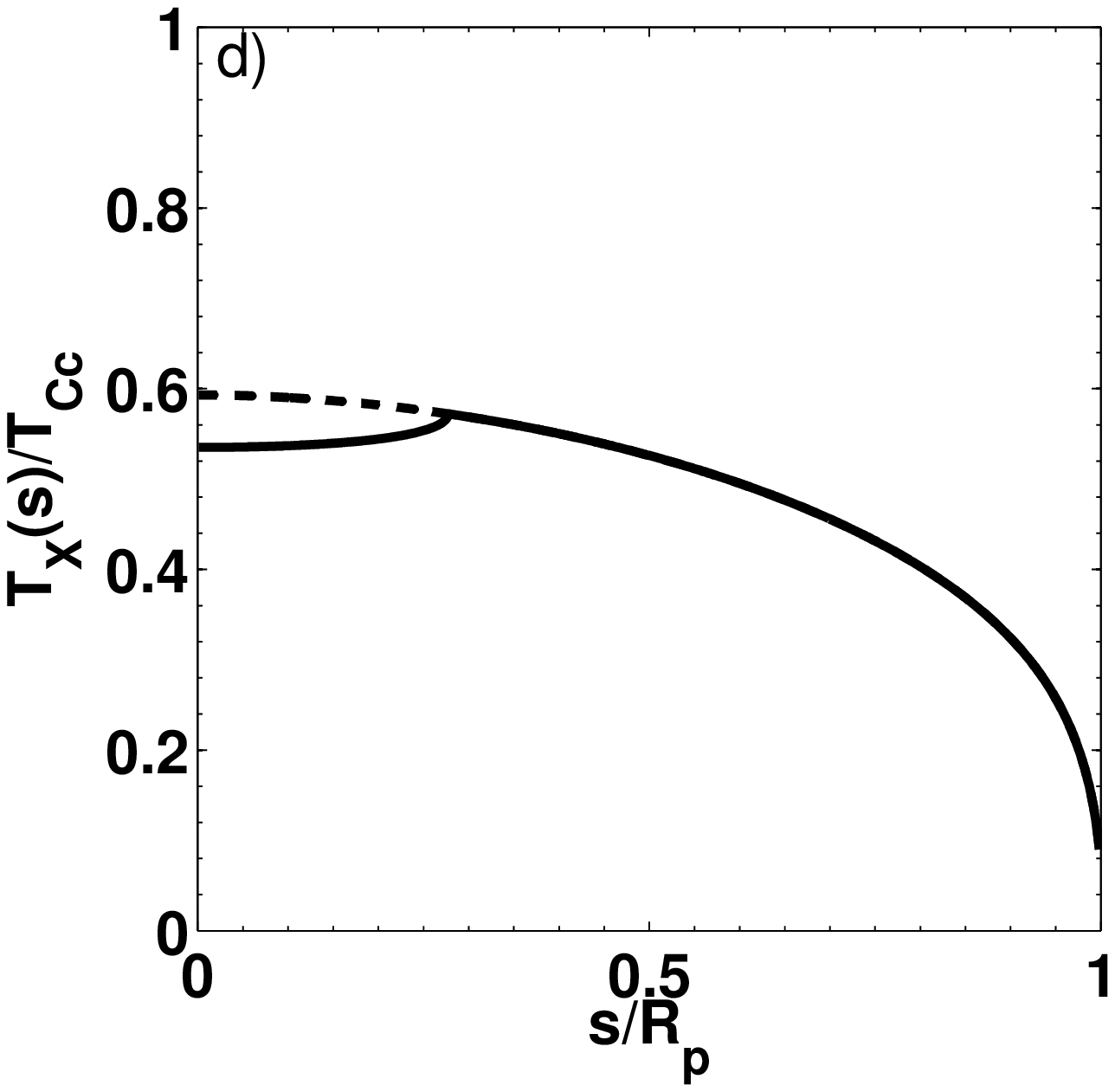}   
\hfill
\includegraphics[height=55mm] {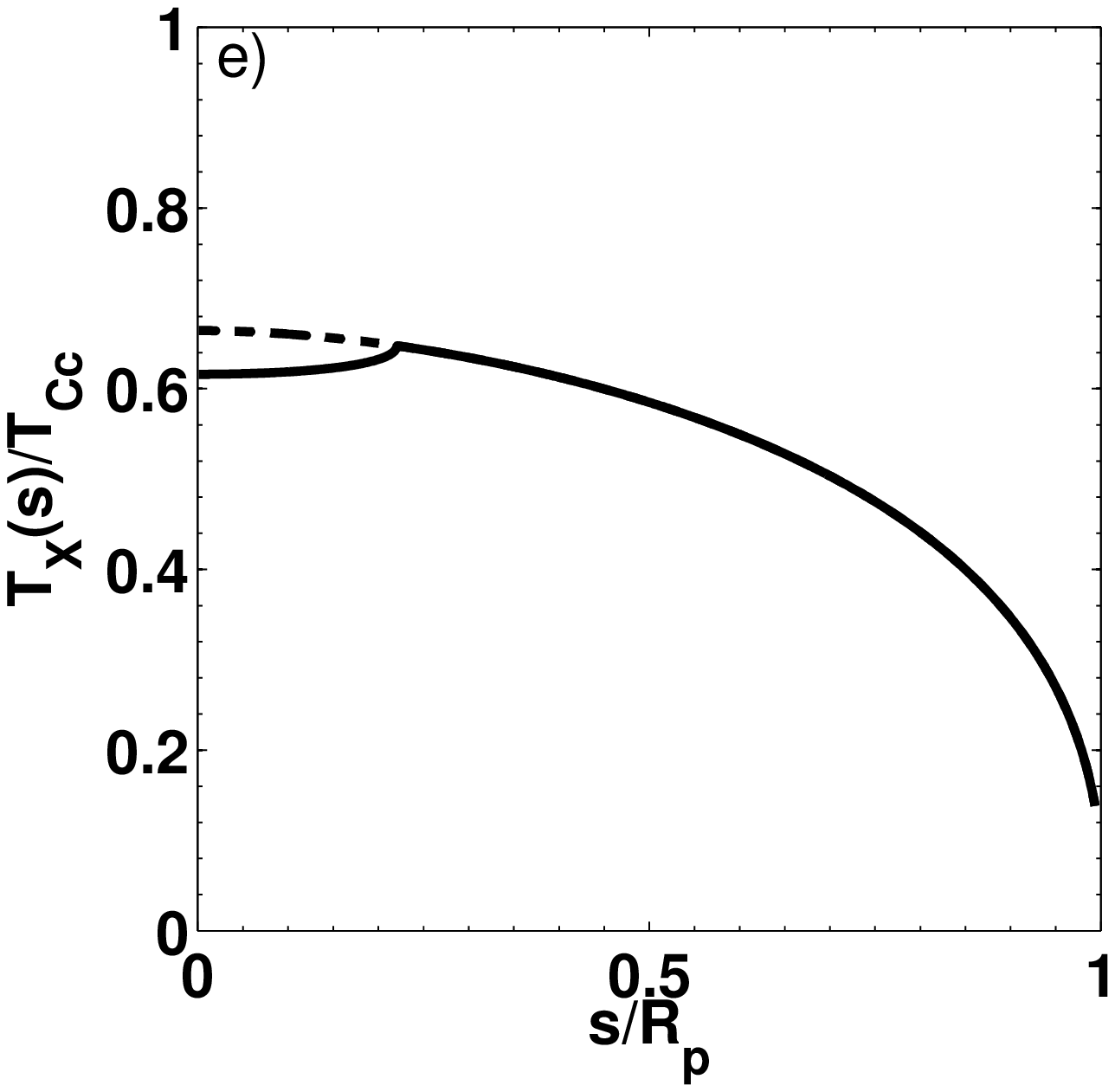} 
\hfill
\includegraphics[height=55mm] {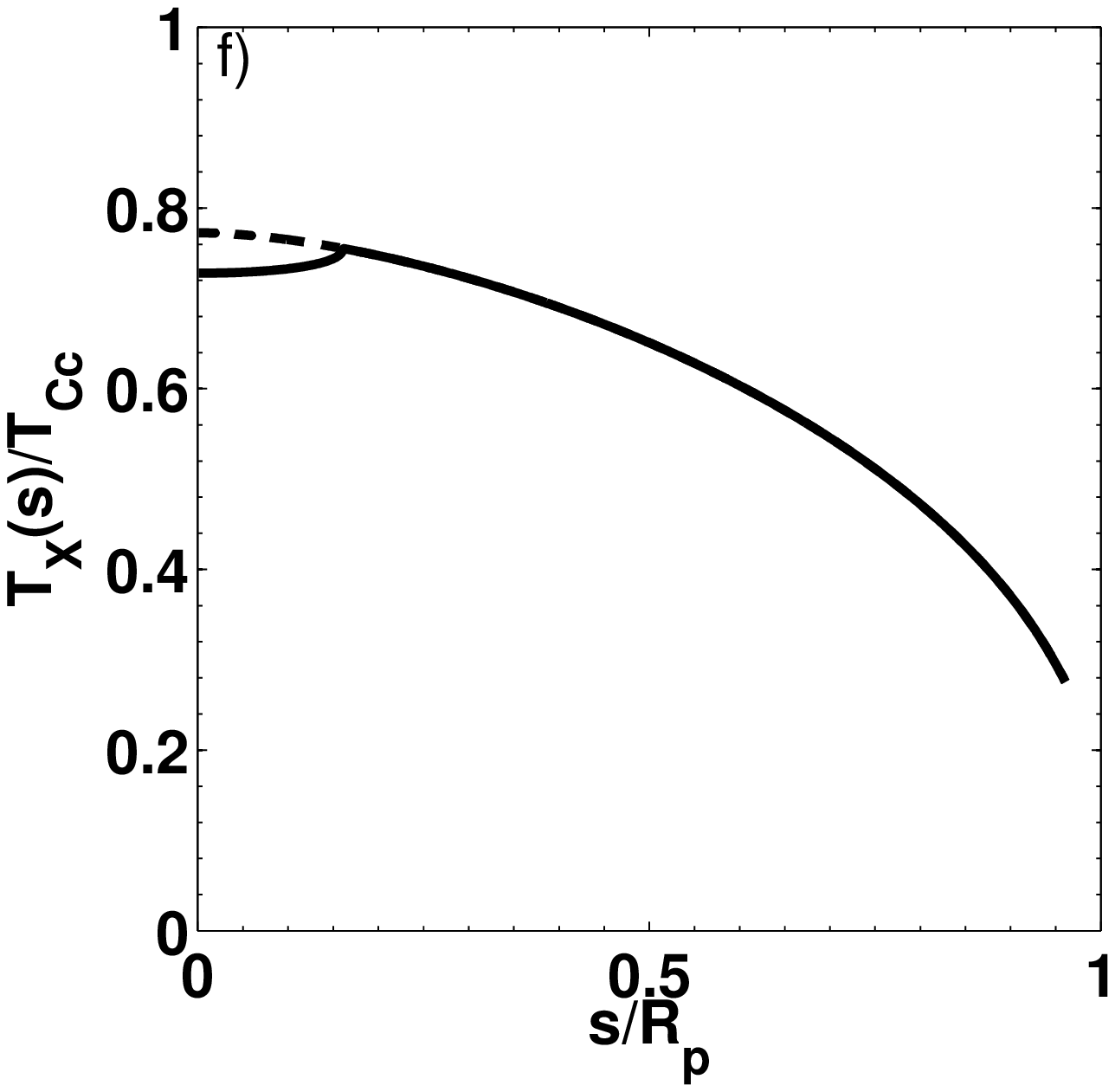}
\caption{Projected  temperatures in the soft band for  the bubbles shown in Fig. \ref{fig:fig7} and Fig. \ref{fig:fig8}. The hottest bubbles tend to present a more uniform profile. However, the ratio of their projected temperatures to the central one is lower. In spite of this, in some cases their projected temperatures can be of the order of 1 keV.  \label{fig:fig9}}
\end{figure*}

  As it was indicated in Paper I, most of  the superbubble X-ray emission is in the soft energy band because of  the  contribution of the outermost layers with temperatures in the range of $T\approx$ 5$\times10^5$--$10^6$ K. However, the projected  temperatures  can be fairly larger and  more similar to those near the centre of the superbubbles. In some cases, they can be of the order of $1$ keV. Superbubbles with hot interiors ($T\approx 10^7$ K) have already been reported by Townsley  et al. (2001, 2003) and Dunne et al.  (2003).  In Fig. \ref{fig:fig9}, we present the   projected temperature  curves corresponding to Fig. \ref{fig:fig8}.     It follows that  according to the models, the spectra of shell-evaporating superbubbles  could be more adequately described by a  two temperature plasma model: a hot component to account for  the central part and most of the bubble area,  and a warm one to  account for the outer layers. These theoretical temperature  maps  are independent from the speed of the geometrical evolution because they were derived  from a ratio, see equation (\ref{PTemp}).  For the hottest bubbles (i.e. when  $T_{\rm Cc}\ga 10^7$ K), $T_{\rm X}(0)$ is  a good measure of the average  projected temperature    inasmuch as  their temperature maps are quite flat   almost up to  the contact discontinuity. Likewise, for  'warm' bubbles ($T_{\rm Cc}\sim 10^6$ K),   the  projected central value is  a good approximation  up to half  of the projected size. Because of this, in what follows we shall indistinctly refer to the  average projected temperature and $T_{\rm X}(0)$ as $T_{\rm XC}$.
 
\begin{figure}
\centering
\includegraphics[height=60mm]{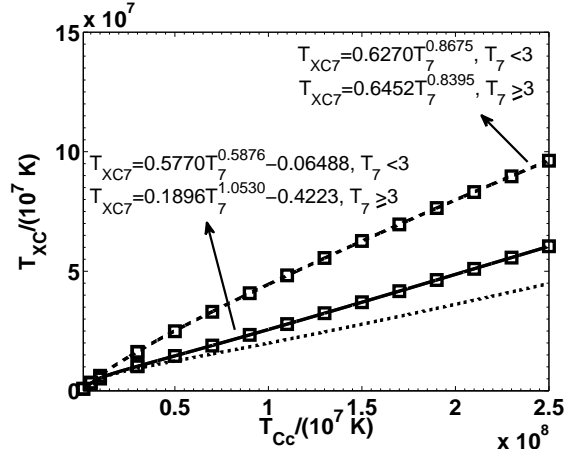}
\caption{ Bubble projected temperature in the soft band  as a function of $T_{\rm Cc}$ and metallicity. Dashed line: $Z_{\rm C}=0\,$Z$_\odot$, solid line: $Z_{\rm C}=$Z$_\odot$ and dotted line: $Z_{\rm C}=5\,$Z$_\odot$. The reverse shock position was neglected. Analytical expressions for the  best-fitting lines  (squares) to  the hydrogen component and to the solar metallicity case  are  shown.   $T_{\rm XC7}$ and $T_7$ are the projected  and  central temperatures   in units of  $10^7$ K, respectively. As expected from the shape of the emissivity function, the projected temperature decreases with increasing metallicity.\label{fig:fig10}}
\end{figure}

   For the C95 model, the temperature map can be obtained from formula (\ref{Pro3})   with $\alpha=1$ normalised by itself  with $\alpha=0$.  The  central value can be obtained in the same way from equation (\ref{Proc}). Its  low and high temperature limits are noteworthy: $T_{\rm XC}\rightarrow ({1}/{3})T_{\rm Cc}$ as   $T_{\rm Cc}\gg T_{\rm cut}$ and $T_{\rm XC}\rightarrow T_{\rm Cc}$ as $T_{\rm Cc}\rightarrow T_{\rm cut}$. Here, the temperature shift    is  an exclusive consequence of the thermodynamical structure established by evaporation. Our weighting scheme  considers additionally the effect of $\Lambda_{\rm X}$  and  predicts  a central temperature (see Appendixes \ref{Bev} and \ref{HT}) 
 
 \begin{equation}
 T_{\rm XC}=\frac{ \displaystyle{ \sum_{\alpha, \vartheta_{\rm i},\vartheta_{\rm f}<\vartheta_{R_{\rm s}}}^{\vartheta_{\rm f}=\vartheta_{\rm cut}} }\left(\frac{1}{3+2\alpha}\lambda_{
 \alpha+1} \xi^{\frac{3}{2}+\alpha}\right)_{\xi=\vartheta_{\rm f}} ^{\xi=\min(\vartheta_{\rm i},\vartheta_{\rm Rs})}}{ \displaystyle{ \sum_{\alpha, \vartheta_{\rm i},\vartheta_{\rm f}<\vartheta_{R_{\rm s}}}^{\vartheta_{\rm f}=\vartheta_{\rm cut}} }\left( \frac{1}{1+2\alpha}\lambda_\alpha \xi^{\frac{1}{2}+\alpha}\right)_{\xi=\vartheta_{\rm f}} ^{\xi=\min(\vartheta_{\rm i},\vartheta_{\rm Rs})}}, \label{TXX}
 \end{equation} 

\noindent where $\vartheta_{\rm i }$ and $\vartheta_{\rm f}$ are  the normalised to $T_{\rm CC }$ upper and lower temperature limits  for which the power law with index $\alpha$ is valid.  For a realistic emissivity the  intervals and  indexes are indicated in Table \ref{Table2}. The sum starts on the interval in which  $T_{\rm Rs}$ falls. This means that  the reverse shock position is embedded in the formula; none the less, given that  the projected temperatures at  intermediate projected radii are alike to the one obtained at the  centre when $R_s$ is neglected (Fig. \ref{fig:fig9}), it is also useful to remove its effect. This is accomplished by setting $\vartheta_{\rm Rs}=1$ (\rm{i.e.} $R_{\rm s}=0$).  Notice that as the coefficients $\lambda_\alpha$ depend on metallicity so does equation (\ref{TXX}). The value of $T_{\rm XC}$ for different central temperatures and abundances is shown in Fig. \ref{fig:fig10}. The best-fitting equations are also displayed.  Note that the value  of $T_{\rm XC}$ decreases as the  metal abundance increases. This occurs because the emission at lower temperatures increases with the metal abundance.

According to our analytics (\rm{i.e.} no observational artefacts)   $T_{\rm XC}$  just depends on the  central temperature. Nevertheless,  it should not be assumed to be $T_{\rm Cc}$,  because  it is  only  a fraction of the later.  It  should not be taken  either as a temperature similar to those  of the outer layers.  This is particularly important for  interpreting observations of hot bubbles with $T_{\rm XC}\approx10^7$ K (see Section \ref{M 17}).

\section{BUBBLES WITHOUT EVAPORATION}\label{Bnev}

If the ISM encircling  the superbubble owns an intrinsic strong magnetic field, the supposed dissemination  of cold material to the hot bubble interior  could  be inhibited (Band \& Liang 1988;   Soker 1994; Dunne et al. 2003). The magnetic field prevails even after the material in question has been trapped in the cold outer shell and their components  tangential to the contact discontinuity could critically  hinder  the evaporative process.
In this case, the expression for the X-ray luminosity during the  sweep-up phase  is very similar to equation (10) in Paper I,  but with a coefficient of 8.05 instead of 9.5 and the proper correction for the reverse and main shock positions with respect to the superbubble completely adiabatic stage.

\begin{figure}
\centering
\includegraphics[height=60mm]{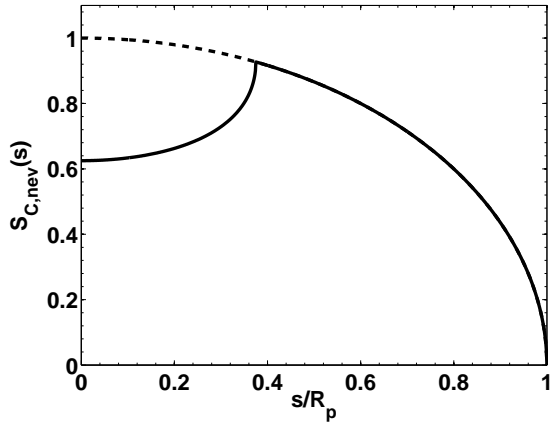}
\caption{Surface brightness spatial profile for a superbubble generated by a cluster of $10^5$  M$_\odot$ when no evaporation occurs. Here,  $t_6=10$ ($\dot E_{38}=30$), $n_0=10$-cm$^{-3}$ and $V_{8}=1$. The solid line traces the  profile considering the  reverse shock position. The dashed line indicates how the inner region would look if the reverse shock is neglected.  A slight limb brightening near the reverse shock is expected together with a smooth transition to  a limb darkening towards the contact discontinuity.
\label{fig:fig11}}
\end{figure}

 In this case, the integral that gives the surface brightness  is elemental
  \begin{equation}
\sigma_{\rm C,nev}( s_{\rm })=A_{\rm C,nev}(t)\Lambda_{\rm X,-23}(\overline T_{\rm C},Z_{\rm C}) 
\left[\left(1-\frac{ s_{\rm }^2}{R_{\rm p}^2}\right)^{1/2} -\right.
\left.\frac{R_{\rm s}}{R_{\rm p}}\left(1-\frac{ s_{\rm }^2}{\max( s_{\rm },R_{\rm s})^2}\right)^{1/2}\right], \label{SXCnev}
\end{equation}

\noindent where
 
   \begin{equation}
A_{\rm C,nev}(t) =  2\times 10^{-23}(1.454)^2\frac{(\gamma+1)^2\dot E^2}{4\pi^2\mu_{\rm n}^2(\gamma-1)^2V_{\rm \infty P}^6} \frac{R_{\rm p}}{R_{\rm s}^4}
= (1.80\times 10^{30}\mbox{erg s$^{-1}$ pc$^{-2}$})\frac{n_0\dot E_{38}}{V_{8}^4}t_{6}^{-1}, \label{AXCnev}
\end{equation}

\noindent and we have used  the Koo \& McKee (1992b) average  post-shock density. In this case, the average temperature of the shocked gas is close to the temperature at the star cluster surface, $\overline T_{\rm C}\approx T_{\rm sc}$, because a fraction of the thermal energy is used to displace the outer shell of swept-up interstellar material.  For very fast winds, this hot plasma is expected to emit  in the hard X-ray band. The corresponding surface brightness  spatial profile, $S_{\rm C,nev}$, is  given by the quantity between square brackets in equation (\ref{SXCnev}). It is presented in Fig. \ref{fig:fig11}.
 
\section{A COMPREHENSIVE X-RAY PICTURE} \label{Pic}

 In Paper I, we used the total  X-ray luminosity to compare the emission of  superwinds and their associated superbubbles.  In this paper,  the corresponding X-ray models have been improved by refining the analytics and incorporating explicitly a realistic emissivity function that  separates the contributions from hydrogen and metals. Here, we give a description of how to proceed in order to obtain the theoretical X-ray picture.  All  the parameters appearing in our formulae are expected to be derived observationally.  The models can handle observational artefacts as long as they could be parameterized as power laws or series in the temperature, as in Mazzotta et al. (2004, see Sections \ref{WindT} and \ref{PTC}).

  The theoretical diffuse X-ray luminosity of  the star cluster core can be obtained from equations (\ref{LXA1}) or (\ref{LXA2}). The  theoretical   X-ray luminosity of the free wind   is $\sim 25$ per cent that of the core and its extension can be estimated from equation (\ref{RTout}).  The superwind X-ray surface brightness  profile is given by equations (\ref{SXA}) and  (\ref{AXAB}) which may  be helpful to extrapolate the  profile of the  observed diffuse emission. Estimators of  the  wind spectroscopic temperature  ($T_{\rm X}$) are given by equations (\ref{TX1}) and (\ref{TX3}). All the estimators  yield that  for solar metallicity  $ T_{\rm X}$  is very close to the core central temperature, which is  given by (\ref{polyTc}). The last  estimator  depends explicitly on $Z$.   As  the polytropic index was left as a free parameter, these  models admit a large variety of wind configurations, as explained in Section \ref{Polywinds}.
  
 The bubble X-ray luminosity can be obtained from  (\ref{LXCW77}) if the Weaver et al. (1977) model is assumed; however our approach is more general.  For bubbles, once that the observational data would have been reduced and analysed, one of   the first things to do for comparing with \emph{any}  model that considers thermal evaporation from the outer shell is to derive the theoretical central temperature, $T_{\rm Cc}$,  from the one obtained for the observed spectrum.   This   can be  done using the formulae  in  Fig. \ref{fig:fig10} or  equation (\ref{TXX}). With $T_{\rm Cc}$,  the shape of the theoretical surface brightness profile is automatically known because it depends just on the former. For a  power-law  X-ray emissivity, the surface brightness profile can be obtained from (\ref{Pro3}). For a more general function,  equations (\ref{Pro2}) and (\ref{Pro4}) can be used. Particularly, we have calculated the profiles that correspond to the X-ray emissivity tabulated by  Strickland \& Stevens (2000).    These profiles are shown in Fig. \ref{fig:fig8} and can be used as a reference. Additionally,  the derived value of $T_{\rm Cc}$ can be helpful  to constrain the parameter space of  the   bubble model  that one is considering, or well,  to derive some parameters in order to compare them with the observations.  In this regard,  $T_{\rm Cc}$, the observed X-ray luminosity and the  model-independent X-ray luminosity given by equation (\ref{LXGMI}) can be used to compare the observationally derived electron density with the one that follows from a shell-evaporating bubble model.   All of this can help to decide whether or not a specific model is adequate to describe the observations.

 The X-ray luminosity of bubbles without evaporation was given in Paper I. The corresponding surface brightness is given by equations (\ref{SXCnev}) and (\ref{AXCnev}). The temperatures of their hot interiors are  similar to the temperatures at the edges of the star clusters that generate them, equation (\ref{TscRA}).

  \begin{figure*}
\centering
\includegraphics[height=60mm]{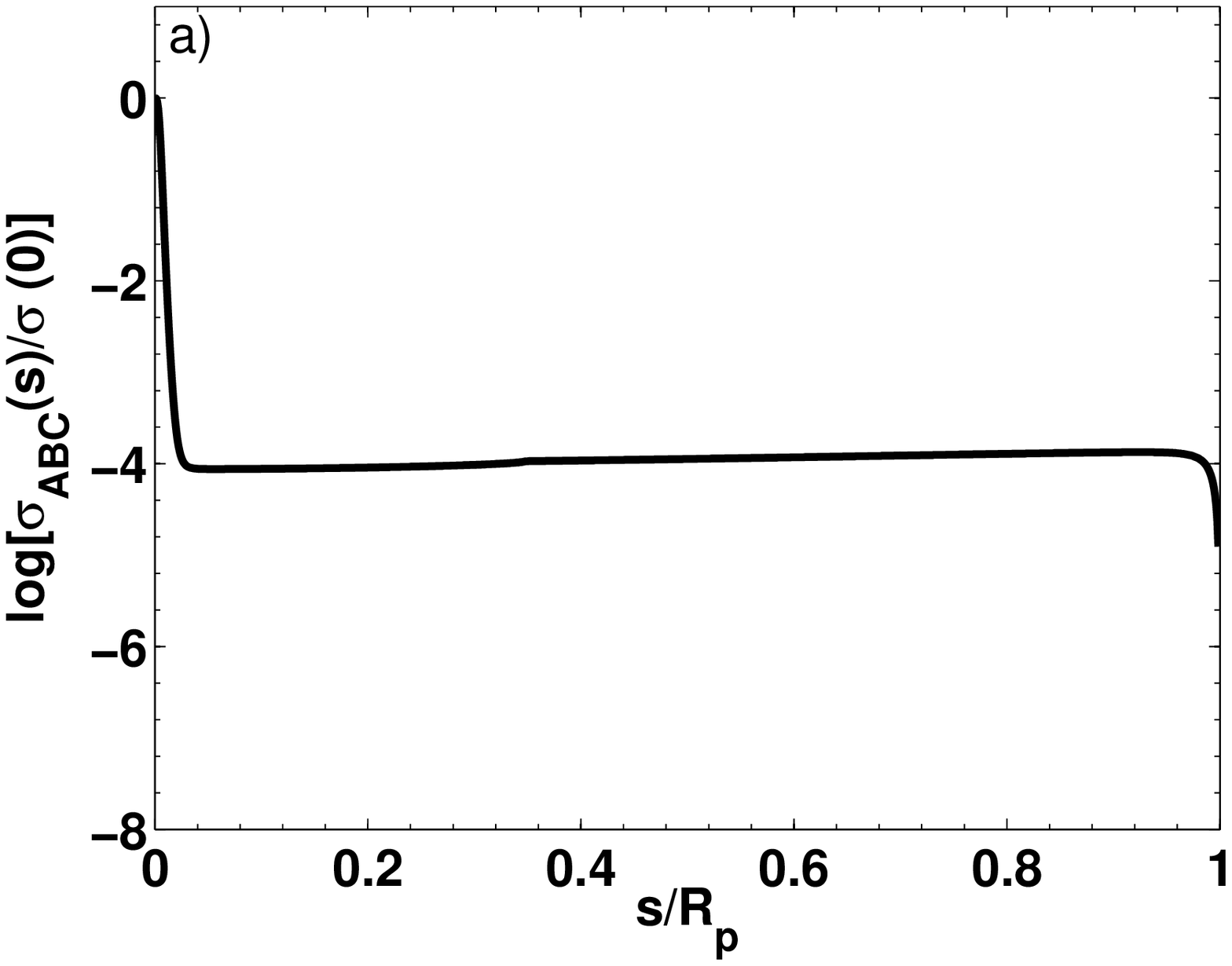}
\hfill
\includegraphics[height=60mm]{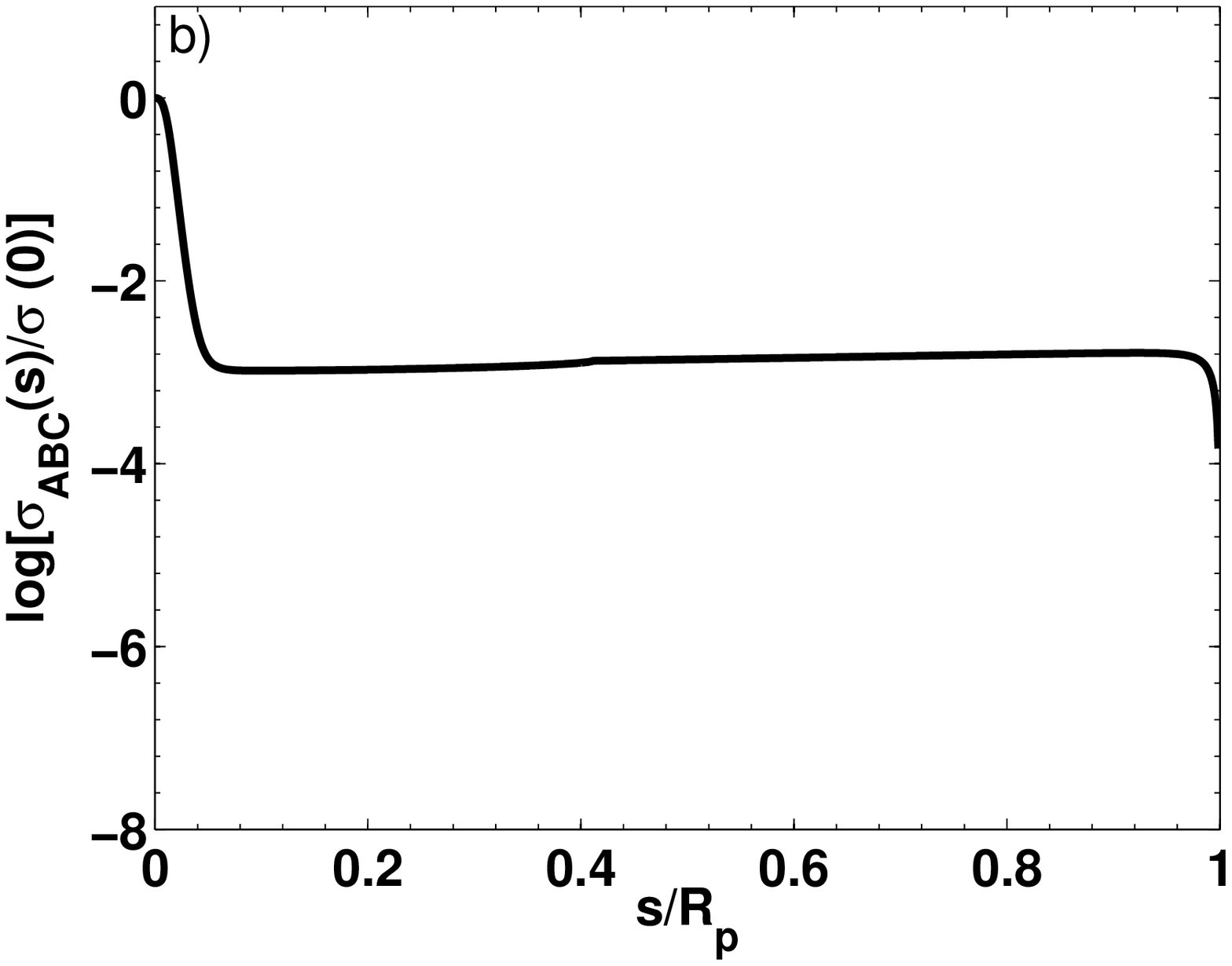}\\
\vfill
\includegraphics[height=60mm]{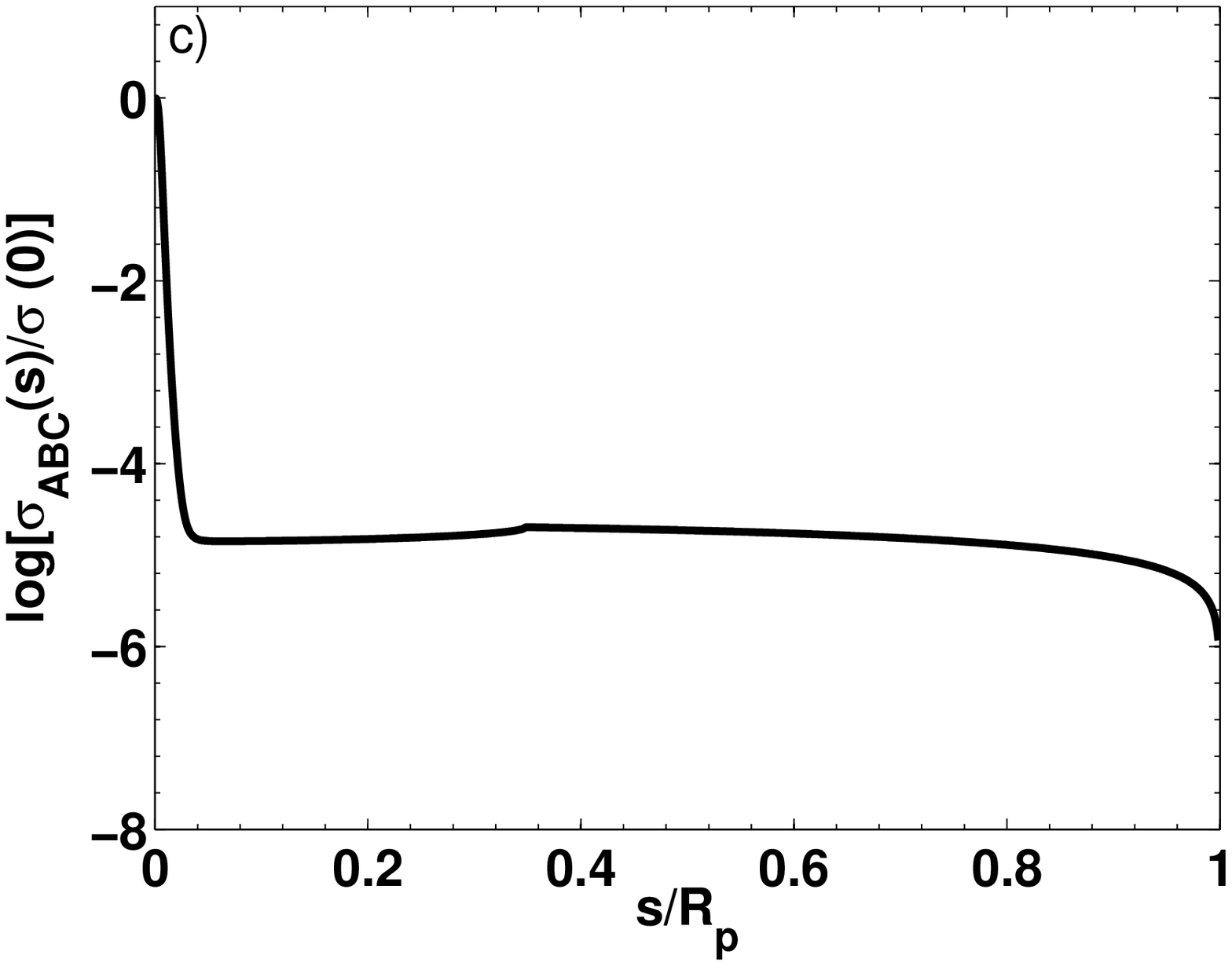}
\hfill
\includegraphics[height=60mm]{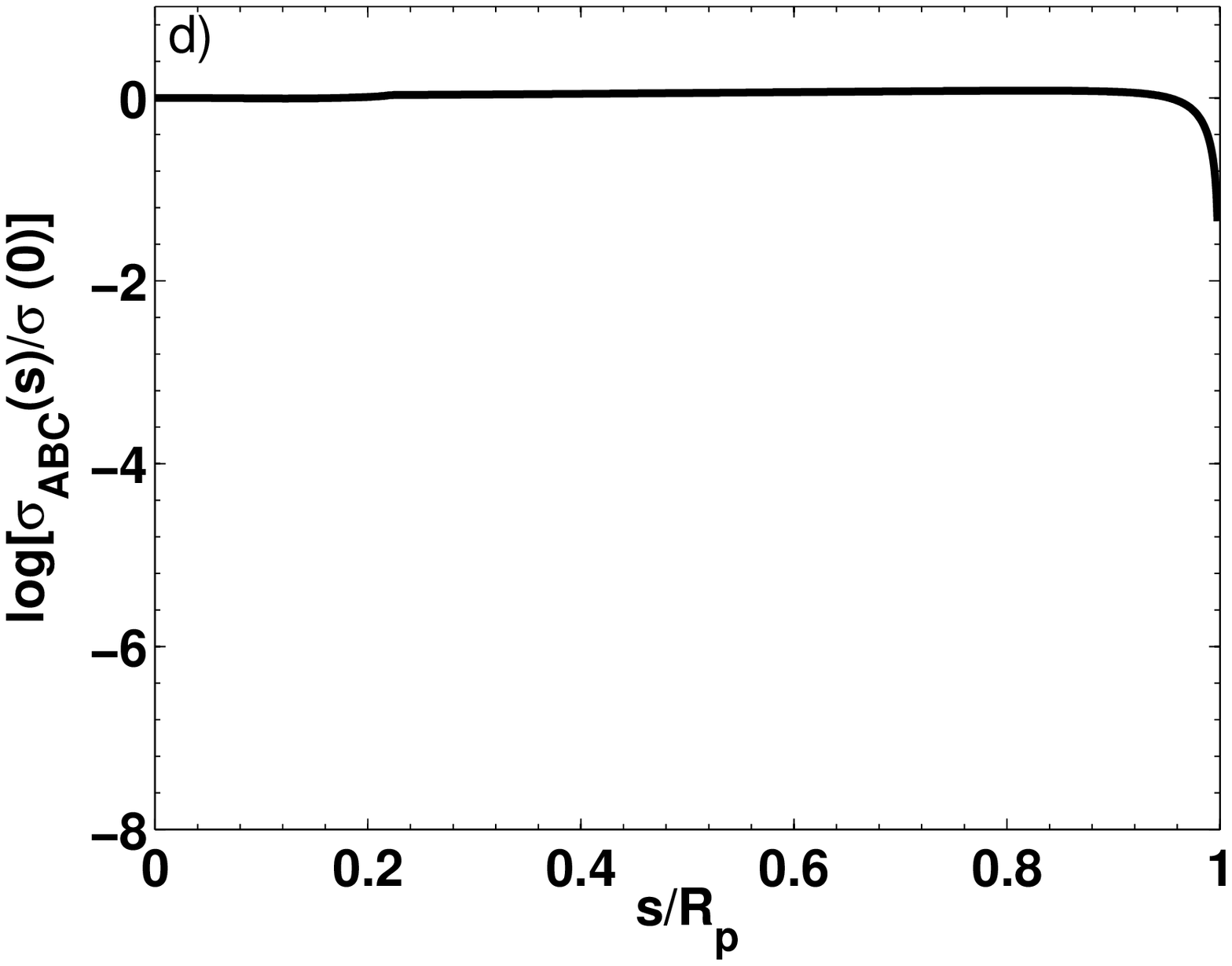}
\caption{  Combined X-ray surface brightness of  regions A, B and C.  In panels (a), (b) and (c) the superbubble is blown by a 1000 km s$^{-1}$ wind generated by a   cluster with a  radius of 1-pc and a mass of $10^5$ M$_\odot$. The ISM density is 10-cm$^{-3}$ and  $Z=Z_{\rm C}=$Z$_\odot$.   Panel (a) shows  a shell-evaporating bubble with $\dot  E_{38}=30$ and  $t_6=3$  ($R_{\rm p} \approx 161$-pc and $T_{\rm Cc}\approx1.08\times 10^7$ K).  This bubble corresponds to line c in Fig. \ref{fig:fig8}, and thus, it exhibits limb-brightening near $R_{\rm p}$. In panel (c) the same $\dot E_{38}$ and $t_6$ are used but evaporation is not present; hence, the bubble has the same size than in panel (a), but it is hotter and exhibits limb-darkening towards $R_{\rm p}$. Its average temperature is $\overline T_{\rm C}\approx4.42 \times 10^7$ K.  Panel (b) shows the effect of age and the smaller energy deposition rate at early times (Leitherer and Heckman 1995). The used parameters are $\dot E_{38}=10$ and $t_6=0.75$. This produces a smaller bubble with $R_{\rm p}\approx56$-pc; however, the bubble temperature is very similar ($T_{\rm Cc}\approx 1.06 \times 10^7$ K) to that of the bubble in panel (a).  In panel (d), the surface brightness  profile  is almost flat because the ISM has a larger density and the star cluster is more extended and  less massive: the cluster has  a mass of  $10^4$ M$_\odot$ and  $R_{\rm sc,pc}=10$,  $V_8=1$, $t_6=3$, $E_{38}=3$,  $Z=Z_{\rm C}=$Z$_\odot$ , $n_0=100$-cm$^{-3}$, $R_{\rm p}=64$-pc and $T_{\rm Cc}=7.3\times 10^6$ K.       \label{fig:fig12}}
\end{figure*}

The main conclusion of our previous paper was  that in X-rays,  the superwind contribution dominates just for compact and powerful SSCs while the bubble dominates in the rest of the cases. Nevertheless, that conclusion was derived from the cumulative X-ray luminosity, which is not directly observable, but estimated from the photon counts on the detectors.  What is directly observed is the surface brightness. Even when the bubble  could dominate by cumulative X-ray luminosity; generally, its  emission would be diluted in a volume much larger than that of the SSC, and as a consequence,  its surface brightness would be  smaller than that of  the superwind (as long as the objects could be spatially resolved).

In Fig. \ref{fig:fig12}, we show the combined  surface brightness  of  a fast  superwind ($V_8=1$)  and  the bubble that it  blows. Panels (a), (b)  and (d) correspond to  an evaporative bubble, and panel (c) to  a bubble for  which evaporation has been suppressed.   In panels (a), (b) and (c),  the cluster mass is $10^5$ M$_\odot$, its radius is $1$-pc and the ISM has a density of  $10$-cm$^{-3}$.  Panels (a) and (b) represent the same bubble at different age.  Its relative surface brightness (with respect to that of the wind) decreases as it  expands. Since  the surface brightness of the superwind goes as  $\dot E^2$  and that of  an evaporative bubble goes  as $\dot E_{38}^{19/35}$  (according to the W77 model), the later effect is enhanced by the increment  of $\dot E$ by a factor of $\sim 3$ after 3-Myr (Leitherer \& Heckman 1995).  Similar arguments apply for bubbles without evaporation.  The parameters used  in panel (a)  are the same that were used in panel (e) of fig.  6 in Paper I, where it  was shown that  the bubble dominates by total X-ray luminosity.\footnote{The X-ray luminosity  estimated in Paper I has to be corrected by a factor of $\sim 2$ to account for the SS00 realistic emissivity function. However, this does not affect the current discussion.}  Here,  we demonstrate that because of the core compactness, the superwind dominates in the X-ray picture.  In panel (c), the bubble has the same parameters that in panel (a), but    evaporation is not present.  The same dynamical model is assumed;  thus, the bubble has also the same size, but it is  hotter due to the thermalisation process at the reverse shock. Finally, in panel (d)  we show the interaction  that results    when the  ISM has a larger density and   the cluster  has  a larger size and a  smaller mass. For the assumed parameters, the bubble dominates and the profile is almost flat.

Thus, for compact (when  $R_{\rm sc}$ is up to several parsecs) and powerful ($\dot E_{38}\gtrsim 1$) clusters, the central emission is dominated by the cluster wind (region A).  Around the core,  there is  a still   bright and   rapidly decreasing component due to the free wind (region B), but most of  the brightness at radii larger than a few times $R_{\rm sc}$  and up to the reverse shock comes from background emission from the superbubble (region C),  as the contribution from the free wind  becomes negligible in comparison.   The rest of the projected emission ($r\ge R_{\rm s}$) comes from the superbubble itself  and adopts  the profiles  shown in Fig. \ref{fig:fig8} as it evolves with age.  

For powerful but extended, or compact but less  powerful clusters,   the brightness transition from one region to the next is smoother   because  the  central peak produced by the wind emission becomes   smaller  and wider. Additionally,  high ISM densities also contribute to smooth out the transition,  since the brightness of region C depends on $n_0^{31/35}$ (W77).  Only  when this circumstances are enhanced, the bubble surface brightness can be comparable to or even higher than that of the superwind.

Nevertheless, there is an additional factor that one must  consider. If the evaporation process is  inefficient,  the bubbles will  show   limb brightening,  but the overall amplitude of  their emissions will decrease (see Sections  \ref{Bevsur} and  \ref{M 17}).  In such a case, the corresponding pictures would look  similar to those shown in panels (a) and  (b) of  Fig. \ref{fig:fig12}. This means that  if the integration times are not large enough,  the observed filling factors will be small, even for young normal clusters (i.e. for clusters that do not qualify as SSCs). This would translate into observing   centre brightening  or  even an apparent spatial disconnection between the X-ray emission and the motion of the outer shell.

\section{COMPARISON WITH OBSERVATIONAL DATA} \label{Comparison}

\subsection{Nearby clusters}

Here, we give a brief summary of the clusters analysed in Paper I. In Table  \ref{Table1},   we give the X-ray luminosities  obtained from formula (\ref{LXA2})  for the same sample.  The first six columns indicate the observed values. The last three give  the results from our analytics. The contributions from hydrogen ($L_{\rm XA,H}$) and metals ($L_{\rm XA,M}$) were separated. The amplitude of the  central surface brightness ($A_{\rm AB}$)  is given in the last column.

{\bf NGC 3603.} Using color-magnitude diagrams for  the massive stars contained within the core, Sung \& Bessell (2004)  determined an age of $\tau_{\rm SC}\approx 1 \pm 1$ Myr for this cluster.  Given the remarkable similarity of the spatial distribution of its stellar content with that  of R136, the core of 30 Dor in the LMC (Moffat, Drissen \& Shara 1994),  we adopt  a  mass of $1.7 \times 10^4$ M$_\odot$ (see Paper I). This corresponds to an energy deposition rate of  $1.7 \times 10^{38}$ erg s$^{-1}$ (Leitherer 
\& Heckman 1995).   Moffat et al. (2002) studied the X-ray emission of this cluster and determined  a core radius of 0.71-pc. They reported a diffuse X-ray emission  of $2\times 10^{34}$ erg s$^{-1}$ in the 0.5-10 keV band. Their best-fitted spectra have temperatures $T_{\rm keV}\approx 3.1$ keV  for  the core and $kT \approx 2.1$ keV  for  the outer region. They also reported an approximately gaussian-shaped X-ray surface brightness, same that is in agreement with equation  (\ref{SXA}). The utility of having  theoretical estimations for  the cumulative  X-ray luminosity and surface brightness profiles of  star clusters is manifest from  that work: its authors used the exterior  profile of the diffuse X-ray emission to extrapolate to  that of the  cluster core, where the high density of point  sources made less reliable  its direct estimation; even after  the use of removal algorithms.

{\bf  Arches Cluster.} The Arches cluster is a very compact and young star cluster. 
Figer et al. (1999b) determined its  age to be  $\tau_{\rm SC}\approx 
 2 \pm 1$ Myr.  Adopting a low-mass cut-off of  1-M$_\odot$, they  estimated a total stellar mass of $M_{\rm SC}\approx 1.1 \times 10^4$ M$_\odot$. This corresponds to an energy deposition rate of  $\dot E_{38}=1.1$ (Leitherer \& Heckman 1995). The cluster radius is  $\sim 0.2$-pc (Figer et al. 1999a). Its X-ray emission was studied by  Law \& Yusef-Zadeh (2004), whom reported a total X-ray luminosity of $(0.5-1.0)\times 10^{35}$ erg s$^{-1}$ with  $Z=(4-5) \,$ Z$_{\odot}$. Just 40 per cent of this is diffuse emission, being the rest due to point sources. As  in Paper I, we adopt   the Law \& Yusef-Zadeh (2004) one-temperature plasma model; thus,  $T_{\rm keV}\approx 1.5$ keV.

{\bf Quintuplet Cluster.} For the 1-pc radius Quintuplet cluster, Figer et al. (1999b) determined  a mass of  $M_{\rm SC}\approx 8.8 \times 10^3$ 
M$_{\odot}$ using again a low-mass cut-off of 1-M$_\odot$.  The cluster  age has been estimated to be $\tau_{\rm SC} = 4 \pm 1$ Myr  (Figer et al. 1999a).  According to the Leitherer \& Heckman (1995) evolutionary models, at this age, supernovae have already started to contribute to the wind energy budget. Hence, we adopt $\dot E=2.64 \times 10^{38}$ erg s$^{-1}$.  Law \& Yusef-Zadeh (2004) studied the X-ray emission of this cluster with the aid of \emph{Chandra}. Their best fitted spectrum for the cluster core corresponds to a single-temperature  solar-metallicity plasma with $T_{\rm keV}=2.4$ keV. They observed  an X-ray luminosity of $\sim 1 \times 10^{34}$ erg s$^{-1}$.

{Thanks  to the additional physics synthesized in  equations (\ref{LXA1})-(\ref{LXA2}), i.e.  the direct incorporation of a realistic X-ray emissivity function and the  projected  temperature, we reach a better agreement between the theoretical and  observed luminosities.  Their  ratios are now reduced to  factors $<2$ for the case of the Arches and Quintuplet clusters. A comprehensive  interpretation of the X-ray emission of these clusters can be found in Wang et al. (2006).  For the case of NGC 3603, the cluster wind X-ray emission  accounts for  $\sim30$ per cent of the observed one. This prediction agrees with the results obtained by Li et al. (2006). Their observed and simulated intensity profiles also   agree with the  general form predicted by equation (\ref{SXA}), although each with different  indexes $\alpha_1$ and $\alpha_2$. The later difference is due to the  exponential stellar distribution  used in their hydrodynamical model. However,   our  predicted  bulk X-ray luminosity is consistent with theirs within a factor of $\sim 2$, since the rapid decrease of the gas density  warrants that most of the X-rays come from within the star cluster core and its vicinity. Also, notice that, as  they have pointed out, non-CIE effects become important in the supersonic wind region.     }

  The luminosity  of the metals surpasses that of hydrogen for the case of the Arches and Quintuplet cluster, whereas for NGC 3603 they are comparable. The  effect of  compactness and a  relative lower temperature can be appreciated in the case of the Arches cluster. This object has a radius just  3.5--5 times smaller than those of  the other clusters in the sample, and its temperature is  about 1--1.5 keV lower. Comparatively, such differences do not affect too much the X-ray luminosity, but they have an important impact on the surface brightness: the Arches Cluster  diffuse  X-ray emission   is  more than 2 orders of magnitude brighter in projection than those of NGC 3603 and the Quintuplet Cluster,  albeit   its total bolometric X-ray luminosity is just $\sim$3 times larger.  This later factor is mainly due to the Arches higher metallicity.

\begin{table*} 
\begin{minipage}{175mm}
\caption{Comparison with Nearby Clusters}
\begin{tabular}{@{}lllllllll}
\hline
 Object$^{\, \rm (a)}$& $T_{\rm c}^{\, \rm (b)}$&$R_{\rm sc}^{\, \rm (c)}$& $\dot E^{\, \rm (d)}$&  $Z^{\, \rm(e)}$&$L_{\rm XA,obs}^{\, \rm (f)}$ & $L_{\rm XA,H}^{\, \rm (g)}$ & $L_{\rm XA,M}^{\, \rm (h)}$ & $A_{\rm AB}^{\,\rm (i)}$\\
 &[keV]&[pc]& $[10^{38}$ erg s$^{-1}]$&[Z$_\odot$]&[$10^{34}\,$erg s$^{-1}$]&[$10^{34}\,$erg s$^{-1}$]&[$10^{34}\,$erg s$^{-1}$]& [$10^{34}\,$erg s$^{-1}$ pc$^{-2}$]\\

 \hline
 NGC 3603&$3.1\pm 0.7$&0.71&1.7&1&$1.6$&$0.22$ & $0.27$&0.80\\
Arches& $1.5\pm 0.2$ &  0.20   &  $1.1$ &4& $2.0-4.0$ & $0.41$ &$3.90$&430\\
Quintuplet& $2.4\pm 0.5 $& 1.0 & 2.6 &1 & $1.0$& 0.71&1.02&1.44\\

 \hline
 
\end{tabular}\label{Table1}
 \medskip {\\
 (a): object name;  (b): central temperature; (c) star cluster radius, (d): mechanical luminosity; (e) metallicity;  (f): observed X-ray luminosity;  (g)-(i):  predicted  X-ray  luminosity contributions from hydrogen and  metals and  amplitude of the central surface brightness. The respective units are indicated between square brackets.} \label{Table3}
 
\end{minipage}

\end{table*}

\subsection{Bubbles: {the case of M 17,   magnetically hampered thermal conduction?}}\label{M 17}

M 17 --also known as the Omega Nebula-- is powered by a young open cluster with an age of  $\sim$1-Myr. The nebula has an overall diameter of  $\sim$10--12 pc, and coinciding with it, diffuse X-ray emission has been detected by \emph{Chandra}  (Townsley et al. 2003).  Dunne et al. (2003) have studied its properties using a previous data set from a \emph{ROSAT} observation. Their data sets have similar parameters. The best-fitting spectra is described by a single temperature plasma with $T_{\rm spec}\approx0.7$ keV, $Z=$Z$_\odot$  and an rms electron density of $n_e\approx9\times 10^{-2}$ cm$^{-3}$.  From the observed bubble size (they assumed a filling factor $f_{\rm f}=0.5$), expansion velocity,  and stellar content, they derived that $n_0\approx$40--60 cm$^{-3}$, $\dot E_{38}=0.1$ and $t_6\approx0.13$. Using these parameters, they made a comparison  with the W77 model. They found that $n_{\rm Cc}\approx$0.2--0.4 cm$^{-3}$, $T_{\rm Cc}\approx$5.0--5.6$\times 10^6$ K and $L_{\rm XC}\approx$(3--5)$\times 10^{35}$ erg s$^{-1}$.

The  predicted X-ray luminosity   differs by a factor of 120-200 with the observed one, $L_{\rm XC,obs }\approx2.5\times 10^{33}$ erg s$^{-1}$. To explain these discrepancies,  Dunne et al.  (2003) proposed the total inhibition of the evaporative process by  the  strong magnetic field  (100--500 $\mu$G) that  permeates M 17 (Brogan et al. 1999). They required the  winds  to be   significantly clumped and homogeneously  mixed  in order to match the predicted post-shock  temperature ($8\times 10^7$ K) with the spectroscopically derived one. An alternative explanation  is that while the magnetic field might render the evaporation ineffective, it  would not be able to completely avoid it, allowing a smaller fraction of cold  gas  to mix   with the hot bubble interior. In this scenario, the extrapolated central temperature  of the bubble  would be   more similar to the post-shock temperature. In fact, from our analysis,  we have that for their value of $T_{\rm spec}$  the  almost central temperature  of the  evaporative bubble is  $\approx$2 $\times 10^7$ K (see Fig. \ref{fig:fig10}),  which makes the discrepancy with the full-evaporative  case predicted  $T_{\rm Cc}$ even worse. 

Nevertheless,  if we ignore the particular  prescribed dynamical evolution and just consider a general shell-evaporating bubble,  we have that $T_{\rm XC}\approx T_{\rm spec}$.  In such a case and according to Fig. \ref{fig:fig10} and equation (\ref{TXX}),  the value of $T_{\rm  XC}$ derived by Dunne et al. (2003) would  correspond to a bubble with  $T_{\rm Cc}\approx 4$--$5\, T_{\rm XC}$   (the upper bound considers the reverse shock position).  This increment of temperature with respect to the one derived from the W77 model has to be compensated by  the same decrement in density.  Since $L_{\rm X}\propto n_{\rm Cc}^2$ this would reduce the predicted X-ray luminosity by a factor of 16--25.  If additionally,  we consider that the C95 X-ray model overestimates the  value of $L_{\rm X}$ by a factor of $\sim$2 (for our derived $T_{\rm Cc}$, see Fig. \ref{fig:fig6}) with respect to the averaged SS00 emissivity, we get a correction factor $\sim$32--50, which would bring $L_{\rm X}$  acceptably close to the observed value. The density near the centre  would be also similar  to the value derived by Dunne et al. (2003).   This would require a decrease of the classical thermal conduction coefficient (Spitzer 1956) of about 2 orders of magnitude.  As it was indicated in Section \ref{Bnev}, such a  drastic change is  possible  when strong  ($>1 \mu G$) magnetic fields  are involved (see Markevitch et al. 2003 and references therein).  Such magnetic fields can hamper  the thermal conduction process and establish steeper  temperature profiles.  Note that this  effect  is independent from the other observationally deduced parameters ($n_0,\dot E_{38}$ and $t_6$) because  the bubble size and  internal pressure are not affected by the correction factors introduced above (see Section \ref{Bubbles}).

The previous analysis remarks the relevance of the projected temperature as  a diagnostic tool, as well as the necessity of a broader-scope assessment of the thermal conductivity  efficiency for the case of  bubbles.

\section{SUMMARY}

Our findings and their immediate consequences are:

\begin{enumerate}
\item A new  and simple star cluster wind model has been developed. Assuming a polytropic equation of state we  reproduced the effect of radiative cooling  on the hydrodynamical profiles, in consonance with    previous numerical calculations. Moreover, the door has been left open to further applications like the parameterization  of other effects,  the constraining of the parameter space through threshold lines and the analysis of  other branches of the solution.
\item The superwind X-ray luminosity has been calculated using a realistic X-ray emissivity  that allowed the separation of the contributions from hydrogen and  metals.  Good agreement was obtained when we compared with observational data. Also, transformation laws to include very general emissivity functions  and completely characterize the global properties of the models have been  given.
\item We have   calculated the corresponding  surface brightness  and shown that   although the adiabatic  superwind model predicts a very extended X-ray emitting volume,  the  X-ray   bright super star cluster  cores are expected to be observed surrounded by   compact and faint haloes  (if  their  bubbles are not considered).    The weighted temperatures associated to the cluster cores were also obtained.
 
\item  An analogue to the  C95  X-ray luminosity model has been obtained considering  a realistic emissivity function.  The model is independent of the assumed dynamical evolution for  bubbles with $R_{\rm s}/R_{\rm p}\le 0.4$. Significant deviations were observed with respect to the C95  model when $T_{\rm Cc}\rightarrow 10^6$ K. It was also found that for under-developed bubbles   (when the ratios of the reverse and leading shocks are $>0.4$) with extrapolated central temperatures on  this range,    the bubble X-ray luminosity  can be drastically reduced by the relative closeness  of the reverse and main shock positions.

\item   The  X-ray surface brightness of  evaporative bubbles has been  obtained in terms of elementary functions by introducing  a pseudo-triangle approximation. It has been shown that   limb-brightening, flat and limb-darkening modes are all possible according to the value of the  central temperature. Because we have separated the self-similar spatial profiles  from  what we call an intensity amplitude and a  weighting emissivity factor,  they are      valid  for all  evaporative models based on the classical thermal conduction theory.

\item It  has been demonstrated  that  moderately hot, well-developed  shell-evaporating bubbles  ($T_{\rm Cc}\approx 10^6$ K) have  X-ray  weighted temperatures  somewhat  similar to their central  ones ($T_{\rm XC}\sim 0.5$--$1 \,T_{\rm Cc}$) in major fractions of their projected areas. On the other  hand, very hot bubbles ($T_{\rm Cc} \ga 10^7$ K)  are expected to have point-wise projected temperatures of about 0.25--0.50$\,T_{\rm Cc}$. Nevertheless,  the hottest  bubbles can  still have projected  temperatures of the order of $1$ keV.  Since $T_{\rm XC}$ is a  measure of the  temperature at which  the plasma is emitting the most,   it  is  important  to account for the difference   between this and the  actual temperature at different radii. This can  be also useful to evaluate the efficiency of the thermal evaporation process.

\item We also studied the case of when the evaporative process is totally inhibited. In this case, the bubbles have very hot interiors since their temperatures are similar to the post-shock temperature. Their surface brightness was also obtained.

\item We have assembled  the X-ray  brightness models and presented a very descriptive and comprehensive  X-ray picture of the interaction of the superwind with the ISM.
\end{enumerate}

All these tools are expected to be helpful  at the time of comparing the models with the observations.

\section{ACKNOWLEDGMENTS}

We wish to thank D. Strickland \& I. Stevens for kindly providing their X-ray emissivity tables. Our  gratitude is also extended to the anonymous referees whose very  valuable suggestions helped to greatly improve this paper. This work has been supported by  CONACYT-M\'exico  research grants 60333 and  82912.  This paper was prepared with the help of the MNRAS  \LaTeX style and class files.

 \appendix

\section{SUPERWINDS X-RAY LUMINOSITY} \label{SWX}

It is convenient to initially adopt a constant X-ray emissivity $\Lambda_{\rm X}=\Lambda_0$ and express the integral (\ref{LX}) in terms of the velocity instead of the radius.  The related quantities will be labeled with the subindex '0'.

\subsection{{Initial formalism}}\label{InitialX}
\subsection*{Region A}
Substituting  (\ref{LX}) and   (\ref{rhoin}) in (\ref{polyin})  and introducing the variables $x={U'}/{U}$,     $z=-3(2\eta +1)U$  and $f=(1-zx)$ we get that $L_{\rm XA0}=X_A  (I_1+I2)$ with

  \begin{equation}
I_1=    \frac{1}{2} U^{3/2}\int_{0}^{1}\Lambda_0 x^{1/2} f^{-\frac{10}{3}} {\rm{d}}x ,\label{I1} \end{equation}
 \begin{equation}
I_2=  -2(2\eta+1)  U^{5/2} \int_{0}^{1} \Lambda_0 x^{3/2} f^{-\frac{13}{3}}{\rm{d}}x, \label{I2} \end{equation}
\noindent and
 \begin{equation}
X_{\rm A}=\frac{D_{\rm 1P}^5\dot E^2}{\pi \mu_{\rm n}^2V_{\rm \infty P}^6 R_{\rm sc}}. \label{XA}
 \end{equation}

In what follows  f will be called  a "base", for reasons that will be evident in brief. The above integrals can be expressed in terms of Gauss hypergeometric functions (see for instance, Whittaker \& Watson 1996  and  Lebedev 1972):

\begin{equation}
_2F_1(a,b;c;z)=\frac{\Gamma(c)}{\Gamma(b)\Gamma(c-b)}\int_{0}^{1} x^{b-1}(1-x)^{c-b-1}(1-zx)^{-a} {\rm{d}}x,
\end{equation}

\noindent where $\Gamma$ is the Gamma function. It follows that $a_1={10}/{3},\,b_1={3}/{2}, c_1= b_1+1$ and $a_2=a_1+1,\,b_2=b_1+1,\,c_2=b_1+2$. Then,

  \begin{equation}
L_{\rm XA0}=X_{\rm A}\Lambda_0U^{3/2}\left[\frac{1}{2b_1}F_1 -2 \frac{(2\eta+1)}{b_1+1}U F_2 \right].\label{LXA0}
 \end{equation}

Above, to simplify the notation we have used the label of each family of indexes to make reference to  the corresponding hypergeometric functions:  $F_1={_2F_1(a_1,b_1;c_1;z)}$ and $F_2={_2F_1(a_2,b_2;c_2;z)}$. We shall keep this custom in what follows. Each $U\in\left [0,{1}/({2\eta+1})\right]$  yields  the value of $L_{\rm XA0}$  at the  radius indicated by (\ref{polyin}). Alternatively, one may use the temperature $T= T_{\rm c}(1-U) \in [T_{\rm c},T_{\rm sc}]$ as the independent variable. 

For  $\eta=\frac{3}{2}$, the total luminosity can be approximated by

  \begin{equation}
L_{\rm XA0,total}=\left(1.12\times 10^{34}\mbox{erg s$^{-1}$}\right) \frac{\Lambda_{\rm X,-23}(\overline T,Z) \dot E_{38}^2}{R_{\rm sc,pc}V_{8}^6} =\left( 9.63\times 10^{33}\mbox{erg s$^{-1}$}\right)  \frac{\Lambda_{\rm X,-23}(\overline T,Z)\dot E_{38}^2}{R_{\rm sc,pc}T_{\rm keV}^3}, \label{LXA0t}
\end{equation}

\noindent where  we evaluated (\ref{LXA0}) at  $U=U_{\rm sc}$  and  took  $\Lambda_0=\Lambda_{\rm X}(\overline T,Z)$. The value of the emissivity function for the reference temperature  $\bar T$ is taken from the SS00 tables.

\subsection*{Region B}

  For   quasi-adiabatic superwinds, $\eta\approx\frac{3}{2}$   for  $r \ge R_{\rm sc}$ and $U\in[U_{\rm sc}, 1)$.  In principle, superwinds with $V_{\rm \infty P}$ of the order of 1000 km s$^{-1}$ have  very extended X-ray envelopes, since the connection between temperature and radius is given by
  
  \begin{equation}
  R=D_{2}\left(1-\frac{T}{T_{\rm c}}\right)^{-1/4}\left(\frac{T}{T_{\rm c}}\right)^{-\eta/2}. \label{RTout}
  \end{equation}
  
  Nevertheless,  the outermost layers meagerly contribute to the luminosity, since most of the emission from this region  comes  from a compact shell of gas around the star cluster. Thus, it suffices to estimate  the external X-ray luminosity and surface brightness of quasi-adiabatic superwinds because those of   superwinds in the  radiative regime would be alike,  despite the effective cropping of their X-ray emitting envelopes by radiative cooling.  Introducing the variables $x={U'}/{U_l}$ and $z=U_l$  we find that  
  \begin{equation}
  L_{\rm XB0}= X_{\rm B}\Lambda_0U_l^{-3/4} \left[\frac{\eta}{2(b_4+1)}UF_3-\frac{1}{4b_4}F_4\right]_{U_l=U_{\rm sc}}^{U_l=U}, \label{LXB0}
  \end{equation}
  
  \noindent  with $a_4=-{\eta}/{2},\, b_4={3}/{4},\,c_4=b_4+1$ and $a_3=a_4+1,\,b_3=b_4+1,  c_3=b_4+2$ and
  
  \begin{equation}
  X_{\rm B}=\frac{\dot E^2 }{\pi D_2 \mu_{\rm n}^2 R_{\rm sc} V_{\rm \infty P}^6}.
  \end{equation}

The velocity at the cut-off radius is  $U_{\rm cut}=\left(1-{T_{\rm cut}}/{T_{\rm c}}\right)$. For fast winds $U_{\rm cut}\approx 1$.  Accordingly, we can replace the hypergeometric functions with their asymptotic limits using the  Euler beta function and the identity $\Gamma(v)\Gamma(1-v)= {\pi}/{\sin(\pi v)}$:

  \begin{equation}
 \lim_{z\rightarrow 1^{-}} \,_2F_1(a,b;c;z)=\frac{\Gamma(c)\Gamma(c-a-b)}{\Gamma(c-b)\Gamma(c-a)} = \frac{\Gamma(c)}{\Gamma(c-a)\Gamma(a)} \frac{\pi}{\sin(\pi a)}.
\end{equation}

  Then, for $\eta=\frac{3}{2}$, $F_3$ and $F_4$ can be substituted by

  \begin{equation}
\begin{array}{ll}
 \lim_{z\rightarrow 1^{-}} F_3&=2^{-3/2} \pi, \\
  \lim_{z\rightarrow 1^{-}}  F_4&=3(2^{-3/2} \pi).
\end{array}
\end{equation}

 Combining these equations with (\ref{LXB0}) we obtain that
  
  \begin{equation}
  L_{\rm XB0}\approx \frac{1}{4} L_{\rm XA0}. \label{LXB/LXA}
  \end{equation}
  
\noindent This luminosity  comes from a very large volume. Since radiative cooling brings the X-ray cut-off temperature closer to $R_{\rm sc}$, the actual emitting region is more compact. Consequently, the  contribution of the free wind is  smaller. This conclusion does not change when the SS00 emissivity is considered, as can be shown using the transformation laws given below and consulting the numerical results of Paper I.   Because of this, we shall focus on region A, but providing a general formalism to account for region B if  it is desired.

\subsection{\bf{Transformation laws}}\label{TLaws}

Before proceeding any further  we shall establish laws for transforming  equations (\ref{LXA0}) and (\ref{LXB0})  such that they can handle more general X-ray emissivity functions with explicit dependence on metallicity. For this reason,  $\Lambda_0$ was kept intentionally inside of the integrals (\ref{I1}) and (\ref{I2}). These laws can be used to calculate other integral properties of the superwind  such as its mass and energy within a certain volume. The laws  also apply for bubbles.

Let $a'$, $b'$, $\alpha$ and  $\Lambda_{a'}$,$\Lambda_{b'}$, $\Lambda_{\rm \alpha}$  be arbitrary constants. The following transformations apply to (\ref{I1}) and (\ref{I2}) and similar integrals like those for region B and C:

{\bf{Law I:}} if  $\Lambda_{\rm X} =\Lambda_{\rm \alpha}f^{\alpha}$  then the transformation $a \rightarrow a-\alpha$ applies for the leading index and $\Lambda_0\rightarrow \Lambda_{\rm \alpha}$ for each term.\footnote{In  the  more general case, $\Lambda_{\rm X}=\Lambda_{\rm X}(f)$, the   Einstein summation convention can be applied to express the function as a series in $f$. Similar arguments are  valid for the other two laws.}

{\bf{Law II:}} if $\Lambda_{\rm X} = \Lambda_{\rm \alpha} f'^{a'}$ with $f'=(1-xz')$ and  $z' $ independent from  $z$, then $_2F_1 \rightarrow F_{\rm A}$, where $F_{\rm A}$ is the Appell hypergeometric function which has two bases $(f,f')$ and two leading  indexes $(a,-a')$:

  \begin{equation}
F_{\rm A}(a,-a',b;c;z,z')=\frac{\Gamma(c)}{\Gamma(b)\Gamma(c-b)} 
\int_{_0}^{1} x^{b-1}(1-x)^{c-b-1} f^{-a}f^{a'}{ \rm{d}}x.
\end{equation}
Again, $\Lambda_0 \rightarrow \Lambda_{a'}$ applies for each term.

{\bf{Law III:}} if $\Lambda_{\rm X}= \Lambda_{b'}x^{b'}$ then we need to use first the transformations  $b\rightarrow b+b'$ and $c\rightarrow c+b'$ and then $\Lambda_0\rightarrow{\Gamma(c-b)\Gamma(b)}/{\Gamma(c)}  \Lambda_{b'}$ for each term.

Since $a$, $b$ and $c$, $\alpha$ and $a'$ are mutually independent, we can apply  the three laws simultaneously if it is required. Because all the hydrodynamical variables are functions of  $f$, $f'$ and $x$  (or $u$) these laws suffice to calculate any integral property. 

In this paper,  we used them to calculate the X-ray emission assuming a realistic $\Lambda_{\rm X}$. This was done  as follows: considering the more general case $\Lambda_{\rm X}= \sum\Lambda_{\rm \alpha} T^{\alpha}$, where $\Lambda_{\rm \alpha}$ are constants that can scale with metallicity (see Table \ref{Table2}) and   each  $\alpha\in[-4,4]$, we applied Law II  to the quantity between square brackets in (\ref{LXA0}) and proceeded to  expand it rationally  around $U=0$  (after dividing  by $T_{\rm c}^{\alpha}$) recurring to  admissible hypergeometric transformations (see the references in Section \ref{InitialX}). We used a (0,2) Pad\'e approximant to make the rational approximation around 0. This warrants  a coincidence of up to the second order with the Taylor expansion of the original function and provides an accurate approximation for large values of $\eta$, since $U_{sc}=1/(2\eta+1)$. To account for small values of $\eta$ and the effect of the exponents of the power laws, we later introduced two curvature matching filters. The next  expression valid  on   $U\in[0,{1}/({2\eta+1})]$ was obtained:

\begin{equation}
\mathcal{F_{A}}(\eta,\alpha,U)=\frac{350\left(1-U\right)^{\frac{\alpha}{30}}\left[1+\frac{\eta^2(8\eta+1)}{5(2\eta+1)}  U^2\right]^{4/10}}{9\left(\frac{350}{3}+\phi_1 U+\phi_2 U^2\right)}, \label{MFA}
\end{equation}

\noindent where

  \begin{equation}
\phi_1(\eta,\alpha)=70(\alpha+28\eta+14),
\end{equation}
\noindent and 
\begin{equation}
\phi_2(\eta,\alpha)=17\alpha^2+\alpha(501+952\eta)+9528\eta(1+\eta)
+2382.
 \end{equation}
 
 The quantity between parenthesis on the numerator  of equation (\ref{MFA}) is a matching filter of the form $(T/T_{\rm c})^b$ that corrects for  the curvature introduced by  $\alpha$. Likewise, the quantity between square brackets is a filter of the form $[1-g(\eta)U^b]^c$ that corrects  for the curvature introduced by $\eta>0.3$. The filter parameters where obtained by minimization of the relative error of the approximation. For the SS00 emissivity function we always obtain an accuracy better than 5 per cent.  In general, the same applies    for  all emissivity functions with $\alpha\in[-4,4]$ provided that $\eta\geq 3/2$. This ensures a good coverage of the radiative wind case. For the most extreme combination of parameters  we are considering, i.e. for $\eta\approx 0.3$ and $\alpha\approx 4$, the accuracy is $\sim15$ per cent.

From  equation (\ref{MFA}) and Law II,  it follows that the cumulative X-ray luminosity is 

\begin{equation}
L_{\rm XA}(\eta,\{\alpha\},U) =X_{\rm A} U^{3/2}\sum_{\rm \alpha} \mathcal{F_{A}}(\eta,\alpha,U)\Lambda_{\rm \alpha}T_{\rm c}^{\alpha},
\end{equation}

\noindent where $X_{\rm A}$ is  given by equation (\ref{XA}).

\section{\bf{ EVAPORATIVE BUBBLES SURFACE BRIGHTNESS}}\label{Bev}

 Just for convenience, we assume  initially   an  X-ray emissivity  function with a power-law dependence on the temperature, $\Lambda_{\rm X}=\Lambda_{\rm \alpha}T^{\alpha}$, where the constant $\Lambda_{\rm \alpha}$ might or might not scale with metallicity.   We then express all the relevant variables of  the evaporative model (Section \ref{Bubbles}) in terms of the ratio  of the point-wise temperature  $T_{\rm c}$ to  the central value:  $\vartheta=T_{\rm c}/T_{\rm Cc}$. From equations (\ref{TCc}) and (\ref{nCc}) one has that  $r=R_{\rm p}(1-\vartheta^{5/2})$, $n_{\rm C}=n_{\rm Cc}\vartheta^{-1}$, $ s_{\rm }=R_{\rm p}(1-\vartheta_{ s_{\rm }}^{5/2})$, $ s_{\rm cut}=R_{\rm p}(1-\vartheta_{\rm cut}^{5/2})$ and $R_{\rm s}=R_{\rm}(1-\vartheta_{\rm Rs}^{5/2})$. The  surface brightness can then be separated into a temporal  amplitude, $A_{\rm C}$; a weighted emission factor, $\lambda_\alpha$; and a  spatial profile  $\Theta(s)$ --between two reference points $ s_{\rm i}$ and $ s_{\rm f}$-- that implicitly but  slowly  changes with time, such that   $\sigma_{\rm C}=A_{\rm C}(t)\lambda_{\rm \alpha}\Theta(\vartheta_{\rm i};\vartheta_{{f}})$ with

\begin{equation}
A_{\rm C}(t)=2 R_{\rm p} n_{\rm Cc}^2 \Lambda_{1}, \label{AXCpre}
\end{equation}

\begin{equation}
\lambda_{\rm \alpha}= \frac{\Lambda_{a} T_{\rm Cc}^{\alpha} }{\Lambda_1}=  \Lambda_{\alpha ,-23}T_{\rm Cc}^{\alpha}, \label{lambdaapre}
\end{equation}
\noindent and

\[
 \Theta(
 \vartheta_{s},\alpha;\vartheta_{{i}},\vartheta_{{f}})=\int_{\vartheta_{{i}} }^{\vartheta_{\rm f}} \frac{\left(\frac{5}{2}\right)\vartheta^{\alpha} (\vartheta^2-\vartheta^{-1/2})}{\sqrt{(\vartheta_{\rm s}^{5/2}-\vartheta^{5/2})(2-\vartheta_{\rm s}^{5/2}-\vartheta^{5/2})}}{ \rm{d}}\vartheta,\]

  \begin{equation}
\label{Abubev}
\end{equation}
 \noindent where $\Lambda_1=10^{-23}$ erg cm$^3$ s$^{-1}$. Notice that $\vartheta_{\rm i}>\vartheta_{\rm f}$ because the bubble temperature gradient is negative.  
  
The  integral  above can be expressed in terms of Appell hypergeometric functions $F_{\rm A}$

\begin{equation}
 \Theta(\vartheta_{s},\alpha;\vartheta_{\rm i},\vartheta_{\rm f})=\left[ \frac{\vartheta^{\frac{1}{2}+\alpha}}{(2\vartheta_{\rm s}^{5/2}-\vartheta_{\rm s}^5)^{1/2}} \right. 
 \left. \left(\frac{\vartheta^{5/2}}{b_1+1}F_{\rm A2}- \frac{1}{b_1}F_{\rm A1}\right)\right]_{\vartheta=\vartheta_{\rm i}}^{\vartheta=\vartheta_{\rm f}},  \label{SXCpre}
\end{equation}

\noindent where $b_1=\frac{1}{5}+\frac{2}{5}\alpha$, $F_{\rm A1}=F_{\rm A}[{1}/{2},{1}/{2},b_1;b_1+1;{\vartheta^{5/2}}/{\vartheta _{\rm  s}^{5/2} },{\vartheta^{5/2}}/({2-\vartheta _{\rm  s}^{5/2}})]$, and similarly, $F_{\rm A2}=F_{\rm A}[1/2,1/2,b_1+1;b_1+2;{\vartheta^{5/2}}/{\vartheta _{\rm  s}^{5/2} },{\vartheta^{5/2}}/({2-\vartheta _{\rm  s}^{5/2}})]$.  For a single power law X-ray emissivity,  the surface brightness spatial profile in terms of the temperature is  

\begin{equation}
\sigma_{\rm C}(\vartheta,t)=A_{\rm C}(t) \lambda_{\rm \alpha}\Theta[\vartheta _{\rm  s},\alpha;\min(\vartheta _{\rm  s},\vartheta_{R_{\rm s}}), \vartheta_{\rm cut}]. \label{sigmaC1a}
\end{equation}

In terms of the projected radius $ s_{\rm }$ we have that

  \begin{equation}
\sigma_{\rm C}( s_{\rm },t)= A_{\rm C}(t) \lambda_{\rm \alpha}  S_{\rm C}[ s_{\rm },\alpha;\max( s_{\rm },R_{\rm s}), R_{\rm cut}].\label{sigmaC1b}
 \end{equation}

 In the last formula, we used the transformations indicated at the beginning of this appendix to  go back to the spatial coordinate; thus, $\Theta(\vartheta _{\rm  s}...\vartheta_{ref})\rightarrow S_{\rm C}(s_{\rm }...s_{Xref})$. For the singular cases --when both arguments of the minimum or  maximum function are the same-- we define $\min(a,a)=\max(a,a)=a$.  For a realistic emissivity expanded as a power series,  equation (\ref{sigmaC1a}) immediately transforms into

\begin{equation}
\sigma_{\rm C}(\vartheta_{\rm s},t)=\left\{\begin{array}{ll}
A_{\rm C}(t) \displaystyle{ \sum_{\alpha, \vartheta_{\rm i},\vartheta_{\rm f}<\vartheta_{\rm R_{\rm s}}}^{\vartheta_{\rm f}=\vartheta_{\rm cut}} }\lambda_{\alpha} 
\Theta[\vartheta_{\rm s},\alpha;\min(\vartheta_{\rm i},\vartheta_{\rm R_{\rm s}} ), \vartheta_{\rm f}],  \mbox{if  $s \le R_{\rm s}$;} \\
A_{\rm C}(t) \displaystyle{ \sum_{ \ \alpha, \vartheta_{\rm i},\vartheta_{\rm f}<\vartheta_{s}}^{\vartheta_{\rm f}=\vartheta_{\rm cut}} }\lambda_{\alpha}
 \Theta[\vartheta_{\rm s},\alpha;  \min(\vartheta_{\rm s},\vartheta_{\rm i} ), \vartheta_{\rm f}], \mbox{otherwise;}  \label{sigmaC2}
\end{array}\right.
\end{equation}

\noindent here, $\vartheta_{\rm i}$ and $\vartheta_{\rm f}$ are  respectively the upper and lower temperature limit for which the power law with index $\alpha$ is valid.

Equations  (\ref{SXCpre})-(\ref{sigmaC2}) and their analytical properties are enough to calculate the surface brightness spatial profile. Nevertheless, we still need closed algebraic formulas for  evaluating them swiftly and accurately. They are obtained in  the next appendix.

\section{HYPERGEOMETRIC PSEUDO-TRIANGLES} \label{HT}
Consider the negative of  the quantity between the big parenthesis in equation (\ref{SXCpre}) evaluated separately at the end-points  while considering a single power-law emissivity and  ignoring temporally the reverse shock position
\begin{equation}
\mathcal{F_{\rm c}}(\vartheta, \vartheta _{\rm  s};\alpha)= \left.\left(\frac{5}{1+2\alpha}F_{\rm A1}- \frac{5}{6+2\alpha}\vartheta^{5/2}F_{\rm A2}\right)\right|_{\vartheta_{ s_{\rm }}}^{\vartheta_{\rm f}},
\end{equation}

\noindent where $\vartheta_{\rm f}$ is a reference temperature below $\vartheta _{\rm  s}$ ($r_{\rm f}> s_{\rm }$). For a single power law  $r_{\rm f}=R_{\rm cut}$. We get then a pair of functions that we express in terms of the variable $\tau=\vartheta^{5/2} _{\rm  s}=\left(1-{ s_{\rm }}/{R_{\rm p}}\right)$

\begin{equation}
\mathcal{F_{C,H}}(\tau;\alpha)=\begin{array}{ll}
\frac{5}{1+2\alpha}F_{\rm A1}\left(", 1,\frac{\tau}{2-\tau}\right)-
 \frac{5}{6+2\alpha}\tau F_{\rm A2}\left(",1,\frac{\tau}{2-\tau}\right), \label{MFCH}
\end{array}
\end{equation}

\begin{equation}
\mathcal{F_{C,C}}(\tau,\tau_{\rm f};\alpha)=\begin{array}{ll}
\frac{5}{1+2\alpha}F_{\rm A1}\left(", \frac{\tau_{\rm f}}{\tau},\frac{\tau_{\rm f}}{2-\tau}\right)-
 \frac{5}{6+2\alpha}\tau_{\rm f} F_{\rm A2}\left(",\frac{\tau_{\rm f}}{\tau},\frac{\tau_{\rm f}}{2-\tau}\right).\label{MFCC}
\end{array}
\end{equation}

At $\tau=\tau_{\rm f}$, the two functions above take the same value,  since $\tau_{\rm f}$ is the cut-off value for the  emission with  temperature slope $\alpha$.  By symmetry, the functions in question also take the same value at  the limit $\tau\rightarrow1^-$.  The   simplest way to prove this  is to find the primitive function of the integrand in equation (\ref{Abubev}) in spatial coordinates and carry out a comparison with equation (\ref{SXCpre}). Proceeding this way, one can find that

\begin{equation}
\mathcal{F_{C,H}}(1^-;\alpha)=\mathcal{F_{C,C}}(1^-,\tau_{\rm f};\alpha)=\frac{1}{b_1}=\frac{5}{1+2\alpha}. \label{centralS}
\end{equation}

\begin{figure}
\centering
\includegraphics[height=60mm]{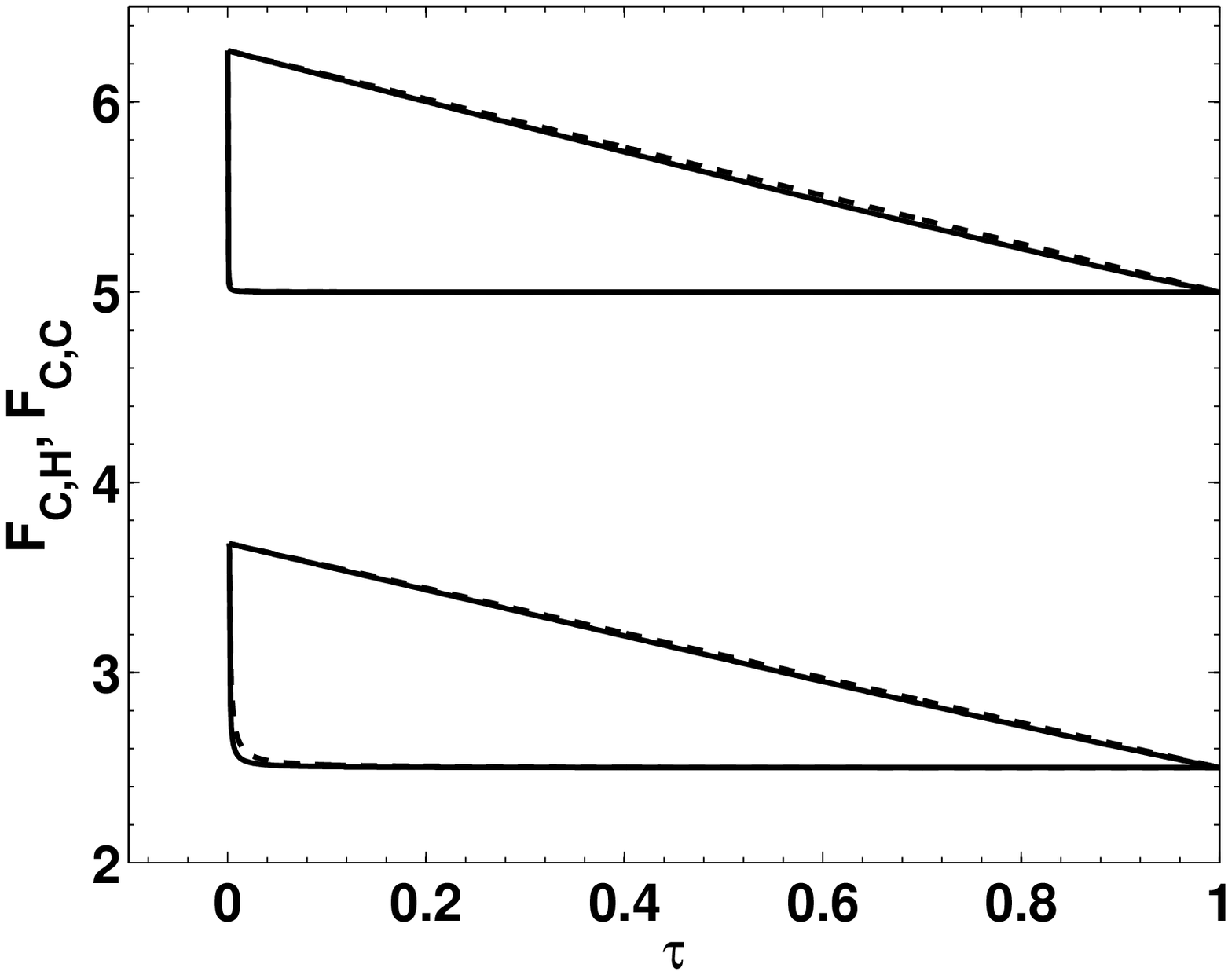}
\caption{Hypergeometric pseudo-triangles. The hypothenuse $\mathcal{F_{C,H}}$ is given by equation (\ref{MFCH}). The function $\mathcal{F_{C,C}}$, that corresponds to  the catheti, is given by equation (\ref{MFCC}). Top:  $T_{\rm Cc}=1.68\times10^7$ K, $\alpha=0$ (Chu et al. 1995 model). Bottom:  $T_{\rm Cc}=6.4\times 10^6$  K, $\alpha={1}/{2}$   (Bremsstrahlung). Solid lines: actual hypergeometric functions. Dashed lines: pseudo-triangle approximations. \label{fig:f13}}
 \end{figure}

Thus, equations (\ref{MFCH}) and (\ref{MFCC}) form a closed path: a pseudo-triangle. This is per se a very interesting property both from the physical and mathematical point of view,  which  complete study  we reserve for a later time. Here,  we give just a  brief discussion that is  general enough to solve the (super-) bubble surface brightness problem. The actual shape of the triangles  depends  on  $\tau_{\rm f}^{2/5}=\vartheta_{\rm f}={T_{\rm f}}/{T_{\rm Cc}}$.  For $T_{\rm Cc} \ga 10 T_{\rm f}$  there is practically no curvature, whereas for $T_{\rm Cc}\approx T_{\rm f}$  the catheti are deformed (Fig. \ref{fig:f13}). For the range of $\alpha$ that we are considering (see Table \ref{Table2}),  a series expansion shows that the hypotenuse is always, for any practical purpose,  a straight line. A convenient point  that one  can use to trace it is $0^+$. There, the second term in (\ref{MFCH}) vanishes. After working with its integrand, the first term on (\ref{MFCH}) can be reduced to  $(1/b_1)\sqrt{\pi}\Gamma(b_1+1){}_1F_{2}[1/2,b; b+1/2;\tau/(2-\tau)]/\Gamma(b_1+1/2)$. At the limit $\tau\rightarrow0^+$,  the previous Gauss hypergeometric function reduces to unity.  As a consequence $\mathcal{F_{C,H}}$ takes the value

\begin{equation}
\mathcal{F_{C,H}}(0;\alpha)= \sqrt{\pi}\frac{\Gamma(\frac{1+2\alpha}{5})}{\Gamma(\frac{1+2\alpha}{5}+\frac{1}{2})}.
\end{equation}

The function $G(x)={\Gamma(x)}/{\Gamma(x+
{1}/{2})}$ can be approximated on the interval $(-{1}/{2}, 1)$ by  the following (1,2) Pad\'e approximant around 0:

\begin{equation}
G(x)=\frac{\Gamma(x)}{\Gamma(x+\frac{1}{2})}\approx\frac{[\pi^2+3\ln(4)^2]x+6\ln(4)}{\sqrt{\pi} x\{[\pi^2 -3 \ln(4)^2]x+6\ln(4)\}}.
\end{equation}

For the case under study, when $x$ falls out of this range, it can always be mapped  back using the relation $\Gamma(x+1)=x\Gamma(x)$.    The equation of the line that traces the hypothenuse is

\begin{equation}
\mathcal{F_{C,H}}(\tau;\alpha)\approx\begin{array}{ll}
 \sqrt{\pi} G\left(\frac{1+2\alpha}{5}\right)+\left[ \frac{5}{1+2\alpha}-\sqrt{\pi} G\left(\frac{1+2\alpha}{5}\right)\right]\tau.
\end{array}
\end{equation}

 \noindent {{The slope of this line,  $m$, just depends on $\alpha$. For $\alpha=0$, it  reduces to $m(0)\approx-{\pi}/{2}=-\mbox{arccsc}(1)=-\mbox{arcsin}(1^{-1})$.  This hints that a secant-like  approximation can be used to obtain an expression for the catheti. Because we have a closed path,  a single secant approximation on $\tau\in[\tau_{\rm f},1]$  just recovers the expression for the hypothenuse. We consider a multi-secant approximation of the form $\sin(\beta)\approx \sin\{[\mathcal{F_{C,C}}-5/(1+2\alpha)]/(1-\tau)\}=g(\tau)$, where $g(\tau)$  accounts for  the variable angle $\beta$ that the   catethi  subtend with the negative direction of the $\tau$-axis.  Consistency requires the closure of the path to be conserved; thus, it is obligatory that   $g(\tau_{f})= \sin(-m)$. Taking into account the catheti dependence on $\tau_{\rm f}/\tau$,  $g(\tau)$ can be written as $g(\tau)=h(\tau_{\rm f}/\tau)\sin(-m)$, where $h(\tau_{\rm f}/\tau)$ is a function that satisfies $h(1)=1$. Many of such functions exist,  for instance, any of the form   $h(\tau_{\rm f}/\tau)/h(1)$. Equation (\ref{MFCC}) diverges as $\sim \tau_{\rm f}/\tau$ as $\tau\rightarrow 0$,  so we use $h(\tau_{\rm f}/\tau)=\tau_{\rm f}/\tau$. This is equivalent to a first order approximation. It suffices for the present discussion. Thus, the expression for the catheti is }}

\begin{equation}
\mathcal{F_{C,C}}(\tau,\tau_{\rm f};\alpha)\approx\left\{\begin{array}{ll}
\frac{5}{1+2\alpha} + (1-\tau)\mbox{arcsin}\left[\frac{\tau_{\rm f}}{\tau}\sin(-m)\right], 
\mbox{if  abs$(m) \le \frac{\pi}{2}$;}\\
\frac{5}{1+2\alpha}+(1-\tau)
\left\{ \frac{\tau_{\rm f}}{\tau}\pi-\mbox{arcsin}\left[\frac{\tau_{\rm f}}{\tau}\sin(-m)\right]\right\}, 
\mbox{otherwise;} \label{catheti}
\end{array}\right.
\end{equation}

\noindent where the last case separation is necessary to  take the correct value  of the arcsin function, specially when machine-evaluated. For a single power law emissivity, equations (\ref{SXCpre})-(\ref{sigmaC1a}) and the expressions for the hypotenuse and the catheti give the next approximation for the spatial profile  as a function of  the normalised temperature

\begin{equation}
\begin{array}{ll}
 \Theta_{\triangle}(\vartheta_{\rm s},\alpha;\xi_{\rm f},\xi_{\rm i})=(2\vartheta_{\rm s}^{\frac{5}{2}}-\vartheta_{\rm s}^{5})^{-1/2} 
 \left[ \xi^{\frac{1}{2}+\alpha} \mathcal{F_{C,C}}(\vartheta_{\rm s}^{5/2},\xi^{5/2};\alpha)\right]_{\xi_{\rm i}= \vartheta_{\rm cut}}^{\xi_{\rm f}=\min(\vartheta_{\rm s},\vartheta_{R_{\rm s}})} , \label{Pro1}
 \end{array}
\end{equation}

\noindent where the dummy variable $\xi$ is meant to be evaluated at the indicated limits as the result of an integral would and $\min(\vartheta _{\rm  s},\vartheta_{R_{\rm s}})$ brings   the reverse shock position back into the game.  In terms of  the 
projected radius, the profile  is

\begin{equation}
\begin{array}{ll}
 S_{\rm C\triangle}(s,\alpha;\xi_{\rm f},\xi_{\rm i})= \left[1-\left(\frac{s}{R_{\rm p}}\right)^2\right]^{-1/2}
\left[ (1-\xi)^{\frac{1+2\alpha}{5}} \mathcal{F_{C,C}}[1-s,1-\xi;\alpha]\right]_{\xi_{\rm i}=\frac{s_{\rm cut}}{R_{\rm p}}}^{\xi_{\rm f}=\max(\frac{s}{R_{\rm p}},\frac{ R_{\rm s}}{R_{\rm p}})}. &   \label{Pro3B}
\end{array}
\end{equation}

\begin{figure*}
\centering
\includegraphics[height=60mm]{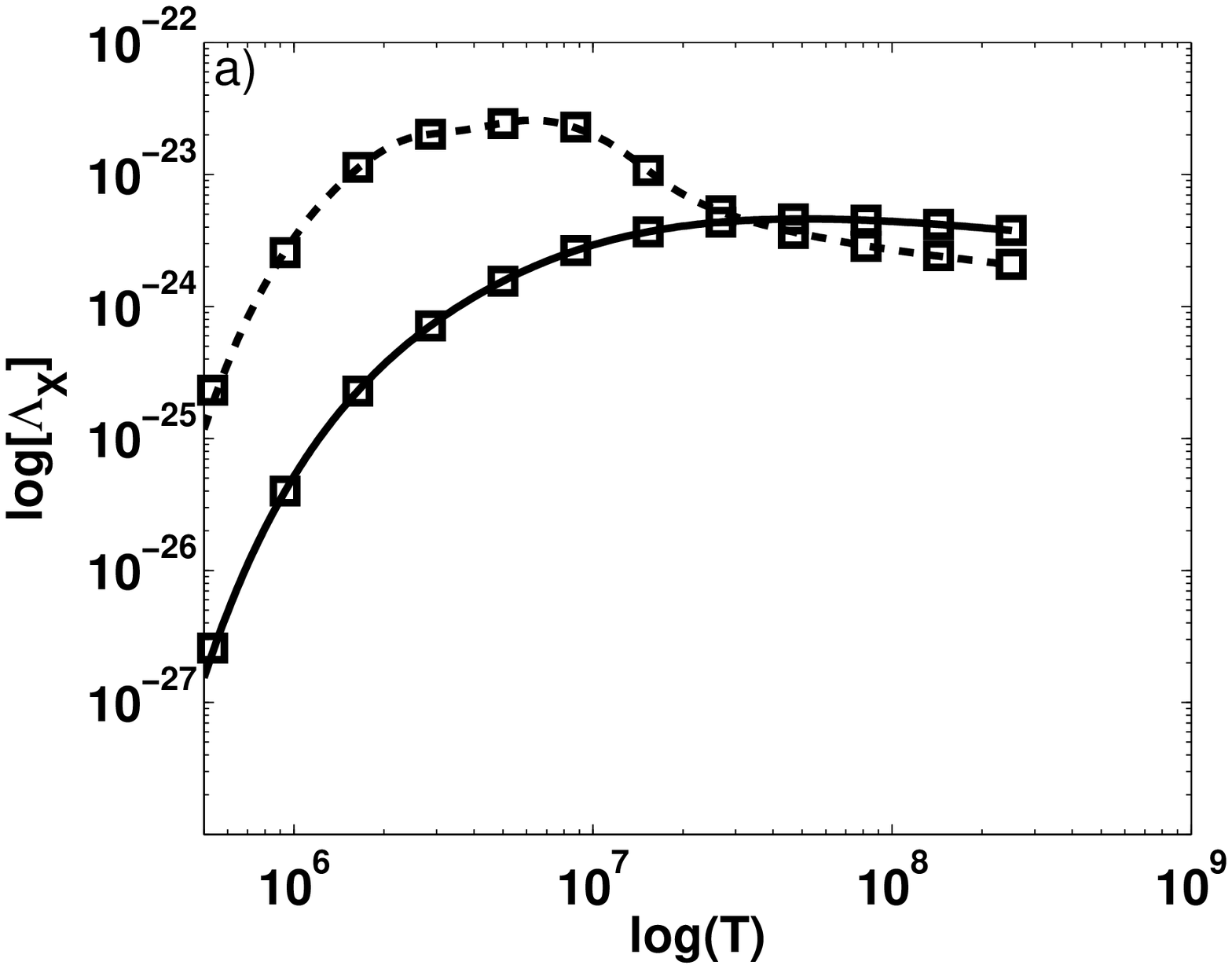}
\hfill
\includegraphics[height=60mm]{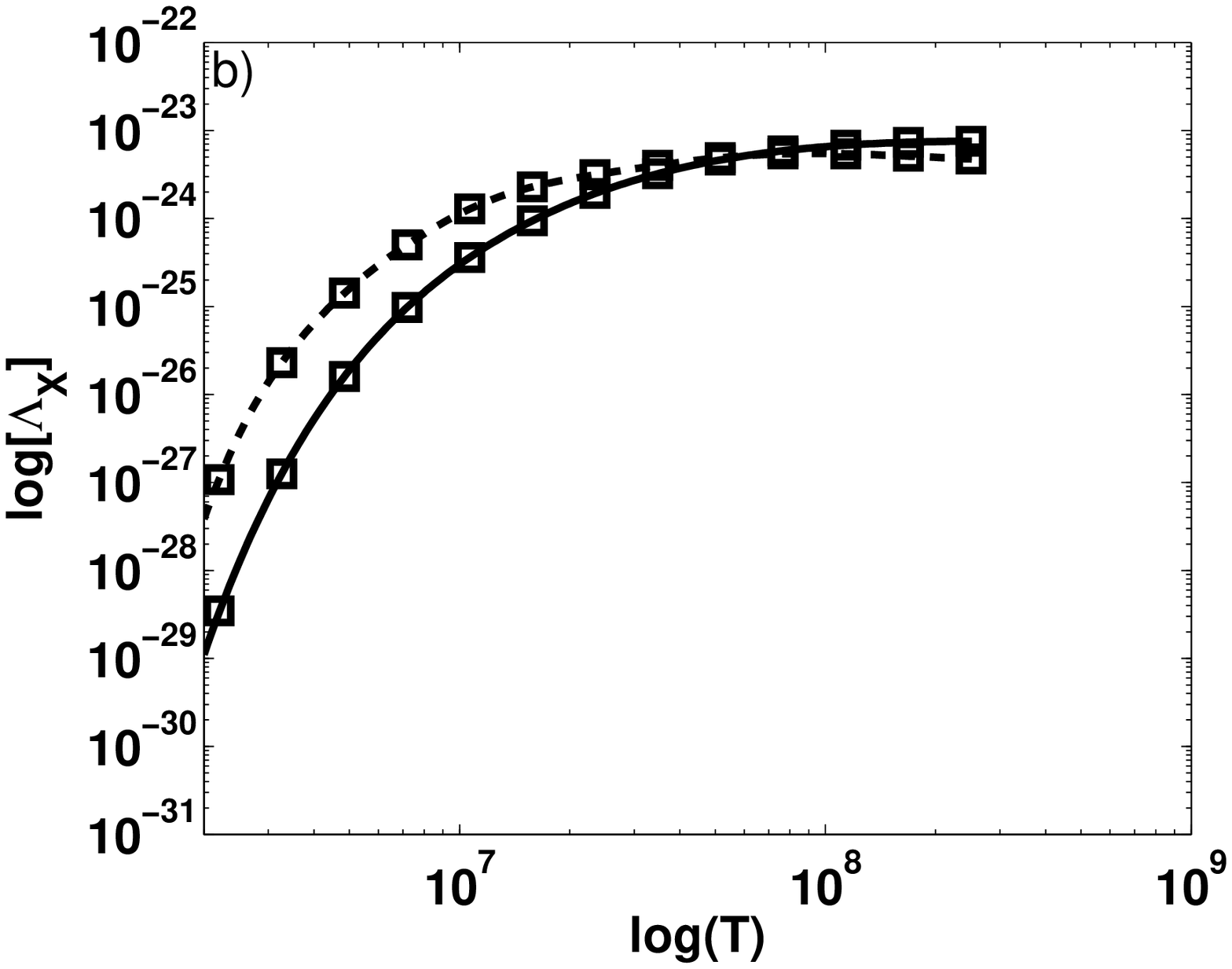}\\
\vfill
\includegraphics[height=60mm]{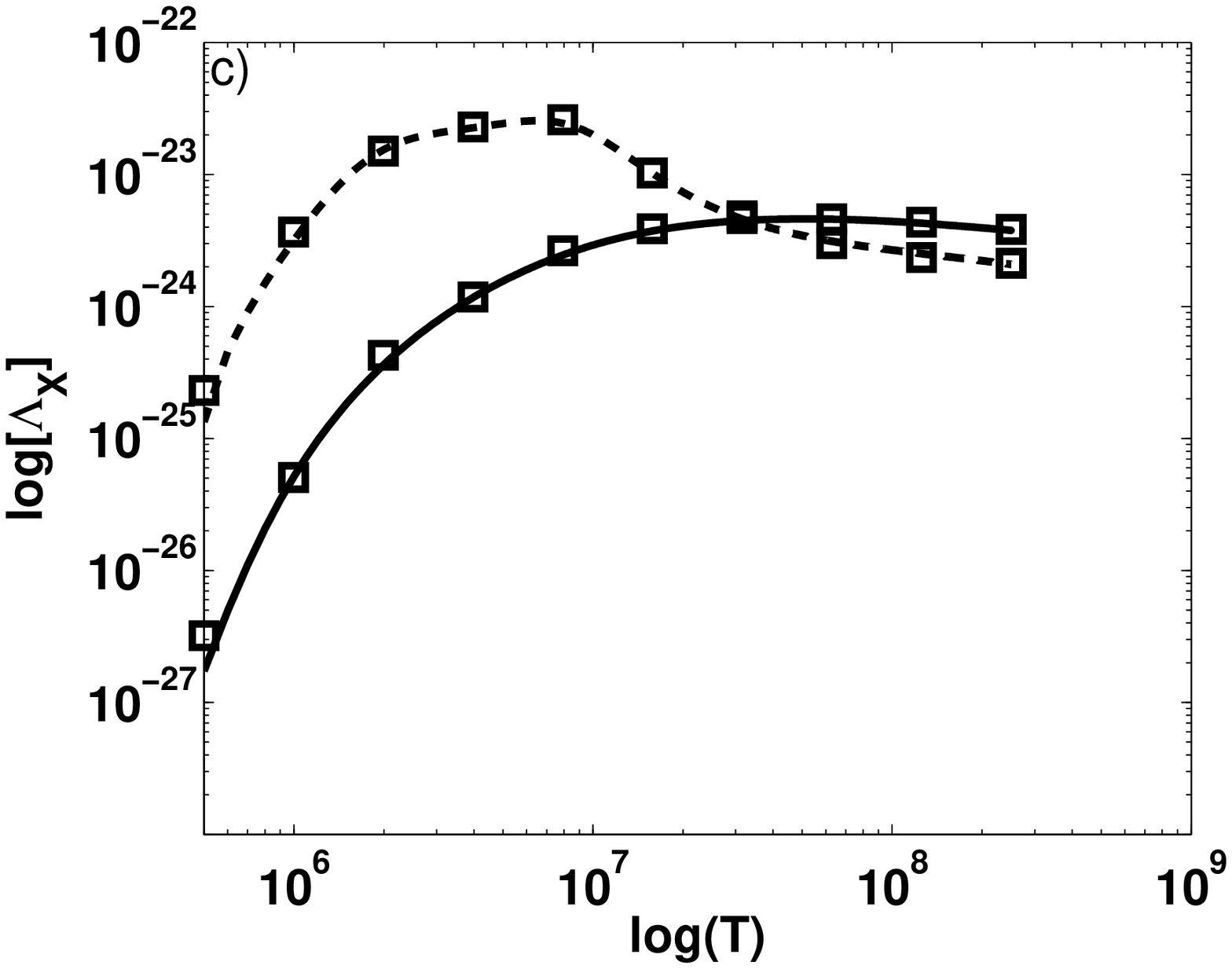}
\hfill
\includegraphics[height=60mm]{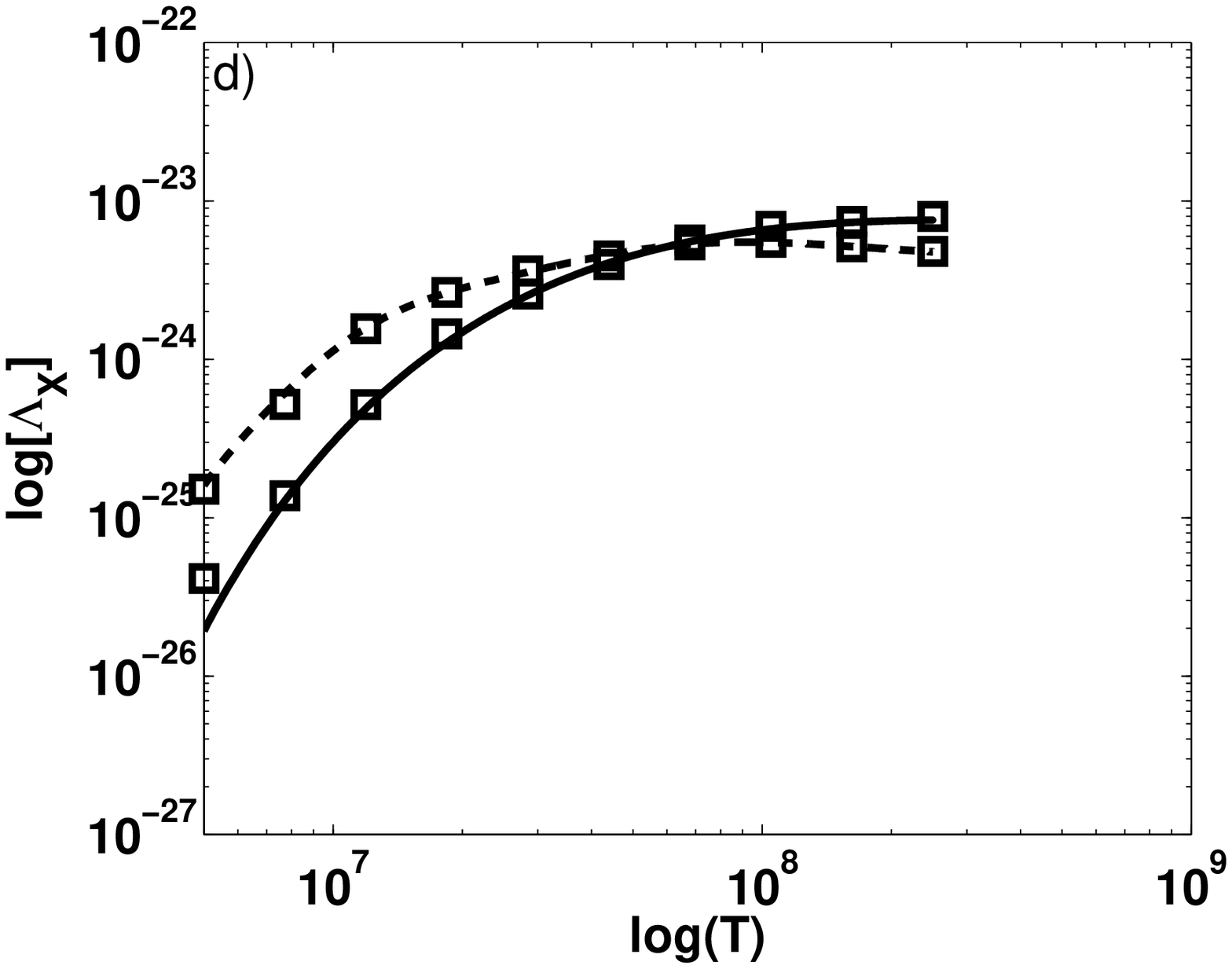}
\caption{  Comparison of the SS00 X-ray emissivity tables for hydrogen (solid lines) and metals (dashed lines) with  the fitted approximations (squares).  Top panels: rational approximations. Bottom panels: power law approximations. Panels (a) and (c): emissivities in the soft band. Panels  (b) and (d): emissivities in the hard band.     \label{fig:figD1}}
\end{figure*}

The expression that corresponds to  a realistic emissivity (see Fig. \ref{fig:figD1} and  Table \ref{Table2})  parameterized as piecewise continuous power laws is obtained from equation (\ref{sigmaC2})

\begin{equation}
 \sigma_{\rm C}(\vartheta_{\rm s},t)=\left\{\begin{array}{ll}
A_{\rm C}(t) \displaystyle{ \sum_{\alpha, \vartheta_{\rm i},\vartheta_{\rm f}<\vartheta_{R_{\rm s}}}^{\vartheta_{\rm f}=\vartheta_{\rm cut}} }\lambda_\alpha 
\Theta_{\triangle}[\vartheta_{\rm s},\alpha; \min(\vartheta_{\rm i},\vartheta_{R_{\rm s}} ),\vartheta_{\rm f}], & \mbox{if  $s \le R_{\rm s}$;}\\
A_{\rm C}(t) \displaystyle{ \sum_{ \
 \alpha, \vartheta_{\rm i},\vartheta_{\rm f}<\vartheta_{\rm s}}^{\vartheta_{\rm f}=\vartheta_{\rm cut}} }\lambda_\alpha 
\Theta_{\triangle}[\vartheta_{\rm s},\alpha;  \min(\vartheta_{\rm s},\vartheta_{\rm i} ), \vartheta_{\rm f}],   & \mbox{otherwise;} \label{Pro2}
\end{array}\right.
\end{equation}

\noindent or, in terms of $ s_{\rm }$

\begin{equation}
\sigma_{\rm C}( s_{\rm },t)=\left\{\begin{array}{ll}
 A_{\rm C}(t) \displaystyle{ \sum_{\alpha,  s_{\rm i}, s_{\rm f}>R_{\rm s}}^{ s_{\rm f}=R_{\rm cut}} }\lambda_\alpha
 S_{\rm C\triangle}[s_{\rm },\alpha; \max( s_{\rm i},R_{\rm s}),  s_{\rm f}],   & \mbox{if  $ s_{\rm } \le R_{\rm s}$;}\\
A_{\rm C}(t) \displaystyle{ \sum_{ \
 \alpha, s_{\rm i}, s_{\rm f}> s_{\rm }}^{ s_{\rm f}=R_{\rm cut}} }\lambda_\alpha
 S_{\rm C\triangle}[s_{\rm },\alpha;\max( s_{\rm }, s_{\rm i} ),  s_{\rm f}], & \mbox{otherwise.}\label{Pro4}
\end{array}\right.
\end{equation}

These formulae give  the surface  brightness  spatial profile for all evaporative bubble models. They  can be also used to calculate the weighted temperature, as we show in Section \ref{PTC}.

\section{AUXILIARY TABLES AND FIGURES}

\begin{table*} 
\begin{minipage}{175mm}
\caption{LIST OF RELEVANT SYMBOLS}
\begin{tabular}{@{}lllllll}
\hline
Symbol (page) &  Meaning& Symbol (page) &  Meaning\\
 \hline
 
indexes 'A', 'B' \& 'C'  & properties  at regions A,B and C, respectively & index 'X' & a property in the X-ray band \\
index '$\rm sc$' & values  at $r=R_{\rm sc}$ & index '$\rm c$' & central values \\
index 'H' & contribution from hydrogen& index 'M' & contribution from metals\\
index '7' & temperature in units of $10^7$ K & $\Pi_*$&  $\Pi_*=\{R_{\rm sc},q_{\rm e},\{q_{\rm m},V_{\infty P}\}\}$\\
$A$ (6)& surface brightness  amplitude & $\dot E  (3)$ & mechanical luminosity\\
$\dot E_{38}$ (2,5) &$\dot E$ in units of $10^{38}$ erg s$^{-1}$&  $F$  (18)& Gauss hypergeometric function ${}_2F_1$ \\
$F_{\rm A}$  (19)& Appell hypergeometric function & $k$ (3) & Boltzmann constant \\ 
  $K$  (3)& polytropic proportionality constant& $L_{\rm X}$ (5)& cumulative X-ray luminosity  \\
 $\dot M (3)$ & mass deposition rate & $n$  (5) &   particle number density \\
  $n_0$ (9)& ISM particle number density &$P (3)$ & pressure \\
$q_{\rm m}$ (3) & mass deposition rate per unit volume  &$q_{\rm e} (3)$ & energy deposition rate per unit volume & \\
$ r $ (3)& radial coordinate & $r_{\rm d} (6)$  &  angular diameter distance\\
$R$ (3)&  a normalised radius: $R=r/R_{\rm sc}$ & $R_{\rm sc} (3)$  & star cluster radius \\
$R_{\rm sc,pc}$  (2,5)& star cluster radius in parsecs& $ R_{\rm cut }$  (6) & X-ray cut-off radius\\
$R_{\rm p}$ (9) & principal shock position & $R_{\rm s}$ (9)& reverse shock position\\
$ s_{\rm }$ (6,7) & projected radius &$S$ (10,21)& surface brightness spatial profile\\
$t$  (9)& time& $t_6$(2,9)& time in Myr \\
$T$  (3)& temperature&$T_{\rm X}$ (7) & X-ray weighted temperature \\
$T_{\rm keV}$ (4,5) & superwind central temperature in  keV &$T_{\rm Rs}$ (10)& temperature at $R_{\rm s}^+$ \\
 $T_{\rm cut}$ (6)& X-ray cut-off temperature & $T_{\rm spec}$ (7) &  X-ray  spectroscopic temperature\\
$u$  (3)&  velocity & $U$ (3) & squared velocity: $u^2$ \\
$V_{\rm \infty A} (3)$ &  adiabatic terminal  speed  & $V_{\rm \infty P}$ (3) & polytropic terminal speed\\
$V_{8}$ (2,5)& $V_{\rm \infty P}$ in units of 1000 km s$^{-1}$  &  $V_{\rm sc}$ (3)& star cluster volume \\
$X$  (5)& X-ray luminosity scaling factor& $Z$ (5)& metallicity \\
 $\alpha$ (5) & X-ray emissivity temperature slope &    $\gamma$ (3)  & adiabatic index\\
 $\Gamma$ (18)&  Gamma function &$\triangle$ (10,21)& pseudo-triangle approximation \\
  $\epsilon$  (3) &  total energy per unit mass & $\epsilon_{\rm BP}  (3)$& region B $\epsilon$ in the polytropic case\\
  $\eta$ (3)& polytropic index & $\lambda_{\alpha}$ (10,20) & weighting X-ray emissivity factor\\
 $\Lambda_{\rm X}$ (5)& X-ray emissivity function &  $\Lambda_{\rm X,-23} (2 )$&$\Lambda_{\rm X}$ in units of $10^{-23}$  erg s$^{-1}$ cm$^3$\\
 $\Lambda_{\rm \alpha}$ (5)&   X-ray emissivity  proportionality constant ($\Lambda_{\rm X} =\sum \Lambda_{\alpha} T^{\alpha}$) & $\overline \Lambda_{\rm X} (9)$ & evaporatively averaged $\Lambda_{\rm X}$ \\
 $\mu_{\rm i}$  (4)& mean mass per particle for a ionized gas & $\mu_{\rm n}$ (5) & mean mass per particle for a neutral gas\\
$\rho$ (3) & density &$\sigma (6)$  & X-ray surface brightness \\
$\vartheta$ (9)& region C normalised temperature & $\vartheta_{\rm cut}$ (9,20)& region C normalised $T_{\rm cut}$\\
$\theta$  (6)& object subtended angle &$\Theta$ (20)& surface brightness profile in terms of $\vartheta$\\
 $\tau$  (20)&  $\vartheta^{5/2}$&$\mathcal{F}$  (5,19,20)& combination of hypergeometric functions\\

 \hline
\end{tabular}
 \medskip {\\ The page in which  each symbol appears for the first time together with pages that contain  important related formulae   are indicated between parenthesis. The list is not exhaustive, but provides the definition of  all relevant symbols. A description of the most used indexes is also given. The physical properties of each region (as  they are defined in Fig. \ref{fig:fig0}) are labeled with the subindexes A, B and C, except when no place for confusion exist. The same applies for the subindex X, which designates an X-ray property. } \label{Table4}

\end{minipage}

\end{table*}

\begin{table*}
\begin{minipage}{175mm}
\caption{Fitted  X-Ray Emissivity Function - POWER LAWS}
%\begin{tabular}{@{}llllllc}
\begin{tabular}{@{}lllllll}
\hline
 BAND & Component &   Temperature  &  Parameters &  Fitting Details\\
 && ($\times10^7$ K) && &\\
 \hline
 
 SOFT &Hydrogen&  0.05-0.1 &$\Lambda_0=0,\, \Lambda_{\rm \alpha}= 5.12\times10^{-27},\,\alpha=4$ & $R^2= 0.991,\, e_{\rm RMS}=0.0014$  \\
 (0.3-2.0 keV )&...  &  0.1-0.75 &  $ \Lambda_0=-0.0329, \,\Lambda_{\rm \alpha}=0.3802\times10^{-7},\, \alpha=1 $  & $R^2=0.998, \, e_{\rm RMS}=0.003049$  \\
... & ...    & 0.75-3.5  &  $\Lambda_0=0.590,\, \Lambda_{\rm \alpha}=-18700 , \alpha= -0.69$ & $R^2=0.9993, \, e_{\rm RMS}=0.001659$\\
... &...& 3.5-25 &  $\Lambda_0=0.488, \, \Lambda_{\rm \alpha}= -4.15\times 10^{-10},\,\alpha=1$ & $R^2=0.9802, e_{\rm RMS}=0.003738$\\
... & Metals &0.05-0.1&$\Lambda_0=0,\, \Lambda_{\rm \alpha}= 3.70\times10^{-25},\,\alpha=4$ & $R^2= 0.995,\, e_{\rm RMS}=0.007$\\
...   & ... & 0.10-0.75 &   $\Lambda_0=4.09,\, \Lambda_{\rm \alpha}=-5630, \alpha=-0.53$ & $R^2=0.987,\, e_{\rm RMS}=0.090$ \\
... &   ... &  0.75-25 &  $\Lambda_0=0.201,\, \Lambda_{\rm \alpha}=1.01\times 10^{11}, \alpha=-1.54$ & $R^2=0.9993,\,e_{\rm RMS}=0.0176$\\

HARD    \\
(2.0-8.0 keV )  & Hydrogen & 0.05-0.5 & $\approx0$ \\
...&...& 0.5-1.1 &  $\Lambda_0=0,\,\Lambda_{\rm \alpha}=1.2\times10^{-21}, \, \alpha=2.767$ & $R^2=0.9961,\, e_{\rm RMS} =0.0008925$  \\
...&...& 1.1-8.5 &  $\Lambda_0= -0.365,\, \Lambda_{\rm \alpha}=2.238\times10^{-4}, \, \alpha=0.4621$ & $R^2=0.9944,\, e_{\rm RMS} =0.0143$\\
...&...& 8.5-25 & $\Lambda_0=0,\,\Lambda_{\rm \alpha}=0.0318, \, \alpha=0.1667$ & $R^2=0.9274,\, e_{\rm RMS} =0.01013$\\
...& Metals &  0.05-0.5 & $\approx0$ \\
...& ...&  0.5-1  &  $\Lambda_0=0,\,\Lambda_{\rm \alpha}=1.54\times10^{-21}, \, \alpha=2.835$ & $R^2=0.9951,\, e_{\rm RMS} =0.00194$\\
...& ...&  1-8.5  &$\Lambda_0= 1.208,\, \Lambda_{\rm \alpha}=-63.1674, \, \alpha=-0.2513$ & $R^2=0.9951,\, e_{\rm RMS} =0.00194$\\
...& ... & 8.5-25 &  $\Lambda_0=0,\,\Lambda_{\rm \alpha}=8.7793, \, \alpha=-0.1502$ & $R^2=0.9799,\, e_{\rm RMS}=0.003425$\\  

BOTH${\star\star}$  &  Hydrogen & $ \ga$ 0.9 & $\Lambda_{\rm H}=1.23\times10^{-4},\,\alpha_{H}=\frac{1}{2}$&$R^2=0.96,\, e_{\rm RMS}=0.04$\\
...& Metals & ... &  $\Lambda_{M1}=4.49\times10^{17},\,\alpha_{M1}=-\frac{5}{2}$&$R^2=0.98,\, e_{\rm RMS}=0.04$\\
&&&$\Lambda_{M2}=0.79,\,\alpha_{M2}=0$\\
\hline
\end{tabular}
\medskip {\\
 Power laws  fitted to the SS00 emissivity tables. On each temperature interval, a normalised X-ray emissivity of the form  $\Lambda_{\rm X,-23} =\Lambda_0+\Lambda_{\rm \alpha} T^\alpha$ was fitted.  For the metals, the values of $\Lambda_0$ and $\Lambda_{\alpha}$ scale linearly with $Z$. The mean square errors correspond to the dimensionless  $\Lambda_{\rm X,-23}$, thus they are also in units of $10^{-23}$ erg s$^{-1}$ cm$^3$. The curves were fitted considering   piecewise continuity as a requirement.}

\label{Table2}
\end{minipage}

\end{table*}

\begin{table*} 
\begin{minipage}{175mm}
\caption{Fitted X-Ray Emissivity Function - RATIONAL FUNCTIONS}
\begin{tabular}{@{}lllllll}
\hline
 BAND & Component &   Temperature  &  Fitted $\Lambda_{\rm X,-23}$ &    Fitting Details\\
&&($\times10^7$ K)&&\\
 \hline
 SOFT &Hydrogen&  0.05-25&  $\frac{0.2442T_{7}^5 +10.13T_{7}^4+7.54T_{7}^3-1.079T_{7}^2+0.05112T_{7}-0.00083}{T_{7}^5+17.91T_{7}^4+21.08T_{7}^3+16.68T_{7}^2+1.063T_{7}+0.1009}$  & $e_{\rm RMS}=1.40\times 10^{-6}$  \\
 (0.3-2.0 keV )&Metals &  0.05-25 & $\frac{21.57T_{7}^4-1.817T_{7}^3+38.26T_{7}^2-0.7704T_{7}-0.04085}{T_{7}^5+83.31T_{7}^4-129.7T_{7}^3+91.13T_{7}^2-19.09T_{7}+1.953}$   & $ e_{\rm RMS}=3.40\times 10^{-6}$  \\

 HARD & Hydrogen& 0.2-0.5& $\frac{0.07841T_{7}^5+0.2146T_{7}^4-0.1441T_{7}^3+0.02784T_{7}^2-0.00113T_{7}-0.0001026}{T_{7}^5+0.891T_{7}^4+0.2492T_{7}^3+1.052T_{7}^2+1.475T_{7}+1.06}$& $ e_{\rm RMS}=3.46\times10^{-6}$ \\
 (2.0-8.0 keV) &...&0.5-25&$\frac{0.5726T_{7}^4+40.17T_{7}^3-46.13T_{7}^2+19.07T_{7}-2.812}{T_{7}^4+39.6T_{7}^3+114.8T_{7}^2+224.1T_{7}-17.12}$ & $  e_{\rm RMS}=1.09\times 10^{-4}$ \\
 ...&Metals& 0.2-0.5& $\frac{0.2084T_{7}^3-0.1051T_{7}^2+0.01765T_{7}-0.0009808}{T_{7}^2-0.16T_{7}+0.3066}$ &  $ e_{\rm RMS}=3.57\times 10^{-6}$\\
 ...&...&0.5-25&$\frac{0.3816T_{7}^5+1.457T_{7}^4+6.059T_{7}^3-0.1862T_{7}^2+3.877T_{7}-1.478}{T_{7}^5-2.751T_{7}^4+32.31T_{7}^3+31.5T_{7}^2-81.11T_{7}+108.5}$ & $ e_{\rm RMS}=2.30\times 10^{-6}$ \\
 
\hline
\end{tabular}
 \medskip {\\ Emissivity function used for numerical calculations. For $T_7<0.2$ the emission on the hard band is  negligible  since it is several orders of magnitude ($\ge 4$) smaller than the emission at the cut-off temperature  in the soft band. Computationally, the evaluation of the rational approximations is faster than interpolating from the tables.   For all cases $R^2\approx 1$ within the used precision.  The mean square errors correspond to the dimensionless  $\Lambda_{\rm X,-23}$. In order to  obtain this table, we associate to each  temperature $T_{\rm i}$ in the SS00 tables  a function $e_{\rm i}=\frac{1}{2}(\Lambda_{\rm X} (T_{\rm i})- R_{jk}(\{a\},\{b\};T_{\rm i})^2$, where $R_{\rm jk}$ is a $(j,k)$ Pad\'e approximant.   Then we  defined ${{E}}={{E}}(e_1,e,2,...)$ and  solved the associated non-linear least-squares problem $\min || E||^2\equiv$ $\min(\sum_{\rm i}  e_{\rm i}^2)$  using  the trust region method  described by Byrd, Schnabel \& Shultz (1988) and Dennis \& Schnabel (1996).  A similar process was carried out for   Table \ref{Table2}.  In the present case, the values of $j$ and $k$ were successively  increased.  The most compact approximations with $R^2>0.9$ and the lowest RMS errors were selected in the end.}  
 
  \label{Table3}

\end{minipage}

\end{table*}

\bsp

\label{lastpage}

\end{document}